\documentclass{aa}

\usepackage{graphicx}
\usepackage[varg]{txfonts}
\usepackage{lastpage}
\usepackage{lipsum}
\usepackage{physics}
\usepackage{amsmath, amssymb}
\usepackage{dsfont}
\usepackage{bbm}
\usepackage{xcolor}
\usepackage{ulem}
\usepackage{xpatch}
\usepackage{import}
\usepackage{makeidx}
\usepackage{changes}
\usepackage{xifthen}
\usepackage{ifthen} 
\usepackage[shortlabels]{enumitem}

\usepackage{color}
\usepackage{natbib,twoopt}
\usepackage[hyphenbreaks]{breakurl}
\usepackage[breaklinks]{hyperref}

\bibpunct{(}{)}{;}{a}{}{,}
\hypersetup{
  colorlinks,
  citecolor=blue,
  linkcolor=blue,
  urlcolor=blue,
}

\newcommand{\app}[1]{App.~\ref{#1}}
\newcommand{\sref}[1]{Sec.~\ref{#1}}
\newcommand{\srefs}[2]{Secs.~\ref{#1}~-~\ref{#2}}
\newcommand{\tab}[1]{Table~\ref{#1}}
\newcommand{\fig}[1]{Fig.~\ref{#1}}
\newcommand{\equ}[1]{Eq.~(\ref{#1})}

\newcommand{\equs}[2]{Eqs.~(\ref{#1})~-~(\ref{#2})}
\newcommand{\bbar}[1]{\bar{{\bar{#1}}}}
\newcommand{\colout}[1]{\bgroup\markoverwith{\textcolor{#1}{\rule[.5ex]{2pt}{0.4pt}}}\ULon}
\newcommand{\cs}{c_{\sf s}}
\newcommand{\Hp}{H_\mathrm{p}}

\newcommand{\Tgas}{T_\mathrm{gas}}
\newcommand{\zB}{z_\mathrm{B}}
\newcommand{\LX}{L_\mathrm{X}}
\newcommand{\xe}{x_\mathrm{e}}
\newcommand{\xeth}{x_\mathrm{e,th}}
\newcommand{\xenth}{x_\mathrm{e,nth}}
\newcommand{\Brs}{B_\mathrm{r}^\mathrm{s}}
\newcommand{\Bphis}{B_\mathrm{\varphi}^\mathrm{s}}
\newcommand{\etaO}{\eta_\mathrm{O}}
\newcommand{\etaA}{\eta_\mathrm{A}}

\makeatletter
\renewcommand*\aa@pageof{, page \thepage{} of \pageref*{LastPage}}
\newcommand{\pder}[2][]{\frac{\partial#1}{\partial#2}}

\newcommandtwoopt{\citeads}[3][][]{\href{http://adsabs.harvard.edu/abs/#3}%
{\def\hyper@linkstart##1##2{}%
 \let\hyper@linkend\@empty\citealp[#1][#2]{#3}}}
\newcommandtwoopt{\citepads}[3][][]{\href{http://adsabs.harvard.edu/abs/#3}%
{\def\hyper@linkstart##1##2{}%
 \let\hyper@linkend\@empty\citep[#1][#2]{#3}}}
\newcommandtwoopt{\citetads}[3][][]{\href{http://adsabs.harvard.edu/abs/#3}%
{\def\hyper@linkstart##1##2{}%
 \let\hyper@linkend\@empty\citet[#1][#2]{#3}}}
\newcommandtwoopt{\citeyearads}[3][][]%
{\href{http://adsabs.harvard.edu/abs/#3}
{\def\hyper@linkstart##1##2{}%
 \let\hyper@linkend\@empty\citeyear[#1][#2]{#3}}}

\makeatother

\newcommand{\Br}[1][]{%
  \ifthenelse{\isempty{#1}}%
    {B_\mathrm{r}}% if #1 is empty
    {B_\mathrm{r,#1}}% if #1 is not empty
}

\newcommand{\Bphi}[1][]{%
  \ifthenelse{\isempty{#1}}%
    {B_\mathrm{\varphi}}% if #1 is empty
    {B_\mathrm{\varphi,#1}}% if #1 is not empty
}

\newcommand{\Bz}[1][]{%
  \ifthenelse{\isempty{#1}}%
    {B_\mathrm{z}}% if #1 is empty
    {B_\mathrm{z,#1}}% if #1 is not empty
}

\newcommand{\ur}[1][]{%
  \ifthenelse{\isempty{#1}}%
    {u_\mathrm{r}}% if #1 is empty
    {u_\mathrm{r,#1}}% if #1 is not empty
}
\newcommand{\uphi}[1][]{%
  \ifthenelse{\isempty{#1}}%
    {u_\mathrm{\varphi}}% if #1 is empty
    {u_\mathrm{\varphi,#1}}% if #1 is not empty
}

\newcommand{\psif}[1][]{%
  \ifthenelse{\isempty{#1}}%
    {\psi}% if #1 is empty
    {\psi_\mathrm{#1}}% if #1 is not empty
}

\makeindex

\begin{document}

\titlerunning{Protoplanetary disks around magnetized young stars with large-scale magnetic fields}
\authorrunning{D.~Steiner et al.}
\title{Protoplanetary disks around magnetized young stars with large-scale magnetic fields I: Steady-state solutions.   } 

\author{
 D.~Steiner\inst{1},
 L.~Gehrig\inst{1},
 M.~Güdel\inst{1}
}
\institute{
 Department of Astrophysics, University of Vienna,
 Türkenschanzstrasse 17, A-1180 Vienna
}

\date{Received ....; accepted ....}

\abstract
{
Describing the large-scale field topology of protoplanetary disks (PPD) faces significant difficulties and uncertainties.
The transport of the large-scale field inside the disk plays an important role in understanding its evolution.
}
{
We aim to improve our understanding of the dependencies that stellar magnetic fields pose on the large-scale field.
We focus on the innermost disk region ($\lesssim 0.1$~AU), which is crucial for understanding the long-term disk evolution. 
}
{
We present a novel approach combining the evolution of a 1+1D hydrodynamic disk with a large-scale magnetic field, consisting of a stellar dipole truncating the disk and a fossil field. 
The magnetic flux transport includes advection and diffusion due to laminar, non-ideal MHD effects, such as Ohmic and ambipolar diffusion.
Due to the implicit nature of the numerical method, long-term simulations (in the order of several viscous timescales) are feasible.
}
{
The large-scale magnetic field topology in stationary models shows a distinct dependence on specific parameters.
The innermost disk region is strongly affected by the stellar rotation period and magnetic field strength.
The outer disk regions are affected by the X-ray luminosity and the fossil field.
Varying the mass flow through the disk affects the large-scale disk field throughout its radial extent.
}
{
The topology of the large-scale disk field is affected by several stellar and disk parameters.
This will affect the efficiency of MHD outflows, which depend on the magnetic field topology.
Such outflows might originate from the very inner disk region, the dead zone, or the outer disk. In subsequent studies, we will use these models as a starting point for conducting long-term evolution simulations of the disk and large-scale field on scales of $\sim 10^6$~years to investigate the combined evolution of the disk, the magnetic field topology, and the resulting MHD outflows.
}


\maketitle

\section{Introduction}
\label{sec:intro}

Protoplanetary disks (PPDs) do not evolve in isolation from their environment. Among other parameters, magnetic fields play a crucial role in understanding the evolution of PPDs and can be present in two disk regions: (i) Strong stellar magnetic fields from the forming, most likely partly- or fully-convective protostar interacting with the very inner disk, and (ii) large-scale magnetic fields threading the whole disk \citep[cf. e.g.][]{ohashi2018,tang2023}, whose origin is subject of current research. Either the field is generated inside the disk by some dynamo effect \citep[e.g.][]{reyes04,brandenburg1995} or the field is already present during the collapse of the protostellar cloud core \citep[e.g.][]{dudorov2014}. The stellar magnetic field as well as the large-scale disk field strongly influence the dynamics of an evolving PPD on a macroscopic scale \citep[cf. e.g.][]{Ferreira06,Bai2013,simon2013}: Angular momentum can be transferred between protostar and disk through torques of the stellar field acting on the inner disk parts. The details of such a coupled star-disk system are complex and involve the inclusion of many parameters, for example, the abundance and efficiency of accretion-powered stellar winds (APSW) \citep[cf.][]{gehrig2023spin}. Additionally, the position of a magnetically truncated inner disk radius \citep[e.g.,][]{hartmann16, Steiner21} or the launch of magnetocentrifugal winds (MCWs) \citep[cf.][]{blandford1982} are also dependent on the large-scale field. Apart from the macroscopic influence, the development of the magneto-rotational instability (MRI), which in some disk regions is believed to play a non-negligible role in driving accretion \citep[][]{balbus91}, relies on the abundance of an initial large-scale seed field. The strength of the large-scale field is also believed to be important for determining the efficiency of MRI in PPDs \citep[e.g.,][]{baistone2011}. One approach to investigate how MRI is altered by a magnetic field is to assume maximally efficient MRI in a steady-state disk, which then determines the magnetic field strength for every radius in the disk \citep[e.g.][]{mohanty2018,delage2022}.

The dipole field strengths of classical T Tauri stars (CTTS) range between 500 G and 1.5 kG \citep[e.g.][]{johnskrull1999, Johnstone14} and are therefore strong enough to disrupt the disk at a certain radius, the so-called magnetic truncation radius $r_\mathrm{trunc}$ \citep[cf.][]{hartmann16}. Although the details of magnetospheric accretion remain very complex to investigate \citep[e.g.][]{Romanova02}, it has been shown by \cite{bessolaz08} that under the assumption of a dipolar stellar magnetic field, the disk is magnetically truncated very close to the equipartition of magnetic pressure and thermal gas pressure \citep[for detailed simulations of the inner disk movement during disk evolution, cf. e.g.][]{Steiner21}. 
Actual observations of large-scale magnetic fields threading a PPD remain challenging, however magnetic fields have been detected on virtually every observationally accessible scale, including fields abundant in molecular clouds \citep[e.g.][]{crutcher2019}, which indicate that large-scale magnetic fields are likely to be present in a newly forming protostellar core and the resulting PPD. Recent observations of outflows and jets originating from the disk \citep[e.g.][]{lee2020} as well as polarization mapping of the surroundings of T Tauri stars \citep[e.g.][]{bertrang2017, brauer2017} further strengthen the assumption of large-scale magnetic fields being abundant in PPDs. Additionally, the observation and high-precision measurement of the Zeeman effect in various emission lines has led to the calculation of an upper limit for some PPDs in the order of $\sim 10$~mG for both poloidal and toroidal field at a $r \approx 40$ AU \citep[e.g.][]{lankhaar2023}. 

The interaction between the disk and the large-scale field is complex. 
Early 1+1D studies found that an initially purely vertical fossil field threading a thin disk is not dragged inwards efficiently due to outward-directed diffusion counterbalancing inward-directed magnetic advection very easily \citep[cf. e.g.][]{Lubow94}.
Taking the vertical velocity profile of the disk into account has modified this result such that a large-scale field can also be dragged inwards efficiently in a thin disk due to a larger radial mass transport and higher ionization near the disk surface. The corresponding magnetic resistivity is then adequately averaged in the vertical direction to represent this enhanced magnetic flux transport \cite{Guilet2012, Guilet2014}. 
However, the source of resistivity is still a subject of investigation. Earlier studies have used turbulent resistivity to describe the non-ideal effects \citep[cf. e.g.][]{Lubow94,Guilet2012,Guilet2014}, but recent studies strongly indicate that PPDs are laminar almost everywhere but the innermost disk region \citep[cf. e.g.][]{dudorov2014, mohanty2018, delage2022}. Laminar, non-ideal diffusion can either be caused by ohmic diffusion (OD), ambipolar diffusion (AD), or the Hall effect (HE).

Apart from 1+1D models, global, long-term evolution models have been conducted, including magnetic fields in 2+1D \citep[e.g.][]{vorobyov2020}. Such models have to deal with the numerical difficulties of conducting MHD simulations over an extended period of time. Especially the incorporation of magnetic diffusion poses challenges to the numerical stability of a scheme. Furthermore, such models are limited with respect to the position of the inner disk boundary, since the Courant-Friedrichs-Levy (CFL) condition \citep[][]{Courant28} severely limits the time-step achievable in a disk simulation inside of $\sim 1$ AU. 
The model of \cite{vorobyov2020} incorporates large-scale magnetic fields, but without considering non-ideal magnetic diffusion due to numerical challenges, which leads to unusually strong magnetic advection and field amplification in their models.

Previous models that study the influence of a large-scale magnetic field fall short in modeling the following cases: (i) Long-term simulations of a disk and large-scale magnetic field including the inner disk, (ii) the effects of a stellar dipole together with a large-scale field on the inner disk and the topology especially in the inner disk, (iii) describing the effects of a MCW on the long-term evolution of a PPD. In this work we aim to investigate the evolution of a large-scale field towards a stationary state for a disk, which is truncated by a stellar dipole. It has been shown that the inner disk treatment is crucial in terms of long-term evolution \citep[e.g.,][]{Vorobyov19,Steiner21,Gehrig2022}. Therefore, a combined treatment of both stellar- and disk components has the potential to alter the disk properties in the inner disk, compared to the common assumption of a torque-free inner disk boundary. A self-consistent evolution of disk and large-scale magnetic field offers the possibility to study the disk field topology and strength not through assuming maximum efficient MRI \citep[cf.][]{mohanty2018,delage2022} but through inward-dragging and amplification of a more physically motivated background fossil field. Furthermore, our simulations are not confined to steady-state models and allow us to follow the disk field evolution over time. 
In this work, we introduce our new model and focus on the validation of our method by comparison with current work \citep[e.g.][]{dudorov2014,mohanty2018,delage2022}. Therefore, we will focus on steady-state models.
By combining the stellar and disk field components, we can study the effects of various parameters on the large-scale disk field, such as the stellar rotation period, magnetic field strength, X-ray luminosity, as well as the accretion rate (\sref{sec:results}).
These results form the basis for investigating MCWs self-consistently with a hydrodynamic accretion disk in the subsequent studies of this series.

\section{Model description}
\label{sec:model_description}

The construction of a long-term evolution simulation of a PPD including a stellar magnetic field and a large-scale magnetic field involves a model describing the hydrodynamic flow of the disk while also covering the temperature structure of the disk and the influence of magnetic fields (cf. \sref{sec:basic_equations}).

\subsection{Basic equations}
\label{sec:basic_equations}

We use TAPIR for our 1+1D simulations \citep[][]{ragossnig20,Steiner21}, which solves the hydrodynamic equations using a fully implicit time integration scheme, hence the CFL condition does not restrict our timestep \footnote{We discuss the timestep-control in our models in \sref{sec:numerical_details}.}. This allows us to also include the very inner disk in our long-term hydrodynamic simulations, which is necessary to incorporate the torque of the stellar dipole and pressure gradients in our model \citep[cf.][]{Steiner21}. The vertical disk extent is assumed to be hydrostatic, allowing for a vertical integration of the hydrodynamic equations \equs{eq:cont}{eq:ene} \citep[analogously to e.g.][]{vorobyov2020,Steiner21}. The effects of a magnetic field $\vec B$ threading the disk is included by adding the non-ideal induction equation \equ{eq:induction}, 
\begin{alignat}{2}
    & \pder[\Sigma]{t} \,  &&+ \nabla \cdot ( \Sigma \, \vec u ) = 0\;, \label{eq:cont} \\
    & \pder[(\Sigma \, \vec u)]{t} &&+ \nabla \cdot (\Sigma \, \vec u : \vec u) - \frac{B_\mathrm{z} \vec B}{2 \pi}\nonumber \\
        & &&+ \nabla P_\mathrm{gas} + \nabla \cdot Q + \Sigma \, \nabla \psi + H_\mathrm{p} \, \nabla \left( \frac{B_\mathrm{z}^2}{4 \pi} \right) = 0 \;, \label{eq:mot} \\
    &\pder[(\Sigma \, e)]{t}  &&+ \nabla \cdot (\Sigma \, \vec u \, e ) + P_\mathrm{gas} \, \nabla \cdot  \vec u \nonumber \\
    & &&+ Q : \nabla \vec u - 4 \pi \, \Sigma \, \kappa_\mathrm{R} \, \left(J - S \right) + \dot E_\mathrm{rad} = 0 \;, \label{eq:ene} \\
    & \pder[\vec B]{t} \, &&- \nabla \times ( \vec u \times \, \vec B ) +  \nabla \times ( \eta_\mathrm{O} \vec J ) \nonumber \\
    & &&-  \nabla \times \left(\frac{\eta_\mathrm{A}}{|\vec B|^2} \vec J \times \vec B \times \vec B \right) = 0\;, \label{eq:induction}
\end{alignat}
where $\Sigma$, $e$ and $\vec u \equiv (u_\mathrm{r}, u_\mathrm{\varphi}, 0)^\mathrm{T}$ denote current density, specific inner energy density and the gas velocity restricted to the planar direction. We use an ideal equation of state (EOS), for which the gas pressure can be written as $P_\mathrm{gas}= \Sigma \, e \, (\gamma - 1)$ with an adiabatic coefficient $\gamma = 7/5$ \citep[e.g.][]{vorobyov2020} and which allows for the definition of the pressure scale height $H_\mathrm{p}$. The terms including the viscous pressure tensor $Q$ describe the viscous torque enabling radial angular momentum transport and the viscous heating in \equ{eq:mot} and \equ{eq:ene}, respectively. The disk is heated and cooled by stellar irradiation and radiative cooling through its surface on both sides of the disk, respectively. Additionally, vertical radiative transport is included in the diffusion limit by assuming local thermal equilibrium (LTE) in vertical direction. Heating, cooling and vertical radiative transport are described by the net heating/cooling rate per unit surface in \equ{eq:ene}, $\dot E_\mathrm{rad}$. In our model radiative transport in radial direction is also included approximately in the diffusion limit, which is described by the term  $\Sigma \, \kappa_\mathrm{R} \, \left(J - S \right)$, where $J$ and $S$ denote the zeroth moment of the radiation field and and the source function, respectively \citep[for a detailed explanation see][]{ragossnig20}.

The viscous-$\alpha$ is implemented analogously to \cite{Steiner21}. We adopt the layered-viscosity model of \cite{gammie96}, which reads as:
\begin{equation}
    \alpha = \alpha_\mathrm{base} + \alpha_\mathrm{surf} + \alpha_\mathrm{deep} \,, \label{eq:alpha}
\end{equation}
where $\alpha_\mathrm{base}$, $\alpha_\mathrm{surf}$ and $\alpha_\mathrm{deep}$ denote the base viscosity in the dead zone (DZ), the viscosity due to active MRI in the disk's surface layer, and the $\alpha$-viscosity in the deeper disk layers. The surface layer is defined on each side of the disk with a thickness corresponding to a column density of $\Sigma_0 = 100 \, \mathrm{g \, cm^2}$ (measured from the disk surface inwards). If $\Sigma > \Sigma_0$, non-thermal ionization cannot penetrate deeper into the disk and a deep layer with a different viscosity $\alpha_\mathrm{deep}$ is assumed,
\begin{align}
    \alpha_\mathrm{surf}(r) &= \alpha_\mathrm{MRI,eff} \left[ s_\Sigma \, \frac{\Sigma_\mathrm{0}}{\Sigma(r)}+(1-s_\Sigma) \right] \;, \\
    \alpha_\mathrm{deep}(r) &= \alpha_\mathrm{MRI,eff} \, s_\Sigma \left(1 - \frac{\Sigma_\mathrm{0}}{\Sigma(r)} \right) s\left( T_0 \right) \;, \\
    s_\mathrm{\Sigma} &= \begin{cases}
        1 & \forall \quad \Sigma \ge \Sigma_0 \\
        0 & \mathrm{otherwise}
    \end{cases} \;,
\end{align}
where $\alpha_\mathrm{MRI,eff}$ denotes the effective $\alpha$ due to MRI. For all radii $r > r_\mathrm{DZ,out}$ (where $r_\mathrm{DZ,out}$ denotes the outer DZ boundary) MRI operates with reduced efficiency due to ambipolar diffusion \citep[cf.][]{delage2022}. We model this effect by applying a reduced $\alpha_\mathrm{AD} < \alpha_\mathrm{MRI}$ in the region $r > r_\mathrm{DZ.out}$ (the values for $\alpha_\mathrm{MRI}$, $\alpha_\mathrm{AD}$ and $\alpha_\mathrm{base}$ used in our reference model are listed in \tab{tab:ref_model})
\begin{align}
    \alpha_\mathrm{MRI,eff} &= \begin{cases}
        \alpha_\mathrm{MRI} & \forall \quad r < r_\mathrm{DZ,out} \\
        \alpha_\mathrm{AD} & \forall \quad r \ge r_\mathrm{DZ,out}
    \end{cases} \;.
\end{align}
The factor $s(T_0)$ determines if the deep layer becomes MRI-active \citep[cf.][]{Steiner21}, which would lead to episodic accretion events \citep[cf. e.g.][]{Steiner21,cecil2024}. In this work the development of episodic accretion is suppressed by setting $s(T_0) = 0$, which is due to the focus on stationary disks in this work (a stationary disk does not have any time-dependent features like episodic accretion).

The net heating/cooling rate $\dot E_\mathrm{rad}$ is written as follows,
\begin{align}
    \dot E_\mathrm{rad} = \Gamma - \Lambda  \;,
\end{align}
where $\Gamma$ and $\Lambda$ describe the heating and cooling rate per unit surface, respectively. The heating rate is a combination of both viscous heating and stellar irradiation on both sides of the disk surface. Viscous heating of the gas happens mostly close to the midplane, where $\rho$ is largest. The corresponding inner energy $e$ is transported vertically up to the disk surface, yielding a temperature $T_\mathrm{surf}$ at the disk surface, if LTE and an optically thick disk is assumed ($\tau \gg 1)$. With the additional assumption of the source function $S$ being constant over the vertical extent of the disk, $T_\mathrm{surf}$ can be written as \citep[e.g.][]{dong2016,ragossnig20},
\begin{align}
    T_\mathrm{surf}^4 =  \frac{8 \, \tau_\mathrm{P} \, \sigma_\mathrm{SB}}{1 + 2\, \tau_\mathrm{P} + \frac{3}{2} \, \tau_\mathrm{R} \, \tau_\mathrm{P}} \left( T_\mathrm{gas,0}^4 + T_\mathrm{amb}^4 + \frac{L_\star}{4 \, \pi \, r^2} f_\mathrm{irr} \, \sin \alpha_\mathrm{irr} \right) \,. \label{eq:surface_temperature}
\end{align}
Here $\tau_\mathrm{R} = \kappa_\mathrm{R} \, \rho_0 \, H_\mathrm{p} = \kappa_\mathrm{R} \, \Sigma \, / \sqrt{2}$ and $\tau_\mathrm{P} = \kappa_\mathrm{P} \, \Sigma \, / \sqrt{2}$ describe the Rosseland and Planck mean opacity, respectively. $\kappa_\mathrm{R}$ has two contributions: (i) a gaseous component $\kappa_\mathrm{R,gas}$ using the method of \cite{ferguson05} with the protostellar abundances of \cite{asplund2021}, and (ii) a dust-dominated component for low temperatures $\kappa_\mathrm{R,dust}$ ($\Tgas < 500$~K). Then $\kappa_\mathrm{R} = \kappa_\mathrm{R,gas} + f_\mathrm{dust} \, \kappa_\mathrm{R,dust}$, where $f_\mathrm{dust}$ denotes the gas-to-dust ratio \citep[cf.][]{pollack85}. The Planck mean opacity is approximated with $\kappa_\mathrm{P} = 2.4 \, \kappa_\mathrm{R}$ and is based on \cite{nakamoto1994,hueso2005}. The Stefan-Boltzmann constant is denoted as $\sigma_\mathrm{SB}$, and $T_\mathrm{amb}$ is the ambient temperature of the interstellar medium (ISM). The parameters $f_\mathrm{irr}$ and $\alpha_\mathrm{irr}$ describe which percentage of the stellar irradiation is reflected at the surface and the impact angle of the irradiation at the surface (with respect to the midplane). 
We note that we do not include the heating of the disk surface due to stellar X-ray irradiation \citep[cf,][and also \sref{sec:model_limitations}]{guedel2015}.
Then $\Gamma$ and $\Lambda$ can be written as \citep[cf.][]{vorobyov2020},
\begin{align}
    \Gamma &= \frac{8 \, \tau_\mathrm{P} \, \sigma_\mathrm{SB}}{1 + 2\, \tau_\mathrm{P} + \frac{3}{2} \, \tau_\mathrm{R} \, \tau_\mathrm{P}} \left( \frac{L_\star}{4 \, \pi \, r^2} f_\mathrm{irr} \, \sin \alpha_\mathrm{irr} + T_\mathrm{amb}^4 \right) \;, \\
    \Lambda &=  \frac{8 \, \tau_\mathrm{P} \, \sigma_\mathrm{SB}}{1 + 2\, \tau_\mathrm{P} + \frac{3}{2} \, \tau_\mathrm{R} \, \tau_\mathrm{P}} T_\mathrm{gas,0}^4 \;.
\end{align}
The stellar magnetic field at the disk surface \citep[which in our case is chosen to be the magnetic surface $\zB$, cf.][explained in detail in \sref{sec:large_scale_disk_field}]{Guilet2012} is modeled as a dipole with the components in cylindrical coordinates $\vec B_\mathrm{\star} = (B_\mathrm{r,\star},B_\mathrm{\varphi,\star},B_\mathrm{z,\star})^\mathrm{T}$ \citep[analogously to][but without ignoring the radial component of the dipole]{Steiner21,gehrig2023spin},
\begin{alignat}{2}
    &B_\mathrm{z,\star}(r,z) &&= B_\star \, R_\star^3 \frac{r^2 - 2\, z^2}{\left(\sqrt{r^2 + z^2}\right)^5} \;, \label{eq:stellar_Bz} \\
    &B_\mathrm{r,\star}(r,z) &&= - B_\star \, R_\star^3 \frac{3 \, r \, z}{\left(\sqrt{r^2 + z^2}\right)^5} \;, \label{eq:stellar_Br} \\
    & B_\mathrm{\varphi,\star} &&\simeq - \alpha_\mathrm{cor} \, B_\mathrm{z,\star}(r,z) \left[ 1- \left(\frac{\Omega_\star}{\Omega(r)}\right)^{\alpha_\mathrm{cor}} \right] \;, \label{eq:stellar_Bphi} \\
        & \alpha_\mathrm{cor} &&\equiv
    \begin{cases}
      +1 & \text{if $r < r_\mathrm{cor}$} \\
      -1 & \text{if $r > r_\mathrm{cor}$} \\
    \end{cases} \nonumber \;, \\
\end{alignat}
where $B_\mathrm{\star}$ denotes the magnetic field strength at the stellar surface and $R_\mathrm{\star}$ the stellar radius. $\alpha_\mathrm{cor}$ is a parameter which ensures that $\mathrm{max(|B_\mathrm{\varphi,\star}|)} \simeq  |B_\mathrm{z,\star}|$. This occurs very close to the star and at a certain distance outside of the corotation radius $r_\mathrm{cor}$ and is physically motivated by the assumption that a stronger winding of the dipole would lead to buoyancy and reconnection of $B_\mathrm{\varphi,\star}$ \citep[cf.][]{Steiner21,gehrig2023spin}.

\subsection{Large-scale magnetic disk field}
\label{sec:large_scale_disk_field}

In this study we make the following assumptions about the large-scale disk magnetic field:
\begin{enumerate}[(i)]
    \item The magnetic field is formally averaged in azimuthal direction and temporally over the dynamical timescale at a certain radius, which means that the large-scale field in our model is axisymmetric. Additionally, the averaging results in only the macroscopic net field being considered and the microscopic fluctuations, which eventually lead to turbulence and to MRI, being modeled in the usual way using the isotropic $\alpha$-viscosity \citep[cf.][]{shakura73} and the turbulent resistivity $\eta_\mathrm{t}$ \citep[e.g.][]{Guilet2012,Guilet2014}.
    \item We do not include a toroidal field component  $B_\mathrm{\varphi}$ in the disk exterior (above the disk surface) in our model to avoid the complexity of starting a MHD wind. Hence, $B_\mathrm{\varphi}^\mathrm{s}$ \footnote{The superscript "s" means that we take the value of the corresponding physical quantity at the disk surface.} is set to zero by means of a Dirichlet boundary condition $B_\mathrm{\varphi}^\mathrm{s} = 0$ \citep[for an in-depth analysis of the possible boundary conditions for $B_\mathrm{\varphi}^\mathrm{s}$ cf.][]{khaibrakhmanov22}. $B_\mathrm{\varphi}^\mathrm{s}$ acts as an outer boundary at the disk surface for a toroidal field inside the disk. This means that $B_\mathrm{\varphi}$ inside the disk can be non-vanishing due to shearing of $\Br$ caused by differential rotation, as long as it is connected to the exterior field at the disk surface (cf. \sref{sec:vertical_vel_profile} for a more detailed treatment). This causes additional diffusion and therefore contributes to the (vertically averaged) magnetic flux transport velocity $\bbar u_\mathrm{\psi}$ \citep[cf.][and \equ{eq:Bphi_internal} below]{Guilet2012,leung2019}.
    \item The magnetic field is assumed to be stationary in vertical direction, compared to its slow evolution in radial direction. This is due to the assumption of a geometrically thin disk ($H_\mathrm{p}/r \ll 1$) \citep[e.g.][for a discussion of the difference between viscous and vertical dynamical timescales in terms of an asymptotic expansion of the basic MHD equations, cf. also \sref{sec:validity_appr}]{Guilet2012}.
    \item No dynamo effect, which might develop in the disk \citep[e.g.][]{reyes04}, is considered in this work.
\end{enumerate}
The poloidal, large-scale magnetic field $\vec B = (B_\mathrm{r}, 0, B_\mathrm{z})^\mathrm{T}$ can be represented by the magnetic flux $\psi$, 
\begin{align}
    \vec B = \nabla \psi \times \vec e_\mathrm{\varphi} \,, \label{eq:def_magnetic_flux}
\end{align}
where $e_\mathrm{\varphi}$ describes the azimuthal unit vector in cylindrical coordinates \citep[cf.][]{Guilet2014}. The induction equation \equ{eq:induction} can then be written as follows \citep[cf.][]{ogilvie2001},
\begin{align}
    &\pder[\psi(r,z)]{t} + \vec u \cdot \nabla \psi(r,z) = \eta(r,z) \, r^2 \, \nabla \cdot \left( \frac{1}{r^2} \nabla \psi(r,z) \right) \label{eq:magn_flux_height_dependent}  \,, \\
    &\eta(r,z) = \eta_\mathrm{O}(r,z) + \eta_\mathrm{A}(r,z) + \eta_\mathrm{t}(r,z) \label{eq:resistivity_combined}\;,
\end{align}
where $\eta$ is the combination of all resistivity mechanisms considered in this work, i.e., turbulent resistivity, OD, AD, which are denoted by $\eta_\mathrm{t}$, $\eta_\mathrm{O}$ and $\eta_\mathrm{A}$, respectively. We note that $\eta_\mathrm{A}$ in \equ{eq:resistivity_combined} only denotes the 'Ohm-like' contribution. AD also has a 'Hall-like' contribution, which is neglected in this work due to a vanishing toroidal field in the disk exterior above the surface. This in turn causes the 'Hall-like' contribution to vanish \citep[cf.][]{leung2019}. The assumption of a thin disk allows for splitting $\psi(r,z)$ into a radial contribution $\psi_0(r)$ and a small contribution, which describes the vertical dependence $\psi_1(r,z)$ and for which $|\psi_1| \ll |\psi_0|$ holds,
\begin{align}
    \psi(r,z) &= \psi_0(r) + \psi_1(r,z) \,, \\
    \Bz &= \frac{1}{r} \pder[\psi_0]{r} = \Bz[d] + \Bz[\star] \,, \label{eq:definition_Bz} \\
    \Br &= -\frac{1}{r} \pder[\psi_1]{z} = \Br[d] + \Br[\star]\,, \label{eq:definition_Br}
\end{align}
where $\Br$ and $\Bz$ denote the combined stellar and large-scale fields in radial and vertical direction, respectively. Our simulations are done in 1+1D; however \equ{eq:magn_flux_height_dependent} is height-dependent and can be evaluated at any $z$ above the midplane, which makes it necessary to average \equ{eq:magn_flux_height_dependent} vertically. Analogously to \cite{ogilvie2001,Guilet2014} a sensible choice is a conductivity-weighted vertical average up to the magnetic surface $\zB$ (for the remainder of this work also referenced as disk surface), which is the height above the disk where magnetic and thermal gas pressure are equal (or equivalently, where the plasma beta $\beta(z = \zB) = 1$). In the case of a hydrostatically stratified disk $\zB$ reads as,
\begin{align}
    \zB &= \Hp \, \sqrt{\ln \left( \frac{2}{\pi} \beta_0 \right)} \,, \\
    \beta_0 &\equiv \frac{8\, \pi \, P_\mathrm{gas}}{B_\mathrm{z}^2} \,,
\end{align}
where $\beta_0$ and $z_\mathrm{B}$ describe the plasma beta at the disk midplane and the magnetic surface, respectively. Averaging this way results in $\bbar \eta$ being dominated by vertical disk layers of high conductivity (low resistivity) and takes into account that the magnetic field is transported inwards more effectively in vertical disk layers of high conductivity. Applying the conductivity-weighted vertical averaging to $\eta$ and the radial gas velocity $u_\mathrm{r}$ yields,
\begin{align}
    \frac{1}{\bar{\bar \eta}(r)} &\equiv \frac{1}{\zB} \int_0^{\zB} \frac{1}{\eta(r,z)} \mathrm{dz} \nonumber \\
    &= \frac{1}{\zB} \int_0^{\zB} \frac{1}{\eta_\mathrm{O}(r,z) + \eta_\mathrm{A}(r,z) + \eta_\mathrm{t}(r,z)} \mathrm{dz} \,, \label{eq:def_averaged_resistivity} \\
    \bar{\bar u}_\mathrm{r}(r) &\equiv \frac{\bar{ \bar \eta}(r)}{\zB} \int_0^{\zB} \frac{u_\mathrm{r}(r, z)}{\eta(r,z)} \mathrm{dz} \,, \label{eq:def_averaged_gas_velocity}    
\end{align}
where the double-bar notation denotes conductivity-weighted averaged quantities. Integrating \equ{eq:magn_flux_height_dependent} vertically between $[-\zB, \zB]$ and applying \equs{eq:definition_Bz}{eq:definition_Br} as well as using \equs{eq:def_averaged_resistivity}{eq:def_averaged_gas_velocity}, we can rewrite the magnetic flux transport equation \equ{eq:magn_flux_height_dependent} in a height-independent form \citep[strictly following][]{Guilet2014},
\begin{align}
    &\pder[\psi_\mathrm{0}(r)]{t} + r \, \bbar u_\mathrm{\psi} \, B_\mathrm{z} = 0  \label{eq:magn_flux_averaged}  \,, \\
    &\bbar u_\mathrm{\psi} = \bbar u_\mathrm{r} + \frac{\bbar \eta}{\zB} \frac{1}{\Bz} \left( \Br(r,z=\zB^-) - \zB \pder[B_\mathrm{z}]{r}  \right) \,, \label{eq:magn_flux_velocity}
\end{align}
where $\zB^-$ denotes the vertical height at the surface, but approached from inside the disk (analogously $\zB^+$ denotes the magnetic surface approached from above the disk). This is important for modeling the connection at the disk surface between an external magnetic field and the large-scale field inside the disk \sref{sec:vertical_vel_profile} \citep[cf.][]{Guilet2014}. The vertical averaging performed in \equ{eq:magn_flux_velocity} has no description of the vertical dependency of $B_\mathrm{r}$ inside the disk, but only its value at the surface  $\Br(r,z=\zB^-)$. The determination of this external field is not trivial and is described in \sref{sec:external_magnetic_field}. Additionally, averaging $\bbar u_\mathrm{r}$ and $\bbar \eta$ requires to know the vertical profiles of $u_\mathrm{r}(r,z)$ and $\eta(r, z)$. Assuming $\beta \gg 1$ inside the disk and a constant viscous $\alpha$ in vertical direction allows us to find an analytic expression for the vertical velocity profiles \citep[cf.][]{Guilet2012,leung2019}, which we briefly summarize in \sref{sec:vertical_vel_profile}). For the vertical profile of $\eta$ we need the ionization fraction $x_\mathrm{e}$, which can be caused by thermal ionization or by various non-thermal ionization sources like X-rays and cosmic rays. We will discuss all ionization sources included in our model in \sref{sec:non_ideal_resistivity}.

\subsection{External magnetic field}
\label{sec:external_magnetic_field}

The radial magnetic field at the disk surface $B_\mathrm{r}^\mathrm{s}$ is connected to the exterior field above and below the disk, which is a superposition of (i) a fossil field $B_\infty$ formally created by currents at infinity, (ii) the stellar dipole field $\vec B_\star$ and (iii) the field created by purely toroidal (due to our assumption of a poloidal field) currents $J_\varphi$ in the disk. Due to the thin-disk approximation used in our model we only need to describe the radial part of the large-scale field's flux function $\psi_\mathrm{0}$ (cf.~\equ{eq:magn_flux_averaged}). The three magnetic flux contributions from the fossil field, stellar dipole, and current density $J_\varphi$ in the disk are denoted by $\psi_\infty$, $\psi_\star$, and $\psi_\mathrm{d}$, respectively,
\begin{align}
    \psi_0 &= \psi_\infty + \psi_{\star} + \psi_\mathrm{d} \,. \label{eq:magn_flux_combined}
\end{align}
The Biot-Savart law is used to find an expression for the large-scale magnetic field and hence for the corresponding magnetic flux $\psi_\mathrm{d}$ of an axisymmetric current distribution $J_\mathrm{\varphi}^\mathrm{s}$ by integrating the field caused by the current loops at every $r$ over the full radial extent of the disk \citep[cf.][]{Jackson1999,Ogilvie1997}, 
\begin{align}
    \psi_\mathrm{d}(r) &= \frac{1}{\pi} \int_{r_\mathrm{in}}^{r_\mathrm{out}} \frac{r \, r'}{r + r'} \left[ \frac{(2 - k^2) \, K(k) - 2\,E(k)}{k^2}  \right] \, J_\mathrm{\varphi}^\mathrm{s}(dr') \, dr' \,, \label{eq:biot_savart} \\
    k^2 &= \frac{4 \, r \, r'}{(r + r')^2} \,, \\
    E(k) &= \int_0^{\pi/2} (1 - k^2 \, \sin^2(x) )^{1/2} \, dx \,, \\
    K(k) &= \int_0^{\pi/2} (1 - k^2 \, \sin^2(x) )^{-1/2} \, dx \,,
\end{align}
where $E(k)$ is the complete elliptic integral of the first kind and $K(k)$ corresponds to the complete elliptic integral of the second kind \citep[cf.][]{Guilet2014}. Applying Stokes' theorem to \equ{eq:def_magnetic_flux} allows for connecting the disk's radial field at the surface $B_\mathrm{r,d}^\mathrm{s}$ to $J_\mathrm{\varphi}^\mathrm{s}$ at the disk surface \citep[cf.][]{Ogilvie1997,Guilet2014},
\begin{align}
    B_\mathrm{r,d}^\mathrm{s}(r) &= 2 \, J_\mathrm{\varphi}^\mathrm{s}(r) \,. \label{eq:stokes}
\end{align}
We note that $B_\mathrm{r,d}^\mathrm{s}$ describes the radial field corresponding to the currents in the disk only, without any contributions from stellar dipole or fossil field.

The magnetic fluxes $\psi_\mathrm{\infty}$ and $\psi_\mathrm{\star}$ are calculated as follows,
\begin{align}
    \psi_\infty &= \frac{r^2 - R_\star^2}{2} B_\infty \,, \label{eq:B_inf} \\
    \psi_{\star,0}  &= B_\star \, R_\star^3 \left(\frac{1}{R_\star} - \frac{1}{r} \right) \;,
\end{align}
which both can be readily calculated, since the stellar dipole field $B_\star$ and the fossil field $B_\infty$ are both parameters of our model. Since \equ{eq:magn_flux_averaged} describes the evolution of $\psi_0$ at a certain time,  we can use \equ{eq:stokes} with \equ{eq:biot_savart} to find a relation between $\psi_\mathrm{d}$ and $B_\mathrm{r,d}^\mathrm{s}$ and then reorganize \equ{eq:magn_flux_combined} as follows,
\begin{align}
    \psi_\mathrm{d} &= \psi_\mathrm{0} - \psi_\infty - \psi_{\star} \,. \label{eq:magn_flux_disk}
\end{align}
Finally, for calculating $B_\mathrm{r,d}^\mathrm{s}$ we interpret the integral of \equ{eq:biot_savart} as a linear operator $\mathcal{L}$, which can then be formally inverted to obtain 
\begin{align}
   B_\mathrm{r,d}^\mathrm{s} = \mathcal{L}^{-1} \, \psi_\mathrm{d} \,. \label{eq:operator_L}
\end{align}
Numerically, $\mathcal{L}^{-1}$ is represented as a matrix describing the relationship between all discretized values of $B_\mathrm{r,d}^\mathrm{s}$ and $\psi_\mathrm{d}$. Its construction and inversion is non-trivial and numerically expensive and is briefly discussed in \app{sec:numerical_details}.

\subsection{Analytical vertical velocity and magnetic field profile}
\label{sec:vertical_vel_profile}

In 1+1D hydrodynamical models usually a density-weighted, vertically averaged radial and rotational velocity, $\bar u_\mathrm{r}(r)$ and $\bar u_\mathrm{\varphi}(r)$, respectively, therefore the actual velocity structure is usually not resolved by assuming a vertically constant velocity $\vec u = (u_\mathrm{r}(r), u_\mathrm{\varphi}(r), 0)^\mathrm{T}$. However, work done by \cite{Guilet2012, leung2019} indicates that radial magnetic flux transport can significantly deviate from mass transport, if the vertical velocity structure is included. This is due to the two different vertical averaging mechanisms for mass and magnetic flux transport (cf.~\sref{sec:large_scale_disk_field} and \equs{eq:def_averaged_resistivity}{eq:def_averaged_gas_velocity}). Using that the vertical dynamical time scale is much shorter than the radial diffusion and viscous time scales, the disk structure in vertical direction appears stationary with respect to viscous evolution, and therefore is treated as such \citep[cf.][]{Guilet2012,dudorov2014,leung2019}.

Strictly following the work of \cite{Guilet2014,leung2019}, the disk can be split into two zones vertically: The disk inside of the magnetic surface $\zB$, where it is assumed that the magnetic field is passive, i.e. $\beta \gg 1$, and a magnetically dominated region outside of $\zB$, where in the absence of outflows the magnetic field can be modeled as force-free. It follows that in the case of a vertically constant $\alpha$ and with a passive large-scale field, the gas velocity profiles are parabolic in vertical direction,
\begin{align}
    u_\mathrm{r} &= u_\mathrm{r,0}(r) + u_\mathrm{r,2}(r) \frac{z^2}{H_\mathrm{p}^2} \,, \label{eq:ur_profile} \\ 
    u_\mathrm{\varphi} &= u_\mathrm{\varphi,0}(r) + u_\mathrm{\varphi,2}(r) \frac{z^2}{H_\mathrm{p}^2} \,,  \label{eq:uphi_profile}
\end{align}
where $u_\mathrm{r,0}$ and $u_\mathrm{\varphi,0}$ denote the radial and toroidal velocities at the disk midplane, respectively, whereas $u_\mathrm{r,2}$ and $u_\mathrm{\varphi,2}$ are functions which describe the vertical dependency of $u_\mathrm{r}$ and $u_\mathrm{\varphi}$. After integration of the vertical, stationary disk profile \citep[cf.][]{Guilet2012},  $u_\mathrm{r,2}$ and  $u_\mathrm{\varphi,2}$ take the following form,
\begin{align}
    u_\mathrm{r,2} &=  \frac{\alpha \, c_\mathrm{s}}{1 + 4 \, \alpha^2} \left( \frac{3 \, H_\mathrm{p}}{r} - 5 \, \pder[H_\mathrm{p}]{r} \right) \,, \label{eq:ur2} \\
    u_\mathrm{\varphi,2} &=  \frac{c_\mathrm{s}}{2} \left( \pder[H_\mathrm{p}]{r} - \frac{3 \, H_\mathrm{p}}{2 \, r} \right) + \frac{\alpha^2 \, c_\mathrm{s}}{1+4 \, \alpha^2} \left(\frac{3 \, H_\mathrm{p}}{r} - 5 \, \pder[H_\mathrm{p}]{r} \right) \,. \label{eq:uphi2}    
\end{align}
We note that \equ{eq:ur_profile} and \equ{eq:uphi_profile} can, strictly speaking, only be written as above for a vertically constant $\alpha$. It has been shown that a more realistic vertical profile of $\alpha$, which is certainly expected in a real-world PPD, has only a minor influence on the large-scale magnetic field in steady-state \citep[cf.][]{Guilet2014}. Therefore the use of those simplified profiles in this work is a viable approximation. However, the diffusion timescale at which the magnetic field evolves may be altered significantly \citep[cf.][]{leung2019}, which makes it necessary to adopt a more realistic vertical velocity profile for future work for cases when the detailed time evolution matters (e.g. the time-dependent long-term disk evolution with an outflow).

The magnetic field profiles $B_\mathrm{r}(r,z)$ and $B_\mathrm{\varphi}(r,z)$ are dependent on the values of the exterior field components $B_\mathrm{r}^\mathrm{s}$ and $B_\mathrm{\varphi}^\mathrm{s}$ at the disk surface. As mentioned above, to avoid the complications of a magnetically induced outflow, no toroidal field at the disk surface is assumed, i.e. $B_\mathrm{\varphi}^\mathrm{s} = 0$, however, inside the disk ($0 \leq z \leq \zB$) a toroidal component can develop \citep[cf.][]{Guilet2012}. The 2-zone model briefly described above has the disadvantage that the intermediate region $\beta \approx 1$ between the passive region $\beta \gg 1$ and the magnetically dominated disk corona $\beta \ll 1$ is infinitesimally small, which leads to a jump between the values of $B_\mathrm{r}(r,z)$ and $B_\mathrm{\varphi}(r,z)$ just below ($z = \zB^-$) and above ($z = \zB^+$) the disk surface, 
\begin{align}
    \Br(z=\zB^-) &= \Br(z=\zB^+) + \Delta \Br \;, \\
    \Bphi(z=\zB^-) &= \Bphi(z=\zB^+) + \Delta \Bphi \;, 
\end{align}
where $\Delta \Br$ and $\Delta \Bphi$ are the jump conditions for the radial and toroidal field, respectively. The boundary conditions for $B_\mathrm{r}$ and $B_\mathrm{\varphi}$ and jump conditions are read as follows \citep[cf.][for an in-depth derivation]{Guilet2012},
\begin{align}
    \Br(\zB^+) &= B_\mathrm{r}^\mathrm{s} + H_\mathrm{p} \, \pder[\Bz]{r} \;, \label{eq:external_Br} \\
    \Delta \Br &= - H_\mathrm{p} \, \pi \, \left. \pder[\Bphi]{z}\right|_{z=\zB^-} \;, \label{eq:jump_condition_Br} \\
    \Bphi(\zB^+) &= 0 \,, \label{eq:external_Bphi} \\
    \Delta \Bphi &= 0 \;. \label{eq:jump_condition_Bphi}
\end{align}
We note that the vertical profile of $\Bphi(r,z)$ is indeed important to obtain the jump condition \equ{eq:jump_condition_Br} for $\Br$. With \equs{eq:external_Br}{eq:jump_condition_Bphi} the final form of the vertical profiles for $\Br$ and $\Bphi$ reads \citep[cf.][but using Ohmic diffusion, 'Ohm-like' AD, ignoring HE, and including hydrodynamical terms]{Guilet2012, leung2019},
\begin{align}
    \Br(r,z<\zB) &= \frac{z}{1 - \pi \, \cs \, H_\mathrm{p} / (2 \, \eta_\mathrm{\zB})} \left[ \frac{\Brs}{\zB} + \pder[\Bz]{r} \right. \nonumber \\
    &\left.  + \left(5 - \frac{3 \, \pi \, \cs \, H_\mathrm{p}}{2 \, \eta_\mathrm{\zB}}\right) \frac{\Bz \, u_\mathrm{r,2} \, \zB^2}{H_\mathrm{p}^2 \, 15 \, \eta_\mathrm{\zB} } - \frac{4 \, \Bz \, \pi \, u_\mathrm{\varphi,2}}{3 \, \eta_\mathrm{\zB} } \right] \nonumber \\
    & - \frac{\Bz \, u_\mathrm{r,2} \, z^3}{3 \, \eta_\mathrm{\zB} \, H_\mathrm{p}^2} \;, \\
    \Bphi(r,z<\zB) &= \frac{\cs}{2 \, \Hp \, \eta_\mathrm{\zB} \, [1 - \pi \, \cs \, H_\mathrm{p} / (2 \, \eta_\mathrm{\zB})]} \Bigg[ \nonumber \\
    & \quad \frac{1}{2} \, \left( \frac{\Brs}{\zB} + \pder[\Bz]{r} \right) \, z \, (z^2 - \zB^2) \nonumber \\
    & + \frac{\Bz \, u_\mathbf{r,2} \, z (z^2 - \zB^2 )}{120 \, \eta_\mathrm{\zB}^2 \, \Hp^2} \left[ \zB^2 \, (3 \, \cs \, \Hp \, \pi - 14 \, \eta_\mathrm{\zB} ) \right. \nonumber \\
    & \left. \qquad \qquad \qquad \quad - ( z^2 \, (6 \, \cs \, \Hp \, \pi - 20 \, \eta_\mathrm{\zB} ) \right] \nonumber \\
    & - \frac{2 \, \Bz \, u_\mathrm{\varphi,2} \, z \, (\zB^2 - z^2) }{3 \, \cs \, \Hp \, \eta_\mathrm{\zB}}(2 \, \eta_\mathrm{\zB} - \, \cs \, \Hp ) \Bigg] \;, \label{eq:Bphi_internal}
\end{align}
where $\eta_\mathrm{\zB} = \eta(z=\zB)$. Taking the value of $\eta$ at $z = \zB$ for $\Br$ and $\Bphi$ is justified both by taking the value of $\eta$ at the transition zone ($z \approx \zB$) and by the fact that the marginally stable mode allowing a single bending of the field line in radial direction is localized around the transition zone as well \citep[cf.][for a thorough analysis of the marginal stability theory]{ogilvie2001,Guilet2012,leung2019}. Finally, the magnetic transport velocity can be calculated by evaluating $\Br$ at $z = \zB^-$ and by using \equ{eq:magn_flux_velocity},
\begin{align}
    \bbar u_\mathrm{\psi} &= \bbar u_\mathrm{r} + \frac{\bbar \eta}{\Bz} \, \frac{\pi \, \cs \, H_\mathrm{p} / (2 \, \eta_\mathrm{\zB})}{1 - \pi \, \cs \, H_\mathrm{p} / (2 \, \eta_\mathrm{\zB})} \, \pder[\Bz]{r} \nonumber \\
    & +\frac{1}{1 - \pi \, \cs \, H_\mathrm{p} / (2 \, \eta_\mathrm{\zB})} \Bigg[  \frac{\bbar \eta}{\zB}\frac{\Brs}{\Bz} \nonumber \\
    & \left. \qquad \qquad + \left(5 - \frac{3 \, \pi \, \cs \, H_\mathrm{p}}{2 \, \eta_\mathrm{\zB}}\right) \frac{ \zB^2}{\Hp^2} \frac{u_\mathrm{r,2} \, \bbar \eta}{15 \, \eta_\mathrm{\zB} } - \frac{4 \, \pi \, u_\mathrm{\varphi,2} \, \bbar \eta}{3 \, \eta_\mathrm{\zB}} \right] \nonumber \\
    & - \frac{u_\mathrm{r,2} \, \zB^2 \, \bbar \eta}{3 \, H_\mathrm{p}^2 \, \eta_\mathrm{\zB}} \;. \label{eq:flux_transport_velocity}
\end{align}
We note that $\bbar \eta$ and $\eta_\mathrm{\zB}$ are found multiple times in \equ{eq:flux_transport_velocity}, which emphasizes again the need for a detailed model of $\eta(r,z)$ (see~\sref{sec:non_ideal_resistivity}).

For solving \equs{eq:cont}{eq:ene}, the usual density-weighted vertical averages are used for the unknowns $\Sigma$, $e$, and $\vec u$ \citep[cf.][]{ragossnig20}. Therefore, using the vertical velocity profiles \equs{eq:ur_profile}{eq:uphi_profile}, $\bar u_\mathrm{r(\varphi)}$\footnote{The index $r(\varphi)$ stands for the radial and toroidal velocity component to abbreviate the otherwise identical formulae.} are obtained as follows,
\begin{align}
    \bar u_\mathrm{r(\varphi)} &= \frac{1}{\Sigma} \int_{-\infty}^{\infty} u_\mathrm{r(\varphi)}(r,z) \, \rho(r,z) \, dz = u_\mathrm{r(\varphi),0} + u_\mathrm{r(\varphi),2} \;, \label{eq:u_vert}
\end{align}
Since $u_\mathrm{r(\varphi),2}$ can be calculated using \equs{eq:ur2}{eq:uphi2}, the remaining unknowns with respect to $\vec u$ are $u_\mathrm{r(\varphi),0}$.

\subsection{Non-ideal resistivity model}
\label{sec:non_ideal_resistivity}

Magnetic resistivity $\eta(r, z)$ occurs due to turbulence, in our case denoted by $\eta_\mathrm{t}$, or laminar, non-ideal diffusion processes like Ohmic diffusion $\eta_\mathrm{O}$, ambipolar diffusion $\eta_\mathrm{A}$ and diffusion to the Hall effect $\eta_\mathrm{H}$. However, $\eta_\mathrm{H}$ is not considered in this work. $\eta_\mathrm{t}$ is connected to the turbulence causing the MRI, and therefore to $\alpha$. The laminar diffusion processes are dependent on the ionization fraction $\xe$ and the vertical temperature structure $\Tgas(r,z)$ and are discussed in this section.

\paragraph{Turbulent resistivity:}The assumption of $\eta_\mathrm{t}$ being caused by the same turbulence as MRI and therefore turbulent viscosity $\nu$ leads to the expectation that the strengths of $\eta_\mathrm{t}$ and $\nu$ are of the same order. This is described by the magnetic Prandtl number $\mathcal{P}$, which is defined as,
\begin{align}
    \mathcal{P} &= \frac{\nu}{\eta_\mathrm{t}} \,,
\end{align}
where $\mathcal{P}$ is fixed in this work to the value $\mathcal{P} = 1$, which leads to
\begin{align}
    \eta_\mathrm{t} &= \nu \,.
\end{align}
\paragraph{Laminar resistivity: }$\eta_\mathrm{O}$ and $\eta_\mathrm{A}$ both depend on $x_\mathrm{e}$ in the disk,
\begin{align}
    x_\mathrm{e} &= \frac{n_\mathrm{e}}{n_\mathrm{tot}} = \frac{n_\mathrm{e}}{n_\mathrm{e} + n_\mathrm{n} + n_\mathrm{i}} \approx \frac{n_\mathrm{e}}{n_\mathrm{n}} \;, \label{eq:ionization_fraction}
\end{align}
where $n_\mathrm{tot}$, $n_\mathrm{e}$, $n_\mathrm{n}$ and $n_\mathrm{i}$ denote the total number density, and the number densities of electrons, neutrals and ions, respectively. By far the most abundant element in PPDs is molecular hydrogen, therefore the assumption of $n_\mathrm{H_\mathrm{2}} \approx n_\mathrm{n} \approx n_\mathrm{tot}$ is valid \citep[cf.][]{mohanty2018}. $\etaO$ is defined as 
\begin{align}
    \eta_\mathrm{O} &= \frac{c^2}{4 \, \pi \, \sigma_\mathrm{O}} \,, \label{eq:ohmic_diffusion} \\
    \sigma_\mathrm{O} &\approx \frac{e^2 \, n_\mathrm{e}}{m_\mathrm{e} \, \langle\sigma v\rangle_\mathrm{en}} = \frac{e^2 \, x_\mathrm{e} \, n_\mathrm{tot}}{m_\mathrm{e} \, \langle\sigma v\rangle_\mathrm{en}}  \,, 
\end{align}
where $\sigma_\mathrm{O}$ denotes the Ohmic conductivity and is directly proportional to $x_\mathrm{e}$ \citep[e.g.][]{bai2009}. The rate coefficient of electron-neutral collisions and electron mass are described by  $\langle\sigma v\rangle_\mathrm{en}$ and $m_\mathrm{e}$, respectively.

$\etaA$ has an additional dependency on both the magnetic field strength and $\Sigma$ \citep[e.g.][]{Lesur2014} and can be written as (again for the assumption of electrons and single-charged ions only),
\begin{align}
    \etaA &= \beta_\mathrm{i} \, \beta_\mathrm{e} \, \etaO \;, \label{eq:ambipolar_diffusion} \\
    \beta_\mathrm{i(e)} &= \frac{e \, B}{m_\mathrm{i(e)} \, c} \frac{1}{\gamma_\mathrm{i(e)} \, \rho_\mathrm{n}} \;. \label{eq:hall_parameter}
\end{align}
Here $\beta_\mathrm{i(e)}$ denote the Hall parameters \citep[cf.][]{mohanty2018} for single-charged ions (index "i") and electrons (index "e"). The variables $m_\mathrm{i}$, $\rho_\mathrm{n}$ and $\gamma_\mathrm{i(e)}$ identify the mean ion mass, density of neutral gas and the drag coefficient for ions/electrons, respectively. With the following definitions of $\gamma_\mathrm{i(e)}$, $\langle\sigma v\rangle_\mathrm{en}$, and the rate coefficient of ion-neutral collisions $\langle\sigma v\rangle_\mathrm{in}$ \citep[e.g.][]{wardle2007,mohanty2018},
\begin{align}
    \gamma_\mathrm{i(e)} &= \frac{\langle\sigma v\rangle_\mathrm{in(en)}}{m_\mathrm{i(e)} + m_\mathrm{n}} \,, \\
    \langle\sigma v\rangle_\mathrm{in} &= 1.6 \times 10^{-9} \, \mathrm{cm^3 \, s^{-1}} \,, \\
    \langle\sigma v\rangle_\mathrm{en} &= 10^{-15} \, \left( \frac{128 \, k_\mathrm{B} \, T_\mathrm{e}}{9 \, \pi \, m_\mathrm{e}} \right) \, \mathrm{cm^3 \, s^{-1}} \,,
\end{align}
the laminar resistivities $\eta_\mathrm{O}$ and $\eta_\mathrm{A}$ can be written as (assuming that the electron temperature $T_\mathrm{e} = T_\mathrm{gas}$),
\begin{align}
    \eta_\mathrm{O}(r,z) &\approx 230 \, \sqrt{T_\mathrm{gas}(r,z)} \frac{1}{x_\mathrm{e}(r,z)} \, \mathrm{cm^2 \, s^{-1}} \,, \label{eq:parametrized_ohmic_resistivity} \\
    \eta_\mathrm{A}(r,z) &\approx 4.95 \times 10^{-15} \frac{|\vec B|^2}{\rho(r,z)^2 \, x_\mathrm{e}(r,z)} \, \mathrm{cm^2 \, s^{-1}} \,. \label{eq:parametrized_ambipolar_resistivity}
\end{align}
We want to point out that we have explicitly stated the radial and vertical dependency of $\Tgas$, $x_\mathrm{e}$, $\rho$, $\eta_\mathrm{O}$ and $\eta_\mathrm{A}$ to emphasize that it is necessary to model these dependencies in a 1+1D model in an approximate way. Gas in a PPD is either ionized thermally (depicted by $x_\mathrm{e,th}$) or non-thermally ($x_\mathrm{e,nth}$),
\begin{align}
    x_\mathrm{e} &= x_\mathrm{e,th} + x_\mathrm{e,nth} \;. \label{eq:ionfrac}
\end{align}

\paragraph{Thermal ionization} A high gas temperature $T_\mathrm{gas}(r,z)$ causes ionizing collisions between atoms, which are counteracted by recombination of those same atoms with electrons. This eventually leads to an equilibrium level of ionization and is described by the Saha equation \citep[e.g.][]{mohanty2018}, in the following shown for potassium (subscript K)
\begin{align}
   \frac{n_\mathrm{e} \, n_\mathrm{+,K}}{n_\mathrm{0,K}} &= 2.41 \times 10^{15} \, \sqrt{T_\mathrm{gas}^3} \, \exp \left( -\frac{5.04 \times 10^4 \, \mathrm K}{T_\mathrm{gas}} \right) \, \mathrm{cm^{-3}} \label{eq:saha}\\
   &= \mathcal{S}_\mathrm{K}(T_\mathrm{gas})  \, \mathrm{cm^{-3}}\,,
\end{align}
where $n_\mathrm{0,K}$ and $n_\mathrm{+,K}$ are the number densities of neutral and ionized potassium, respectively. The function $\mathcal{S}_\mathrm{K}(T_\mathrm{gas})$ denotes the right-hand side of \equ{eq:saha} and is dependent on $T_\mathrm{gas}$ only. Following \cite{mohanty2018}, the thermal ionization fraction $x_\mathrm{e,th}$ can then be written as,
\begin{align}
    x_\mathrm{e,th} &= \frac{-1 + \sqrt{1 + 4\,x_\mathrm{K} \, (n_\mathrm{tot} / \mathcal{S}_\mathrm{K}(T_\mathrm{gas}))}}{2 \, n_\mathrm{tot} / \mathcal{S}_\mathrm{K}(T_\mathrm{gas})} \;, \label{eq:thermal_ionfrac}
\end{align}
where $x_\mathrm{K}$ is the ionization fraction of potassium and is set to $x_\mathrm{K} = 1.97 \times 10^{-7}$ \citep[cf.][]{mohanty2018}. The formulation of $x_\mathrm{e,th}$ as in \equ{eq:thermal_ionfrac} covers an important edge case: In a stratified disk atmosphere, density significantly drops in the surface layers ($n_\mathrm{tot} \rightarrow 0$) which leads to a high $x_\mathrm{e}$ (cf. \equ{eq:ionization_fraction}). In the case of an equilibrium ionization this eventually leads to $x_\mathrm{e,th} \approx x_\mathrm{K}$.

For properly describing thermal ionization, not only the radial gas temperature profile, but also an approximation in vertical direction $\Tgas(r,z)$ is needed. At the midplane $\rho(r,z=0) = \rho_0(r)$ is largest, which leads to effective viscous heating at $z=0$. The resulting thermal radiation is transported radially and vertically through the disk according to \equ{eq:ene}. Additionally, the disk is both heated by stellar radiation on its surface (on both sides) and also cools radiatively across its surface (cf. \equ{eq:surface_temperature}). Assuming LTE and using the Eddington approximation, the vertical temperature distribution can be approximated in the optical thick case ($\tau \gg 1)$ as \citep[cf.][]{hubeny1990},
\begin{align}
    \Tgas^4(r,z) &= T_\mathrm{gas,0}^4 - \frac{\tau_\mathrm{R}^-(r,z)}{\tau_\mathrm{R}} ( T_\mathrm{gas,0}^4 - T_\mathrm{surf}^4 ) \;, \label{eq:vertical_Tgas}
\end{align}
where $\tau_\mathrm{R}^-(r,z) \equiv \kappa_\mathrm{R} \, \Sigma \, \mathrm{erf}(z / ( \sqrt{2} \, \Hp )) / \sqrt{2 \, \pi} $ and $T_\mathrm{gas,0}$ are the optical depth from the midplane up to the height $z$ and the midplane gas temperature, respectively. With \equ{eq:vertical_Tgas} $\xe$ can be calculated in vertical direction (cf.~\equ{eq:thermal_ionfrac}). 

Thermal ionization is very likely to only be significant in the inner disk ($r \lesssim 0.2$~AU), which has two reasons: (i) viscous heating is more effective in the inner regions with large $\Sigma$, as long as the disk is optically thick, and (ii) in regions very close to the star, the disk becomes optically thin and the irradiation from the star suffices for the gas to get thermally ionized, i.e. after heating by absorption.

\paragraph{Non-thermal ionization:} The main sources for non-thermal ionization are galactic cosmic rays and stellar X-ray as well as FUV radiation. Additionally, radioactive decay (RA) plays a role in some regions close to the disk midplane. The cosmic-ray ionization rate is $\xi_\mathrm{CR}$ is adapted from \cite{Lesur2014},
\begin{align}
    \xi_\mathrm{CR} &= 10^{-16} \, \exp{- \frac{\Sigma}{9.6 \times 10^2 \, \mathrm{g \, cm^{-2}}}} \, \mathrm{s^{-1}}  \,, \label{eq:ionrate_CR}
\end{align}
whereas the X-ray ionization rate $\xi_\mathrm{XR}$ can be modeled according to a fit done by \cite{bai2009,delage2022}, using a X-ray temperature of $T_\mathrm{XR} = 3 \, \mathrm{keV}$. X-ray photons can either directly ionize the gas $\xi_\mathrm{XR,dir}$ or indirectly through scattering $\xi_\mathrm{XR,sca}$,
\begin{align}
    \xi_\mathrm{XR,dir} &= \frac{L_\mathrm{XR}}{10^{29} \, \mathrm{erg \, s^{-1}}} \left( \frac{r}{1 \, \mathrm{AU}} \right)^{-2.2} \\
    & \xi_\mathrm{XR,dir,0} \left[ \exp{-\left(\frac{\Sigma^+}{\Sigma_\mathrm{XR,dir}}\right)^{0.4} } + \exp{-\left(\frac{\Sigma^-}{\Sigma_\mathrm{XR,dir}}\right)^{0.4} } \right] \,, \label{eq:xr_direct_ion_rate} \\ \nonumber
    \xi_\mathrm{XR,sca} &= \frac{L_\mathrm{XR}}{10^{29} \, \mathrm{erg \, s^{-1}}} \left( \frac{r}{1 \, \mathrm{AU}} \right)^{-2.2} \\
    & \xi_\mathrm{XR,sca,0} \left[ \exp{-\left(\frac{\Sigma^+}{\Sigma_\mathrm{XR,sca}}\right)^{0.65} } + \exp{-\left(\frac{\Sigma^-}{\Sigma_\mathrm{XR,sca}}\right)^{0.65} } \right] \,,   \label{eq:xr_scatter_ion_rate}   
\end{align}
where $\Sigma_\mathrm{XR,dir} = 3.5 \times 10^{-3} \, \mathrm{g \, cm^{-2}}$ and $\Sigma_\mathrm{XR,sca} = 1.64 \, \mathrm{g \, cm^{-2}}$ are the penetration depths for direct and scattered X-ray ionization \citep[cf.][]{delage2022}, respectively. $\Sigma^+ = \int_\mathrm{z}^\infty \rho(r,z) \, dz$ and $\Sigma^- = \Sigma - \Sigma^+$ denote the column densities above and below a certain height $z$, respectively.  Additionally, $\xi_\mathrm{XR,dir,0} = 6 \times 10^{-12} \, \mathrm{s^{-1}}$ and $\xi_\mathrm{XR,sca,0} = 10^{-15} \, \mathrm{s^{-1}}$ describe the undamped direct and scattered X-ray ionization, respectively. Finally, a rather simple radioactive decay ionization rate $\xi_\mathrm{RA}$ is adopted according to \cite{Lesur2014}, ignoring radial variation and the effects of varying dust-to-gas ratio \citep[cf.][]{delage2022},
\begin{align}
    \xi_\mathrm{RA} &= 10^{-19} \, \mathrm{s^{-1}} \,.
\end{align}
The effects of FUV radiation are neglected in this work, as the penetration depth of FUV is of order $10^{-2} \, \mathrm{g \, cm^{-2}}$ and yields an ionization front in the uppermost surface layer of a PPD. Therefore FUV ionization becomes significant when modeling MCWs or photoevaporative winds \citep[cf.][]{cecil2024}, but is not likely to play an important role in modeling the resistivity for magnetic flux transport of a large-scale, purely poloidal field inside the disk. The total ionization rate $\xi$ is then written as
\begin{align}
    \xi &= \xi_\mathrm{CR} + \xi_\mathrm{XR,dir} + \xi_\mathrm{XR,sca} + \xi_\mathrm{RA} \;.
\end{align}
In an equilibrium state, which is necessary in assuming a steady-state in vertical direction, ionization is balanced by radiative recombination $\alpha_\mathrm{r}$ and recombination on dust particles $\alpha_\mathrm{d}$ in the following way,
\begin{align}
    (1 - x_\mathrm{e,nth}) \, \xi &= \alpha_\mathrm{r} \, x_\mathrm{e,nth}^2 \, n_\mathrm{tot} + \alpha_\mathrm{d} \, x_\mathrm{e,nth} \, n_\mathrm{tot} \;, \label{eq:nonthermal_ionfrac}
\end{align}
where we have assumed that the number density of neutrals is very close to the total number density $n_\mathrm{n} \approx n_\mathrm{tot}$ and that the temperature regimes of thermal and non-thermal ionization are non-overlapping. This is a good approximation, because thermal ionization becomes dominant above $T_\mathrm{th,active} \gtrsim 1000\,K$ and has a very narrow transition temperature range. Below $T_\mathrm{th,active}$ non-thermal ionization dominates, hence the non-thermal ionization fraction $x_\mathrm{e,nth}$ can be calculated separately from $x_\mathrm{e,th}$ \citep[cf.][]{vorobyov2020}. We note that dissociative recombination is not considered in this work due to the fact that radiative and dust recombination yield an upper and lower limit for $x_\mathrm{e,nth}$, respectively \citep[cf.][]{dudorov2014}, whereas dissociative recombination leads to an ionization fraction between those limits. 

The radiative recombination rate $\alpha_\mathrm{r}$ can be approximated by \citep[cf.][]{spitzer1978, vorobyov2020}
\begin{align}
    \alpha_\mathrm{r} &= 2.07 \times 10^{-11} \, T_\mathrm{gas}^{-1/2} \, \mathrm{cm^3 \, s^{-1}} \;, 
\end{align}

whereas for the dust recombination rate $\alpha_\mathrm{d}$ we adopt a parametrization described by \cite{dudorov2014}, which is depicted in \tab{tab:dust_model} and is valid for a grain size $a_\mathrm{d} = 0.1 \, \mathrm{\mu m}$ and a dust-to-gas ratio $f_\mathrm{dust} = 0.01$.
\begin{table}
    \caption{Dust recombination rate for the dust grain size $a_\mathrm{d} = 0.1 \, \mathrm{\mu m}$,  and different gas temperature regimes.}    
    \label{tab:dust_model}
    \centering
    \begin{tabular}{l r}
        \hline\hline 
        T [K] & dust recombination rate $\alpha_\mathrm{d}$ [$\mathrm{cm^3 \, s^{-1}}$] \\
        \hline
         $<150$        &  $4.5 \times 10^{-17}$ \\
         $150 - 400$   &  linear decrease from $4.5 \times 10^{-17}$ to $3.0 \times 10^{-18}$ \\
         $400 - 1500$  &  $3.0 \times 10^{-18}$ \\
         $1500 - 2000$ &  linear decrease from $3.0 \times 10^{-18}$ to $0$ \\
         $>2000$       &  $0$ \\         
        \hline                                            
    \end{tabular}
\end{table}
Formally $\alpha_\mathrm{d}$ is defined as
\begin{align}
    \alpha_\mathrm{d} &= X_\mathrm{g} \, \langle \sigma_\mathrm{g} \, V_\mathrm{ig} \rangle \;, \label{eq:def_dust_recomb}
\end{align}
where $X_\mathrm{g}$, $\sigma_\mathrm{g}$ and $V_\mathrm{ig}$ denote the number density fraction of dust grains, the grain cross-section and ion-grain relative velocity \citep[cf.][]{dudorov2014}. In the case of a fixed $f_\mathrm{dust}$ the following relations hold,
\begin{align}
    X_\mathrm{g} &= \frac{n_\mathrm{d}}{n_\mathrm{tot}} \propto a_\mathrm{d}^{-3} \,, \\
    \sigma_\mathrm{g} &\propto a_\mathrm{d}^2 \;, \label{eq:dust_cross_section}
\end{align}
which, according to \equ{eq:def_dust_recomb}, leads to the overall scaling of $\alpha_\mathrm{d}$,
\begin{align}
    \alpha_\mathrm{d} \propto a_\mathrm{d}^{-3} \, a_\mathrm{d}^{2} \propto a_\mathrm{d}^{-1} \;. \label{eq:dust_recomb_scaling}
\end{align}
Therefore a different grain size $a_\mathrm{d}$ can be used by taking the values of \tab{tab:dust_model} for $a_\mathrm{d}$ and then by scaling it according to \equ{eq:dust_recomb_scaling} \citep[cf.][]{dudorov2014}.

\subsection{Solution procedure}
\label{sec:solution_procedure}

Our 1+1D model aims to solve the hydrodynamical equations \equs{eq:cont}{eq:ene} together with a poloidal large-scale field as described in \equ{eq:magn_flux_averaged} (cf.~\sref{sec:basic_equations} and \sref{sec:large_scale_disk_field}). The vertical structure of the disk is crucial in understanding the radial magnetic flux transport of the large-scale field, hence we have provided a description of the vertical velocity profile \citep[analogously to][]{Guilet2012,leung2019} (cf.~\sref{sec:vertical_vel_profile}) and a model for the vertical ionization fraction and resistivity (cf.~\sref{sec:non_ideal_resistivity}. The large-scale field is assumed to be seeded by an external fossil field, which is taken to be force-free (in the absence of outflows) and which serves as a boundary at the disk surface for the large-scale field in the disk interior (cf.~\sref{sec:external_magnetic_field}). The vertical velocity and resistivity profiles need to be vertically averaged to fit into a 1+1D approach.

The solving procedure follows the following steps:
\begin{enumerate}
    \item Calculate the vertical velocity contributions $\ur[2]$ and $\uphi[2]$ (cf.~\equs{eq:ur2}{eq:uphi2}).The radial contributions $\ur[0]$ and $\uphi[0]$ are solved through the hydrodynamic simulation. 
    \item Determine the density-weighted vertical averages $\bar u_\mathrm{r}$ and $\bar u_\mathrm{\varphi}$ (cf.~\equ{eq:u_vert}).
    \item Obtain $\psi_\mathrm{d}$ by subtracting the magnetic flux contributions of stellar dipole and external field as in \equ{eq:magn_flux_disk}.
    \item Calculate $\mathcal{L}$ (cf.\equ{eq:biot_savart} and \equ{eq:operator_L}) and inversion to obtain $\Brs$ (cf.~\app{sec:numerical_details} for details).
    \item Determine $\xe$ (cf.~\equ{eq:ionfrac}, \equ{eq:thermal_ionfrac} and \equ{eq:nonthermal_ionfrac}), the vertical temperature profile (cf.~\equ{eq:vertical_Tgas}) as well as $\etaO$ and $\etaA$ (cf.~\equ{eq:parametrized_ohmic_resistivity} and \equ{eq:parametrized_ambipolar_resistivity}, respectively) for all radii at five different heights ($z = 0$, $z = H_\mathrm{p}$, $z = H_\mathrm{p} + \frac{1}{3}(\zB - H_\mathrm{p})$, $z = H_\mathrm{p} + \frac{2}{3}(\zB - H_\mathrm{p})$, and $z = \zB$).
    \item Calculate the conductivity-weighted vertically averaged $\bbar \eta(r)$ between $-\zB \le z \le \zB$. The necessary integration over $\eta(r,z) = \etaO(r,z) + \etaA(r,z)$ is done by piece-wise numerical integration using the values for the heights $z$ calculated in the preceding step (cf.~\equ{eq:def_averaged_resistivity}).
    \item Determine the conductivity-weighted vertically averaged radial velocity $\bbar u_\mathrm{r}$ (cf.~\equ{eq:def_averaged_gas_velocity}) by using $\bbar \eta$.
    \item Determine the magnetic flux transport velocity $\bbar u_\mathrm{\psi}$ by using $\bbar \eta$, $\bbar u_\mathrm{r}$ and $\eta(r,z=\zB)$ (cf.~\equ{eq:flux_transport_velocity}).
    \item Finally, solve the hydrodynamical equations \equs{eq:cont}{eq:ene} and \equ{eq:magn_flux_averaged} with TAPIR \citep[cf.][]{ragossnig20}.
\end{enumerate}
The most expensive operation (in the context of simulation time) is the inversion of the operator $\mathcal{L}$. The details of this inversion and the steps taken to maintain adequate performance in solving the MHD equations are explained in \sref{sec:numerical_details}.

\subsection{Validity of our approximations for modeling the vertical disk and magnetic field structure}
\label{sec:validity_appr}

We use a 1+1D disk model and assume a geometrically thin disk ($H_\mathrm{p}/r \ll 1$), which usually means that the radial and toroidal velocity components, $u_\mathrm{r}(r)$ and $u_\mathrm{\varphi}(r)$, respectively, are taken to be constant in the vertical direction. 
However, describing the evolution of a large-scale magnetic field requires to model magnetic advection and diffusion, which both are strongly dependent on the ionization degree and velocity structure in vertical direction \citep[cf. e.g.][see also \sref{sec:large_scale_disk_field} and \sref{sec:vertical_vel_profile}]{Guilet2012, leung2019, delage2022}. 
Hence, we discuss the validity of the essential approximations for describing the vertical disk- and large-scale magnetic field structure. The key points are (for a detailed discussion cf. \app{sec:validation}):
\begin{itemize}
    \item The focus on Class II PPDs justifies the use of the thin-disk approximation.
    \item A comparison of the viscous timescale $t_\mathrm{visc}$ and the dynamical timescale $t_\mathrm{dyn}$ justifies the use of stationary vertical velocity- and large-scale magnetic field profile (cf. also \sref{sec:remarks_stationarity}).
    \item The assumption of a dominant vertical magnetic field is valid for most of the disk, as long as the large-scale magnetic field remains weak and the inclination angle does not exceed $i > 30^\circ$. The validity is only questionable in the innermost disk region, where magnetic compression likely plays a role.
    \item The somewhat arbitrary choice of vertical layers for the calculation of $\xe$ as well as OD and AD in the vertical direction is justified through a comparison with a profile with different choices of vertical layers. It can be concluded that our approach with 5 layers is a good compromise between numerical cost and reasonable accuracy.
    \item For describing the thermally ionized region in the inner disk a vertical temperature profile is used. It is compared to and in good agreement with 2D simulations of the vertical disk structure \citep[cf. e.g.][]{jankovic2021}.
\end{itemize}
Not only the validity of our approximations, but also the comparison with former work is important.
To this end we compare our reference model with the analytic profiles of \cite{dudorov2014} and find very good agreement with their radial magnetic field profiles (cf. \fig{fig:fid_magnetic_field} in \sref{sec:reference_model}). 
We have also dedicated \sref{sec:comparison} to compare our results with the work of various authors. 
We find that our results are consistent with the models of \cite{mohanty2018} and \cite{delage2022}.
Especially the calculation of the ionization degree and resistivity profiles are in very good agreement with the work of those authors. 
The work of \cite{jankovic2021} also agrees well with our model of the inner disk and the vertical temperature structure and thermally ionized region. 
The sum of all those comparison strengthens the confidence that our approximate disk model is capable of reproducing the results from previous studies \citep[cf. e.g.][]{dudorov2014, Guilet2014} but also has the ability to provide additional scientific insights.
We nevertheless discuss remaining fundamental limitations of our model 
methodology in \sref{sec:model_limitations}. 
Some of these can be overcome in future work. 
However, we do not consider them to be of crucial importance in the framework of our present study that focuses on the investigation of the poloidal large-scale magnetic field topology threading a stationary disk.

\section{Results}
\label{sec:results}

We split our results into three main categories:
\begin{enumerate}[(i)]
    \item The evolution of a background fossil field towards a steady state (cf.~\sref{sec:magnetic_field_evolution}).
    \item The in-depth study of the solution obtained after the large-scale magnetic field in the disk has become stationary (cf.~\sref{sec:reference_model}).
    \item A parameter study of how the resulting stationary system consisting of a disk, star, and large-scale magnetic field changes for different parameters (cf.~\srefs{sec:variation_stellar_parameters}{sec:variation_field_properties}).
\end{enumerate}

\subsection{Evolution of the large-scale magnetic field towards a stationary state}
\label{sec:magnetic_field_evolution}

For validation of the large-scale magnetic field structure obtained in this paper, it is useful to compare it to similar, already available work done on this topic \citep[e.g.][]{Guilet2014, mohanty2018, delage2022}, which use a stationary magnetic field topology. A (magneto) hydrodynamic treatment of the large-scale field requires evolving the field and disk sufficiently long to eventually reach a stationary state. 
This is done with TAPIR in two stages for our reference model (see \tab{tab:ref_model}):

\begin{table}[ht!]
\centering
\caption{Parameters used for the reference model.}              
\label{tab:ref_model}  
\begin{tabular}{l r | l r}         
\hline\hline
\multicolumn{2}{l}{Stellar parameters} & \multicolumn{2}{| l}{Disk parameters} \\
$M_\mathrm{\star}$ [$M_\mathrm{\odot}$]\tablefootmark{a} & 0.7 & $\dot M_\mathrm{init}$ [$M_\odot \, \mathrm{yr}^{-1}$]\tablefootmark{e} & $5.0  \cdot 10^{-9}$ \\
$R_\mathrm{\star}$ [$R_\mathrm{\odot}$]\tablefootmark{a} & 2.096 & $\alpha_\mathrm{MRI}$\tablefootmark{f} & $10^{-2}$ \\
$L_\mathrm{\star}$ [$L_\mathrm{\odot}$]\tablefootmark{a} & 1.09 & $\alpha_\mathrm{AD}$\tablefootmark{g} & $5 \cdot 10^{-3}$ \\
$P_\star$ [d]\tablefootmark{b} & 4.0 & $\alpha_\mathrm{base}$\tablefootmark{h} & $10^{-4}$ \\
$B_\star$ [kG]\tablefootmark{c} & 1.0 & & \\
$L_\mathrm{X}$ [$\mathrm{erg \, s^{-1}}$]\tablefootmark{d} & $10^{30}$ & & \\
\hline
\multicolumn{2}{l | }{Fossil field parameters} & & \\
$\beta_0(r_\mathrm{out})$\tablefootmark{i} & $10^2$ & & \\
\hline                                      
\end{tabular}
\tablefoot{
    \tablefoottext{a}{Stellar parameters $M_\mathrm{\star}$, $R_\mathrm{\star}$ and $L_\mathrm{\star}$ are taken from pre-main sequence (PMS) isochrones created by \cite{Baraffe15} for a T Tauri star with an age of $t = 1$~Myr.} \\
    \tablefoottext{b}{The rotation period $P_\mathrm{\star}$ is chosen to lie in the expected range between 1 and 10 days \citep[cf. e.g.][]{irwin2009}.}  \\
    \tablefoottext{c}{$B_\mathrm{\star}$ for Class II objects is believed to lie between $0.1$~kG and $1.0$~kG \citep[cf. e.g.][]{johnstone2014}.} \\
    \tablefoottext{d}{X-ray luminosities of classical T Tauri stars lie between $10^{28} \, \mathrm{erg \, s^{-1}}$ and $10^{32} \, \mathrm{erg \, s^{-1}}$ \citep[cf. e.g.][]{preibisch2005}.} \\
    \tablefoottext{e}{Accretion rates for classical T Tauri stars lie between $10^{-10}\,\mathrm{M_\odot \, \mathrm{yr}^{-1}}$ and $10^{-7}\,\mathrm{M_\odot \, \mathrm{yr}^{-1}}$ \citep[cf.][]{Vorobyov17c}} \\
    \tablefoottext{f}{Viscosity due to MRI is believed to result in $\alpha$ between $10^{-4}$ and $10^{-1}$ \citep[cf. e.g.][]{zhu07,Vorobyov09}.} \\
    \tablefoottext{g}{AD reduces the effectivity of MRI, which results in a smaller $\alpha_\mathrm{AD}$ \citep[cf.][]{delage2022}.} \\
    % \tablefoottext{h}{The dead zone is not MRI-active, hence $\alpha_\mathrm{DZ} \ll \alpha$. } \\
    \tablefoottext{h}{The base viscosity is much lower than the viscosity caused by MRI, hence $\alpha_\mathrm{base} \ll \alpha_\mathrm{MRI}$. } \\    
    \tablefoottext{i}{A plasma beta $\beta_0(r_\mathrm{out})$ in our reference model corresponds to $B_\infty \approx 3.8 \cdot 10^{-3}$~G. This is in the range of [$10^{-5}$~G, $10^{-1}$~G] from studies of the fossil field in protostellar cores \citep[cf.][]{crutcher2004, masson2016}. }
}
\end{table}
\begin{enumerate}[(i)]
    \item The first stage consists of evolving the disk with a prescribed stellar field and a prescribed large-scale disk field, analogously to \citet{Steiner21}. 
    The stationary disk is obtained by self-consistently calculating the position of the inner disk boundary $r_\mathrm{in}$, taking into account the accretion rate $\dot M(r_\mathrm{in})$ and the initial stellar magnetic field strength $B_{\star,0}$. 
    Additionally, we start with a radially constant, purely vertical, large-scale disk magnetic field $B_\mathrm{d,0}(r) = B_\mathrm{\infty}$ (see \equ{eq:B_inf} in \sref{sec:large_scale_disk_field}) threading the disk at time $t = 0$. 
    This field is superimposed with the stellar magnetic field to provide an initial star-disk magnetic field threading the disk $B_\mathrm{0} = B_\mathrm{\star,0} + B_\mathrm{d,0}$ (cf.~\sref{sec:large_scale_disk_field}) and is kept constant during the first stage. 
    This essentially means that only the disk "feels" the impact of the magnetic field, whereas the magnetic field dragging due to radial mass transport is not considered during this first stage.
    \item The second stage then relaxes the constraints of a temporally constant large-scale field by allowing for the evolution of the magnetic field from the initial configuration $B_\mathrm{0}$ towards a stationary field topology by interacting with the accretion disk (using the parameters of \tab{tab:ref_model}).
    In this phase, the disk and magnetic field are evolving as a coupled system.%, resulting in a stationary disk and magnetic field profile (e.g., the variables of \equs{eq:ene}{eq:induction}).
    Starting with the stationary disk obtained from the first stage, the magnetic field is allowed to advect inward, caused by radial mass transport through the disk. 
    Due to non-ideal MHD effects (in this study turbulent resistivity, Ohmic diffusion, and AD are considered, see \sref{sec:non_ideal_resistivity}) the magnetic field advection is counteracted by diffusion, which at some point in time balances the inward-directed advection velocity $u_\mathrm{adv}$, which then yields a stationary magnetic field.
\end{enumerate}

\begin{figure}[ht!]
    \centering
    \resizebox{\hsize}{!}{\includegraphics{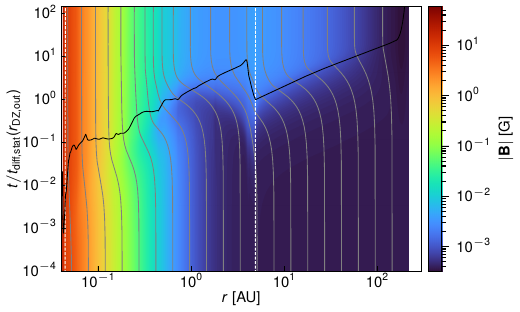}}
    \caption{
    Time evolution of the magnetic field strength over time (normalized in units of the diffusion timescale of the stationary model at the outer DZ boundary, $t_\mathrm{diff,stat}(r_\mathrm{DZ,out})$), where the color coding represents the magnetic field strength $|\vec{B}|$ at every point in space and time.
    The gray lines show magnetic flux transport over time, using the effective, vertically averaged magnetic field transport velocity $\bbar u_\mathrm{\psi}$. 
    The dashed white lines denote the inner and outer boundary of the DZ, respectively, whereas the black full line denotes the diffusion timescale $t_\mathrm{diff,stat}$ of our stationary model. 
    The inner DZ boundary is located very close to the inner disk rim.}
    \label{fig:fid_evolution_time}
\end{figure}

In \fig{fig:fid_evolution_time}, we present the large-scale magnetic field evolution with a focus on the timescales until it reaches its final stationary state. 
The relevant timescale describing the radial magnetic flux transport is the diffusion timescale $t_\mathrm{diff} \equiv r^2 / \bbar \eta$. As a reference, we adopt the value of $t_\mathrm{diff,stat}$, corresponding to the stationary large-scale field, at the outer DZ boundary $r_\mathrm{DZ,out}$. 
In the very inner disk ($r \lesssim 0.1$~AU), the stellar dipole field is dominating the large-scale field, which is represented by a low $\bbar u_\mathrm{\psi}$ and a relatively high $\vec B$ in the inner disk for the whole disk evolution. Interestingly, despite a reduced radial mass transport $\dot M$ in the DZ (between the two vertical white dashed lines in \fig{fig:fid_evolution_time}), the large-scale field in the DZ is effectively advected inwards\footnote{We want to emphasize that the gray lines in \fig{fig:fid_evolution_time} shall not be confused with field lines, but serve as an indicator, where magnetic flux transport is occurring during field evolution.}. This is due to significant magnetic flux transport in the upper disk layers, which is taken into account in our model by the conductivity-weighted vertical averaged resistivity $\bbar \eta$ (cf. \sref{sec:reference_model}) and results in a considerably larger $\bbar u_\mathrm{\psi}$ compared to $\bar u_\mathrm{r}$ in the DZ. The sharp decrease of $t_\mathrm{diff,stat}$ in the vicinity $r_\mathrm{DZ,out}$, together with the inward-directed magnetic field transport, leads to accumulation of magnetic flux $\psi$ early on during magnetic field evolution (starting at $t \sim 10^{-1} \, t_\mathrm{diff,stat}(r_\mathrm{DZ,out})$). This cause a steep radial gradient in $\psi$ and therefore an increase in $B_\mathrm{z,d}$ just inside the outer DZ boundary, revealing a locally increased field strength $|\vec B|$ (cf. color-coding of \fig{fig:fid_evolution_time} around $r_\mathrm{DZ,out}$). In the outer disk, the large-scale field is not transported as effectively as in the inner region, which is due to AD being the dominant non-ideal MHD effect over the whole vertical extent in the outer disk. Eventually, this yields a larger $\bbar \eta$ in the outer disk and therefore a smaller $\bbar u_\mathrm{\psi}$.
For $t \gtrsim 10 \, t_\mathrm{diff,stat}(r_\mathrm{DZ,out})$ the large-scale field becomes essentially stationary everywhere in the disk. 
This means that vertically averaged magnetic advection and magnetic diffusion compensate each other, the resulting magnetic flux transport velocity $\bbar u_\mathrm{\psi} = 0$ (cf. \equ{eq:magn_flux_velocity}), and the gray lines in \fig{fig:fid_evolution_time} become vertical. 

\subsection{Remarks on the validity of stationary models}
\label{sec:remarks_stationarity}

In this work most of our conclusions are drawn from the analysis of stationary PPD models. Since disks are evolving over time, it is important to discuss the validity of using stationary models. 
In our 1+1D models we use a vertically averaged viscous-$\alpha$ \citep[cf. \equ{eq:alpha}, based on the layered viscosity model of][see also \cite{Steiner21}]{gammie96}, which implies that viscous accretion varies in vertical direction between different layers. 
A localized perturbation changing $\alpha(r,z)$ at some radius $r$ and some height $z$ would lead to an increased or decreased local density $\rho(r,z)$ and therefore would cause a deviation from a hydrostatic vertical equilibrium at $r$. 
Additionally, the vertically averaged $\alpha$ would change at $r$, which over time would lead to a bump or gap in $\Sigma(r)$. The buildup of such gaps or bumps would occur on the local viscous timescale $t_\mathrm{visc}$, whereas the return to a local vertical hydrostatic equilibrium would happen on the local dynamical timescale $t_\mathrm{dyn}$ \citep[e.g.,][]{delage2022}.
Therefore, an approximate stationary model can be understood as a time-averaged disk, where such local perturbations are smoothed out under the assumption that the relaxation to a hydrostatic equilibrium happens much faster than the development of a bump or gap in $\Sigma$, or $t_\mathrm{dyn}(r) \ll t_\mathrm{visc}(r)$.

Additionally, we have to discuss the validity of a stationary large-scale magnetic field, which can be argued similarly in comparing the timescale of global magnetic field evolution to the timescale at which such perturbations typically occur. It has been argued by \cite{dudorov2014} that the magnetic field in vertical direction also changes on $t_\mathrm{dyn}$. In contrast, the large-scale magnetic field in radial direction evolves on the local diffusion timescale $t_\mathrm{diff}$, which requires $t_\mathrm{dyn} \ll t_\mathrm{diff}$ for a valid temporal averaging of local large-scale magnetic field perturbations.

Finally, PPDs typically disperse after $\sim 1 - 4$~Myrs \citep[cf. e.g.][]{delage2022,gregory2025}, which raises the question if such disks live long enough to allow both disk and large-scale magnetic field to evolve toward a stationary state (i.e. steady-state accretion with $\dot M = \mathrm{const.}$).
This means that we also have to verify that the global evolution timescales of both disk and large-scale magnetic field, $t_\mathrm{visc}$ and $t_\mathrm{diff}$, respectively, are shorter or in the order of typical disk lifetimes \citep[cf.][]{delage2022}. 
\begin{figure}[ht!]
    \centering
    \resizebox{\hsize}{!}{\includegraphics{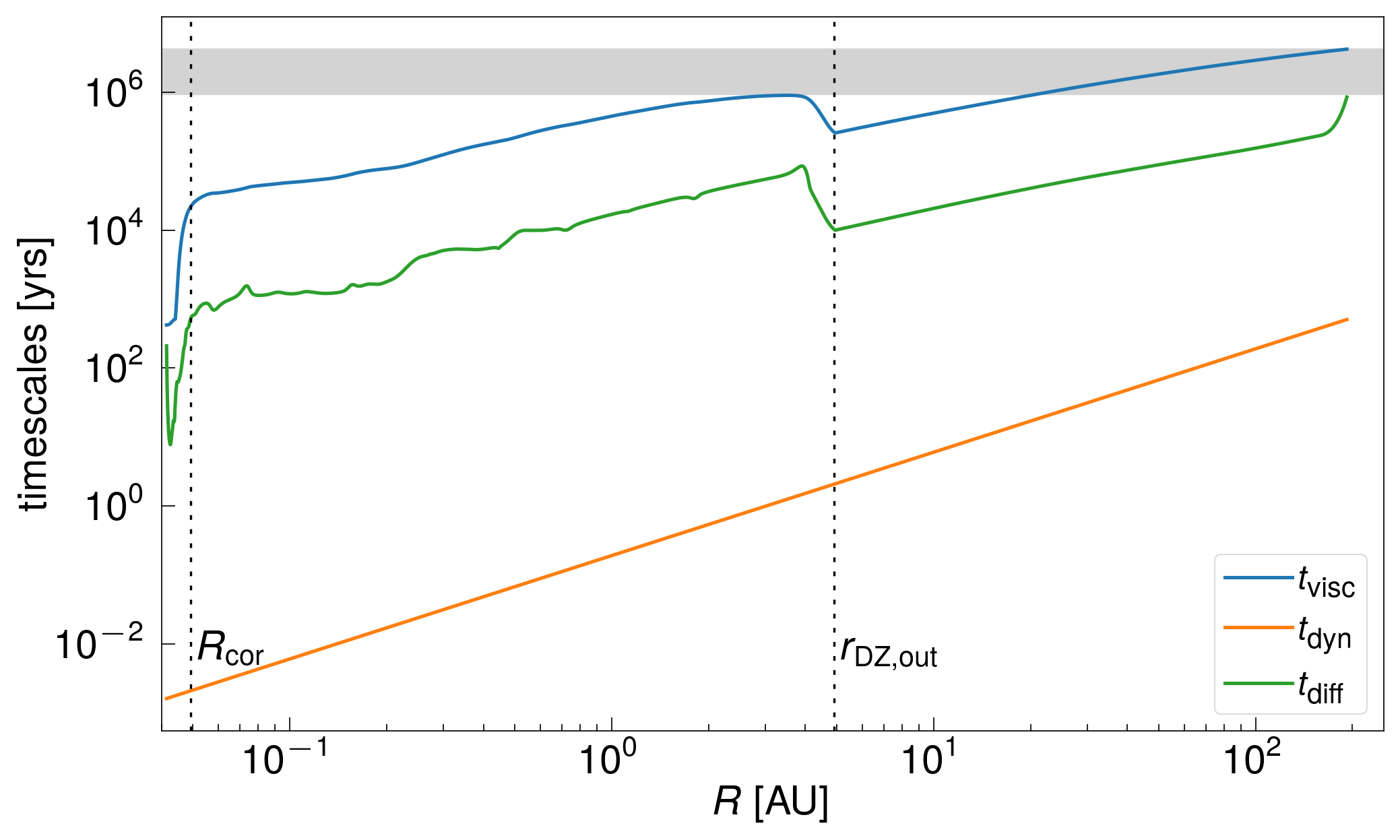}}
    \caption{
        Timescales of our reference model (cf.~\tab{tab:ref_model}) relevant to the long-term evolution of a PPD and a large-scale magnetic field. The solid blue, orange and green lines correspond to the viscous timescale $t_\mathrm{visc}$, the dynamical timescale $t_\mathrm{dyn}$ and the magnetic diffusion timescale $t_\mathrm{diff}$, respectively. The vertical, dashed black lines correspond to the corotation radius $R_\mathrm{cor}$ and the outer DZ boundary $r_\mathrm{DZ,out}$. Finally, the gray area marks the region between $1$~Myr and $4$~Myrs and depicts the expected lifetime of a PPD.
    }
    \label{fig:fid_timescales_stationarity}
\end{figure}
In \fig{fig:fid_timescales_stationarity} we can see that the conditions $t_\mathrm{dyn} \ll t_\mathrm{visc}$ and $t_\mathrm{dyn} \ll t_\mathrm{diff}$ are both fulfilled everywhere in the disk, which allows us temporally average over local perturbations of the vertical disk structure and vertical large-scale magnetic field profile. Furthermore, \fig{fig:fid_timescales_stationarity} shows that $t_\mathrm{diff}$ and $t_\mathrm{visc}$ are indeed shorter or of the order of a typical PPD lifetime, which justifies the assumption that the disk is likely to establish (near)-steady-state accretion and therefore stationarity over most of its radial extent before it dissolves \citep[cf.][for a similar argumentation]{delage2022}.

\subsection{The reference model}
\label{sec:reference_model}

In this section, we describe the details of our reference disk model, especially the large-scale disk magnetic field topology. The field is strongly dependent on the resistivity profile $\bbar \eta(r,z)$, which in turn is dependent on the ionization fraction $\xe(r,z)$. Therefore we investigate in more detail the vertical and radial profile of  $\xe$ and $\eta$, which then can be used to construct the conductivity-weighted vertical resistivity average $\bbar \eta$ counteracting advection. 

\begin{figure}[ht!]
    \centering
    \resizebox{\hsize}{!}{\includegraphics{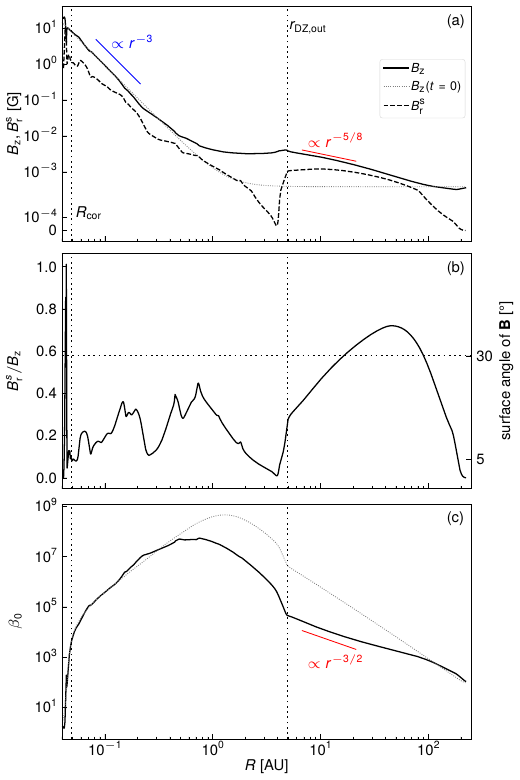}}
    \caption{The poloidal, stationary magnetic field profile of our reference model (cf.~ \tab{tab:ref_model}). Panel (a) shows both the vertical and radial magnetic field component at the surface, $\Bz$ and $\Brs$, respectively. Panel (b) depicts the ratio of $\Brs / \Bz$, which is a measure for the inclination of the field lines, whereas panel (c) shows the magnetic plasma beta $\beta_0$ at the disk midplane. The vertical dashed line correspond to the corotation radius $R_\mathrm{cor}$ and $r_\mathrm{DZ,out}$, respectively. The dotted, gray lines in panels (a) and (c) describe the corresponding values at $t=0$, whereas the horizontal dotted line in panel (b) denotes a surface angle of $30^\circ$ with respect to the midplane normal.}
    \label{fig:fid_magnetic_field}
\end{figure}

\paragraph{Large-scale magnetic field profile:} We want to present and analyze the stationary magnetic field structure.
As already discussed in \sref{sec:magnetic_field_evolution}, the field in the inner disk is dominated by the stellar dipole field (cf. \equs{eq:stellar_Bz}{eq:stellar_Bphi}). Since $B_\mathrm{z,\star} \propto r^{-3}$, the vertical magnetic field $\Bz$ in the inner disk also follows this power law (cf. (a) in \fig{fig:fid_magnetic_field}). The field evolution in the outer disk is strongly influenced by AD, which according to an analytical model from \citep[][]{dudorov2014}, should lead to a dependency $\Bz \propto r^{-5/8}$ in the case of an irradiated disk, which we can verify with our model (cf. (a) in \fig{fig:fid_magnetic_field}). Amplification of the fossil field in the outer region is roughly $\Bz / \Bz(t=0) \approx 10$, whereas in the inner disk, due to the dominant stellar field component, almost no amplification is observed. However, the disk field is advected inwards and a radial field component at the disk surface $\Brs$ develops, which leads to an inclination of the large-scale field with respect to the disk normal (cf. (b) in \fig{fig:fid_magnetic_field}). 
The value of $30^\circ$ shown in (b) of \fig{fig:fid_magnetic_field} corresponds to the critical field inclination, for which purely magneto-centrifugally driven disk outflows can be launched \citep[cf. e.g.][and \sref{sec:implications_inclination}]{Ogilvie1997, ogilvie2001}.

\begin{figure}[ht!]
    \centering
    \resizebox{\hsize}{!}{\includegraphics{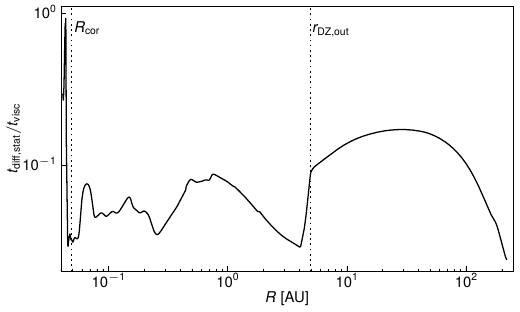}}
    \caption{Comparison of viscous timescale $t_\mathbf{visc}$ and the diffusion timescale $(t_\mathrm{diff,stat})$. The vertical dashed lines close to the inner rim and at $\sim 4$~AU represent $R_\mathrm{cor}$ and $r_\mathrm{DZ,out}$, respectively. }
    \label{fig:fid_timescales}
\end{figure}

\begin{figure}[ht!]
    \centering
    \resizebox{\hsize}{!}{\includegraphics{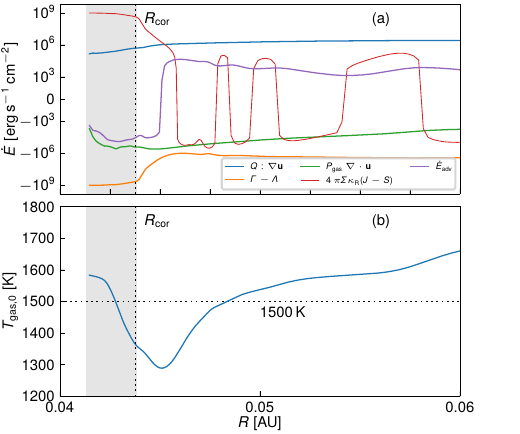}}
    \caption{In panel (a), the energy contributions to the overall inner energy budget for every disk annulus at radius $r$. $Q : \nabla \vec u$,  $\Gamma - \Lambda$, $P_\mathrm{gas} \, \nabla \cdot \vec u$, $4\,\pi \, \Sigma \, \kappa_\mathrm{R} ( J - S)$ and $\dot E_\mathrm{adv}$, denote energy rates per unit volume due to viscous heating,  heating/cooling over the surface, work done by gas pressure per unit time, radial radiation flux and net inner energy advection, respectively (cf.,~\equ{eq:ene}).
    In panel (b), the midplane gas temperature $T_\mathrm{gas,0}$ is plotted. 
    The vertical dashed line in all panels denotes the corotation radius $R_\mathrm{cor}$ and the gray area marks optically the thin inner region in the disk.}
    \label{fig:fid_temperature_profile}
\end{figure}

Inside the DZ close to $r_\mathrm{DZ,out}$ the radial field $\Brs$ decreases significantly compared to its value outside the DZ. 
This can be explained by comparing the viscous timescale $t_\mathrm{visc}$ to $t_\mathrm{diff,stat}$: If $t_\mathrm{diff,stat}$ is significantly shorter than $t_\mathrm{visc}$, then the large-scale field diffuses outwards before any significant inward-directed advection takes place. 
As a consequence, field lines are only weakly inclined and $\Brs$ stays small compared to $\Bz$ (cf. zone around $t_\mathrm{diff, DZ}$ in \fig{fig:fid_timescales}). 
Only very close to the inner rim $t_\mathrm{diff,stat}/t_\mathrm{visc} \approx 1$, which means that there $\Brs \approx \Bz$ (cf. spike in $\Brs$ close to the inner rim in (b) of \fig{fig:fid_magnetic_field} and in \fig{fig:fid_timescales}).
This steep increase in $t_\mathrm{diff,stat}$ close to $r_\mathrm{in}$ is a result of three effects, which occur in a very narrow radial region close to the inner rim: (i) At the transition from optically thin to optically thick gas a dip in $T_\mathrm{gas}$ (for details cf. below paragraph about the temperature profile and (b) of \fig{fig:fid_temperature_profile} leads to a reduced $\xeth$ and consequently to a larger $\bbar \eta$, which is equivalent to a shorter $t_\mathrm{diff,stat}$. (ii) However, just slightly further inwards, the stellar irradiation starts to dominate the disk temperature, which leads to rising $T_\mathrm{gas}$ (cf. (b) of \fig{fig:fid_temperature_profile}) and a steep decrease of $\bbar \eta$ once more (or equivalently to a sharp increase of $t_\mathrm{diff,stat}$, cf. \fig{fig:fid_timescales}) (iii) Closest to $r_\mathrm{in}$, $\bbar \eta$ rises again (equivalently $t_\mathrm{diff,stat}$ becomes shorter) slightly towards $r_\mathrm{in}$ which is due to small $\Sigma$ and relatively large $|\vec B|$ provided by the stellar dipole field (we recall that $\eta_\mathrm{A} \propto |\vec B|^2 \, \rho^{-2}$, also cf. \equ{eq:ambipolar_diffusion}).

The magnetic plasma beta at the midplane $\beta_0$ describes the relative importance of the large-scale field in the disk. We start the time evolution with a fossil field, which corresponds to $\beta_0(r=r_\mathrm{out}) = 10^2$ at the outer disk rim $r_\mathrm{out}$. In the AD-dominated outer disk, the model of \cite{dudorov2014} predicts a radial dependency $\beta_0 \propto r^{-3/2}$ in the case of an irradiated, optically thick disk, which our results also reflect (cf. (c) in \fig{fig:fid_magnetic_field}).

\begin{figure}[ht!]
    \centering
    \resizebox{\hsize}{!}{\includegraphics{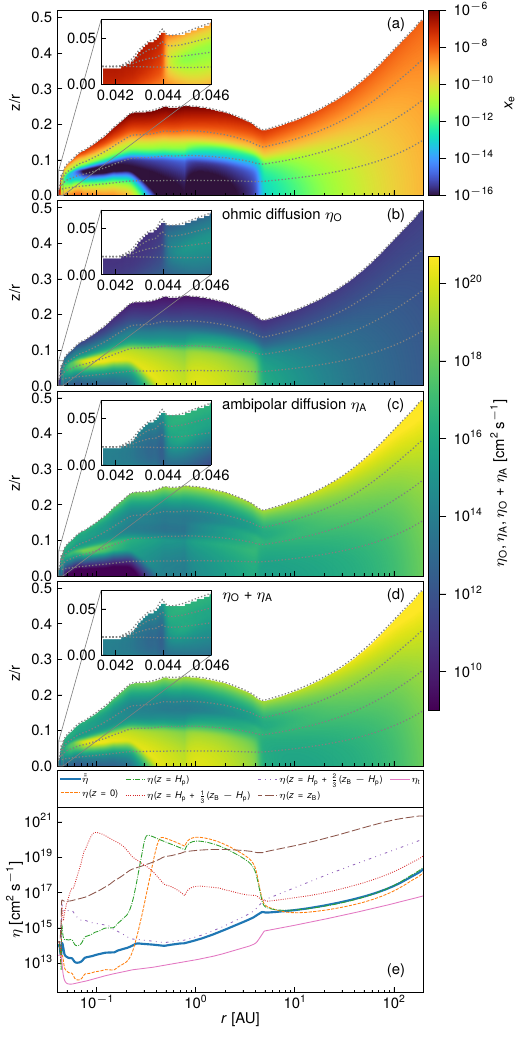}}
    \caption{For better understanding of the regimes of thermal and non-thermal ionization, the radial and vertical profiles of (a) ionization fraction $x_\mathrm{e}$, (b) the contribution of ohmic resistivity $\eta_\mathrm{O}$, (c) the contribution of ambipolar diffusion $\eta_\mathrm{A}$ and (d) the combined resistivity $\eta(r,z) = \eta_\mathrm{O}(r,z) + \eta_\mathrm{A}(r,z)$ are plotted for our reference model, which is explained in \sref{sec:magnetic_field_evolution} (cf. \tab{tab:ref_model}). The dashed gray lines in (a)-(d) correspond to specific heights above the disk midplane ($z = 0$, $z = H_\mathrm{p}$, $z = H_\mathrm{p} + \frac{1}{3}(z_\mathrm{B} - H_\mathrm{p})$, $z = H_\mathrm{p} + \frac{2}{3}(z_\mathrm{B} - H_\mathrm{p})$, and $z = z_\mathrm{B}$). In (e) the combined resistivity is plotted for those specific heights, as well as the resulting vertically averaged resistivity $\bbar \eta$ (thick blue line). For comparison the turbulent resistivity $\eta_\mathrm{t}$ corresponding to a Prandtl number of $1$ is also plotted, showing that $\eta_\mathrm{t} \ll \bbar \eta$ almost everywhere except for the very inner region.}
    \label{fig:fid_resistivity_profile}
\end{figure}

\paragraph{Resistivity profile:} The ionization fraction $x_\mathrm{e}(r,z)$ profile in radial and vertical direction is necessary to calculate the corresponding Ohmic and ambipolar resistivity profiles, $\eta_\mathbf{O}$ and $\eta_\mathrm{A}$, respectively. Both thermal and non-thermal ionization are taken into account (cf. \sref{sec:non_ideal_resistivity}) and show that the innermost region ($r \leq 0.044$~AU) is optically thin due to low $\Sigma$ and the gas temperature $\Tgas$ is dominated by stellar irradiation. Hence, the very inner disk has a high $\xe$; however, at $\sim 0.044$~AU the disk becomes sufficiently dense to turn optically thick for stellar radiation and the primary heating process transitions from stellar irradiation to viscous heating. Additionally, the disk has a DZ between $0.06 \, \mathrm{AU} \lesssim r \lesssim 5$~AU, where the inner part of the DZ close to the midplane $r \lesssim 0.2$~AU is subject to thermal ionization due to viscous heating (cf. region of large ionization in the inner disk region close to the midplane in (a) of \fig{fig:fid_resistivity_profile}). Above $z \approx H_\mathrm{p}$, thermal ionization is not effective anymore, since the disk is hydrostatically stratified and $\rho$ above $H_\mathrm{p}$ becomes very small. Finally, the upper disk is penetrated mainly by X-rays and cosmic rays, yielding an expected significant $\xenth$ in the surface layer.

Ohmic diffusion $\eta_\mathrm{O}$ is only significant inside the DZ, which is also in agreement with predictions made by other authors \citep[cf. e.g.][]{Bai2013,delage2022}. Interestingly, the DZ also spans above (and below) the thermally ionized region in the inner disk, which means that the very weakly ionized gas is shifted to intermediate heights, where thermal ionization cannot be maintained due to low gas density and non-thermal ionization sources cannot penetrate deep enough to ionize the gas. The disk surface layer is sufficiently well ionized by non-thermal processes, such that $\eta_\mathrm{O} \propto x_\mathrm{e}^{-1}$ is very inefficient in the disk surface (cf. (b) of \fig{fig:fid_resistivity_profile})

Ambipolar diffusion $\eta_\mathrm{A}$ is effectively quenched in the thermally ionized disk region, which is due to both high gas density $\rho$ and large $x_\mathrm{e}$ (we recall that $\eta_\mathrm{A} \propto |\vec B|^2 \, \rho^{-2} \, x_\mathrm{e}^{-1}$, cf. \equ{eq:parametrized_ambipolar_resistivity}). Inside the DZ $\eta_\mathrm{A}$ is also negligible compared to $\eta_\mathrm{O}$ due to its strong dependency on $\rho$. In the surface layer as well as in the outer disk; however, AD becomes the dominant diffusion process. The magnetic field strength $|\vec B|$ is roughly independent of height $z$, therefore the relative importance of $|\vec B|$ compared to $\rho$ increases with $z$ (cf. (c) of \fig{fig:fid_resistivity_profile}).
The combined resistivity profile $\eta = \eta_\mathrm{O} + \eta_\mathrm{A}$ is plotted in panel (d) of \fig{fig:fid_resistivity_profile}. 
The regions of low $\eta$ are the thermally ionized disk region, a thin layer at intermediate height $z$ inside the DZ and in the outer disk close to the midplane. 
Recalling \equ{eq:def_averaged_resistivity}, the vertical averaging process of $\bbar \eta$ favors layers of low $\eta$, as in those regions magnetic flux transport $\bar u_\mathrm{\psi}$ is more efficient. 
As explained in \sref{sec:non_ideal_resistivity} it is important for our model to vertically average the resistivity profile $\bbar \eta$ due to constraint of a 1+1D model. 
Therefore we evaluate $(\eta_\mathrm{O} + \eta_\mathrm{A})(r,z)$ at 5 different heights, including at the midplane, the pressure scale height and the magnetic surface, $z=0$, $z = H_\mathrm{p}$ and $z_\mathrm{B}$, respectively. 
Additionally, for $\bbar \eta$ to also incorporate the effects of the DZ above the thermally ionized region and of the disk layer of comparatively low resistivity between the DZ and the surface layer, $\bbar \eta $ is evaluated at two additional heights, $z = H_\mathrm{p} + \frac{1}{3} ( z_\mathrm{B} - H_\mathrm{p})$ and  $z = H_\mathrm{p} + \frac{2}{3} ( z_\mathrm{B} - H_\mathrm{p})$ to adequately describe that in different radial sections different vertical layers contribute dominantly to $\bbar \eta$ (cf. (e) of \fig{fig:fid_resistivity_profile}):
\begin{itemize}
    \item In the thermally ionized region ($r \lesssim 0.3$~AU) $\eta$ is dominated by the vertical disk layers close to the disk midplane (the dashed orange line and dash-dotted green line are closest to the blue full line for $r \lesssim 0.3$~AU, cf. panel (e) of \fig{fig:fid_resistivity_profile}).
    \item In the region inside the DZ, but further out than the thermally ionized region ($0.3~\mathrm{AU} \lesssim r \leq r_\mathrm{DZ,out}$), $\eta$ is dominated by a region above the vertical extent of the DZ but deep enough below the disk surface for non-thermal radiation not ionizing the gas. In this intermediate region, AD is the dominant non-ideal diffusion process (the purple dash-dotted line is very close to the full blue line in this region, cf. panel (e) of \fig{fig:fid_resistivity_profile}).
    \item Outside of the DZ $r \ge r_\mathrm{DZ,out}$ the disk is dominated by AD, which is least effective close to the disk midplane and therefore most dominant for calculating the average $\bbar \eta$ (the dashed orange line and dash-dotted green line are very close to the blue full line, cf. panel (e) of \fig{fig:fid_resistivity_profile}).
\end{itemize}
For comparison, the turbulent resistivity $\eta_\mathrm{t}$ resulting from a magnetic Prandtl number $\mathcal{P} = 1$ is also included in the vertical average of $\bbar \eta$ and is taken to be vertically constant, which means that $\eta_\mathrm{t}$ is added at every height to $\eta(r,z)$. It has been pointed out by other authors \citep[e.g.][]{dudorov2014} that $\eta_\mathrm{t}$ is only important in the innermost hot disk with high $x_\mathrm{e}$ (cf. the magenta line in panel (e) of \fig{fig:fid_resistivity_profile}).

\paragraph{Temperature profile:}The disk temperature profile $T_\mathrm{gas}(r)$ is important for determining the thermal gas pressure $P_\mathrm{gas}$ as well as the pressure scale height $H_\mathrm{p}$ and the magnetic surface $z_\mathrm{B}$. All of those quantities are important to get the magnetic flux transport and the large-scale magnetic field topology $(B_\mathrm{r}^\mathrm{s}(r),\, 0,\, B_\mathrm{z}(r))^\mathrm{T}$. The optically thick regions of the disk are dominated by viscous heating, whereas $T_\mathrm{gas}$ in the optically thin regions is determined by stellar irradiation and the ambient temperature \citep[cf. for implementation details][]{ragossnig20,Steiner21}.

We want to highlight one important temperature feature close to the inner edge of the disk, resulting in a distinct dip in the temperature of the disk (cf. (b) of \fig{fig:fid_temperature_profile}).
At the transition between the optically thick (white) and the optically thin (grey) region, we compare the individual energy contributions.
The pressure work rate $P_\mathrm{gas} \nabla \cdot \vec u$ has a negative sign, which means that pressure aids the gas to be pushed inwards due to $- \Delta P_\mathrm{gas} < 0$ in this area. 
Additionally, the net energy advected by the radial gas flow into a disk annulus just outside $R_\mathrm{cor}$ is positive, because of the stellar torques acting on the inner disk. However, for $r < R_\mathrm{cor}$ the stellar dipole brakes the gas and gas starts to fall radially faster towards the star. This leads to  $\dot E_\mathrm{adv} < 0$, because more material exits than enters any disk annulus at $r < R_\mathrm{cor}$ (cf. (a) of \fig{fig:fid_temperature_profile}). Combined $P_\mathrm{gas} \nabla \cdot \vec u + \dot E_\mathrm{adv} < 0$ for $r \approx R_\mathrm{cor}$, which leads to a net drain of inner energy $e$ from the disk annuli in this area and which is equivalent to a lower $T_\mathrm{gas,0}$. For our reference model, this leads to a drop below $T_\mathrm{gas,0} < 1500$~K (cf. (b) of \fig{fig:fid_temperature_profile}).
At this temperature, some dust can still survive \citep[cf. e.g.][]{ferguson05}.

\subsection{Variation of stellar parameters}
\label{sec:variation_stellar_parameters}

In this section we investigate the influence of stellar magnetic field strength $|\vec B_\star|$, rotation period $P_\star$ and the X-ray luminosity $L_\mathrm{X}$ on the large-scale magnetic field topology. All stellar parameters are held constant throughout the magnetic field evolution toward the steady-state solution.
% , hence no star-disk interaction is considered in this work. 
The parameter range for $P_\star$ is varied between 4 and 6 days, which is based on observations of young stellar objects in the Orion Nebular Cluster \citep[][]{herbst02}. We choose $|\vec B_\star|$ to range from $200$~G to $1$~kG by referring to the analysis of observed magnetic maps published by \cite{Johnstone14}.
We also study the effects of an varying X-ray luminosity $\LX$ by lowering and increasing our reference value of $\LX = 10^{30} \, \mathrm{erg \, s^{-1}}$ by a factor of 10, which is well within the limits of observations made by \cite{preibisch2005}.

\subsubsection{Influence of stellar rotation period}

\begin{figure}[ht!]
    \centering
    \resizebox{\hsize}{!}{\includegraphics{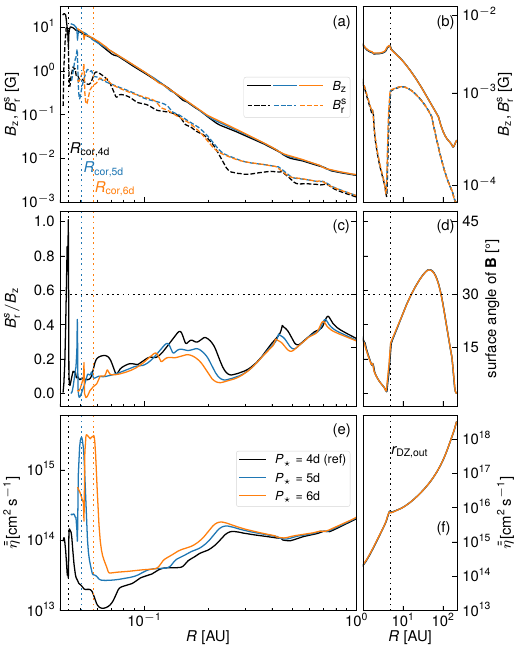}}
    \caption{Comparison of the magnetic field topology between our reference model ($P_\star = 4$~days, black) and two stars with $P_\star = 5$~days (blue) and $P_\star = 6$~days (orange). The vertical dashed lines mark the positions of $R_\mathrm{cor}$ and are color-coded accordingly. Panels (a) and (b) show the vertical and radial magnetic field profiles in the inner and outer disk, respectively. Panels (c) and (d) depict the conductivity-weighted, vertically averaged resistivity $\bbar \eta$, whereas panels (e) and (f) describe the ratio $B_\mathrm{r}^s / B_\mathrm{z}$. The scale on the right hand side of (c) and (d) denotes the angle of the magnetic field lines to the midplane normal, which corresponds to $\tan^{-1}(B_\mathrm{r}^s / B_\mathrm{z})$. The dashed horizontal line marks a surface angle of $30$°, which is an indicator for efficient MHD winds (see text).}
    \label{fig:stellar_period}
\end{figure}

A longer stellar rotation period $P_\star$ yields $R_\mathrm{cor}$ and $r_\mathrm{in}$ moving radially outwards \citep[cf.,][and all panels of~\fig{fig:stellar_period}]{Gehrig2022}. 
Different rotation periods $P_\star$ do not have any significant influence on the disk structure or the magnetic field topology outside of $r \gtrsim 0.3$~AU (cf. (b), (d) and (f) of \fig{fig:stellar_period}).
In the inner disk, however, the disk and large-scale magnetic field have a dependency on $P_\star$. 
The stellar magnetic field just outside of $R_\mathrm{cor}$ accelerates the disk gas, therefore counteracting viscous accretion, which is usually referred as 'propeller effect' \citep[cf.][]{Steiner21}. This leads to a pile-up of mass in the area outside of $R_\mathrm{cor}$ in the disk region $R_\mathrm{cor} \leq r \lesssim  0.3~\mathrm{AU}$. 
A higher $\Sigma$ diminishes $\xenth$ and therefore increases $\eta(r,z)$ in the disk layers at intermediate heights.
However, those intermediate-height disk layers are essentially defining $\bbar \eta$ in the disk for $R_\mathrm{cor} \leq r \lesssim 0.3~\mathrm{AU}$ (cf. (e) of \fig{fig:fid_resistivity_profile}), hence $\bbar \eta$ is increased in this region (cf. (e) of \fig{fig:stellar_period}). Close to $R_\mathrm{cor}$, we observe that $\bbar \eta$ increases significantly more for slower rotating stars. This is due to lower $T_\mathrm{gas,0}$ for larger $r$, leading to lower $\xe$ and therefore to larger $\bbar \eta$.  In panel (b), we can then see the expected result that a larger $\bbar \eta$ causes less inclined field lines.
For $r < R_\mathrm{cor}$ the narrow and steep increase of $\Brs / \Bz$ (cf. \sref{sec:reference_model}) becomes less pronounced for larger $P_\star$ (cf. (c) of \fig{fig:stellar_period}). 
This is because not only $R_\mathrm{cor}$ but also the optically thin region close to $r_\mathrm{in}$ shifts towards larger radii. There, the disk is colder (compared to faster rotating stars), which results in less effective thermal ionization and a less pronounced dip of $\eta$ in the innermost disk region.

\subsubsection{Influence of stellar magnetic field strength}

\begin{figure}[ht!]
    \centering
    \resizebox{\hsize}{!}{\includegraphics{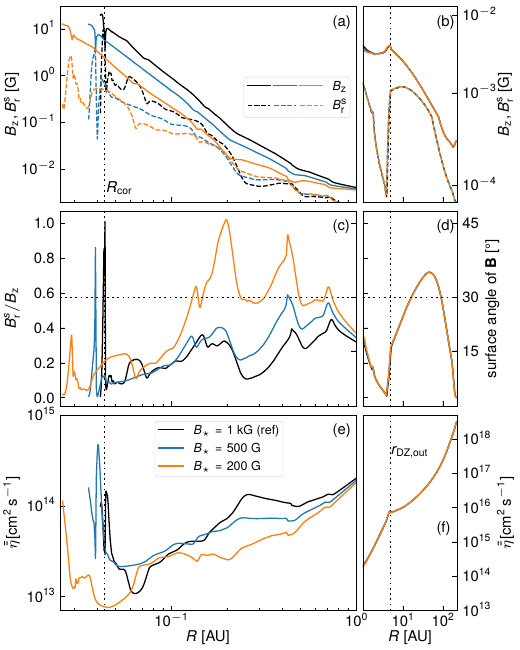}}
    \caption{Same as \fig{fig:stellar_period}, but for different $|\vec B_\star|$. The stellar field strength $|\vec B_\star|$ of $1$~kG, $500$~G, and $200$~G corresponds to black, blue, and orange lines, respectively.}
    \label{fig:cmp_Bstar}
\end{figure}

Different stellar magnetic field strengths $|\vec B_\star|$ do not alter the large-scale field topology in the outer disk for $r \gtrsim 1~\mathrm{AU}$ (cf. (b), (d) and (f) of \fig{fig:cmp_Bstar}).
The inner radius $r_\mathrm{in}$ moves closer to the star for weaker stellar dipole fields \citep[cf.][and \fig{fig:cmp_Bstar}]{Steiner21}, hence the disk close to $r_\mathrm{in}$ gets hotter.
This means that for weaker $|\vec B_\star|$ even during the dip in $T_\mathrm{gas,0}$ (cf. \sref{sec:reference_model}) the gas stays hot enough for efficient thermal ionization and leads to a less distint spike in $\Brs / \Bz$ for weaker dipole fields (cf. (c) of \fig{fig:cmp_Bstar}).
The magnetic field is also altered inside the DZ, which is caused by a different mechanism as in the very inner disk. A larger $|\vec B_\star|$ also leads to a larger large-scale field strength $B_\mathrm{d}$ in the radial range $0.1~\mathrm{AU} \lesssim r \lesssim 1~\mathrm{AU}$ (cf. (a) of \fig{fig:cmp_Bstar}). 
% alternative
In this region, ambipolar diffusion $\eta_\mathrm{A}$ dominantly contributes to $\bbar \eta$ (\sref{sec:reference_model}).
As $\eta_\mathrm{A} \propto B^2$, a small value of $B_\star$ results in a smaller resistivity and the large-scale field is dragged inward more efficiently.
Consequently, the surface angle increases (cf. (b) and (c) of \fig{fig:cmp_Bstar}).
This effect is even amplified by the fact that a weaker field provides less resistance against bending, causing the difference between different stellar magnetic field strengths to be even more significant. 

\subsubsection{Influence of stellar X-ray luminosity}

\begin{figure}[ht!]
    \centering
    \resizebox{\hsize}{!}{\includegraphics{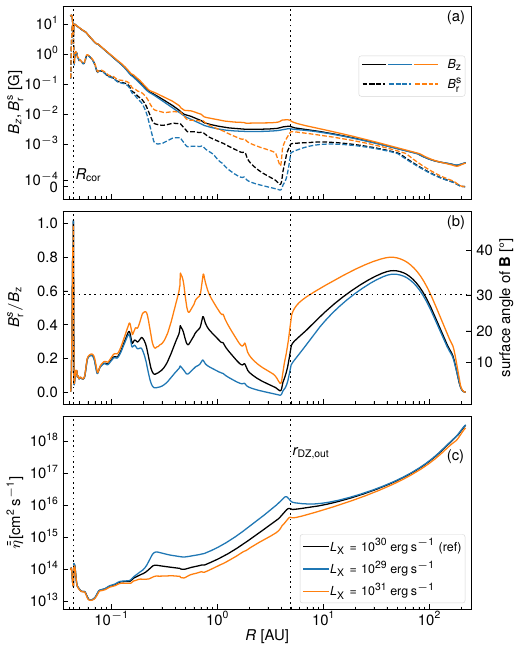}}
    \caption{Same as in \fig{fig:stellar_period}, but for different stellar X-ray luminosities $L_\mathrm{X}$ reaching from $10^{29} \, \mathrm{erg \, s^{-1}}$ to $10^{31} \, \mathrm{erg \, s^{-1}}$.}
    \label{fig:cmp_LX}
\end{figure}

The stellar X-ray luminosity $L_\mathrm{X}$ is directly proportional to the non-thermal X-ray ionization, both direct ionization \equ{eq:xr_direct_ion_rate} and through scattering \equ{eq:xr_scatter_ion_rate}. Therefore it is expected that an increase of $L_\mathrm{X}$ causes a higher non-thermal ionization fraction $x_\mathrm{e,nth}$ in the upper disk layers and hence leading to a reduced vertically averaged resistivity $\bbar \eta$. 
This is indeed the case for our models (cf. \fig{fig:cmp_LX}); however in the inner disk region $r \lesssim 0.15$~AU the dominant ionization mechanism is thermal ionization, which explains the lack of variation in this disk region (cf. \fig{fig:cmp_LX}). 

A lower $\bbar \eta$ means a more effective coupling of the large-scale field to the disk, which leads to more magnetic flux transport towards the star and therefore a stronger large-scale disk field (cf. (a) in \fig{fig:cmp_LX}), which in turn increases $\eta_\mathrm{A}$ and so lessens the effect of a stronger $L_\mathrm{X}$ to a certain degree, but remains of minor significance compared to the effect of increased $\xenth$. More effective magnetic field dragging also results in a stronger radial large-scale field component $B_\mathrm{r}^\mathrm{s}$ as well as a larger ratio $B_\mathrm{r}^\mathrm{s} / B_\mathrm{z}$, which corresponds to a larger surface angle of the resulting field (cf. (b) and (c) of \fig{fig:cmp_LX}). Contrary to the variation of $P_\star$ and $B_\star$, also the magnetic field in the outer parts of the disk is influenced by a varying $L_\mathrm{X}$. This is also expected, as the dominant ionization process in the outer disk is non-thermal ionization. 
In summary, the effects of $L_\mathrm{X}$ on the large-scale field topology lead to a stronger large-scale field for larger values of $L_\mathrm{X}$ and additionally allow for a significantly stronger inclined large-scale field for all radii $r \gtrsim 0.2$~AU.

\subsection{Influence of the mass transport rate}
\label{sec:influence_dotm}

We investigate how the variation of $\dot M$ influences the magnetic field in the disk. 
In the picture of a steady-state disk and the viscous $\alpha$ description, the mass transport rate through the disk defines the surface density profile.
A higher $\dot M$ results in a larger DZ, which extends further outwards due to a larger $\Sigma$ and therefore to lower non-thermal ionization in the layers below the disk surface \citep[cf.,][]{Steiner21,delage2022}. 
The outer boundary of the disk $r_\mathrm{out}$ is defined by the condition of $\Sigma = 1 \, \mathrm{g \, cm^{-2}}$ \citep[cf.,][]{vorobyov2020} and hence also shifts towards larger radii for a more massive disk.

\begin{figure}[ht!]
    \centering
    \resizebox{\hsize}{!}{\includegraphics{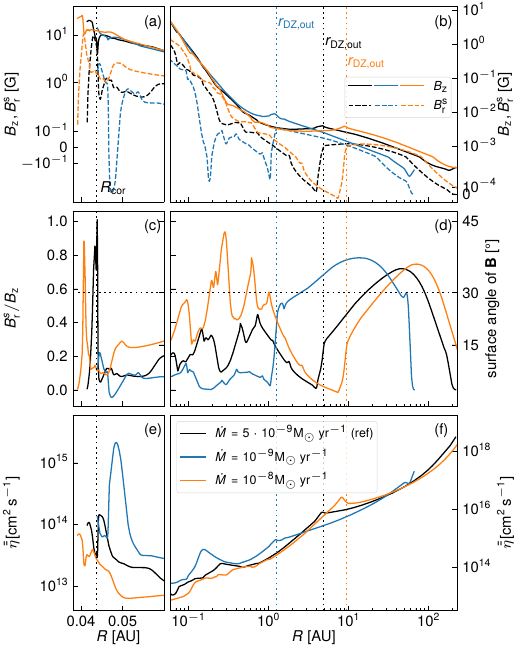}}
    \caption{Same as in \fig{fig:stellar_period}, but for different $\dot M$. The color-coded vertical dashed lines correspond to the outer dead zone boundaries $r_\mathrm{DZ,out}$ of the respective disk model. }
    \label{fig:cmp_Mdot}
\end{figure}

Having a more massive and larger disk, we expect the large-scale magnetic field to react in the outer disk in the following way: 
A larger disk can drag (and amplify) the fossil field inwards from further out.
Hence, for a given $r$ in the outer disk (sufficiently far away from the outer disk boundary to rule out any boundary effects), the large-scale field is expected to be slightly stronger. (ii) A lower $\xe$ due to a larger $\Sigma$ and therefore reduced $\xenth$ results in less magnetic flux transport and therefore leads to a less inclined field.
We can confirm both of those expectations with our results (cf., $r \sim 10$~AU in (b) and (d) of \fig{fig:cmp_Mdot})
The increased field inclination in the DZ for a more massive disk can be explained in the following way: A more massive disk is more extended in the vertical direction (larger $\Hp$), which means that the vertical disk layers contributing most to magnetic flux transport are at larger $z$.
As stated in \sref{sec:model_description}, in our model $\ur \propto z^2$ (cf. \equ{eq:ur_profile}) and for the vertical average $\bbar u_\mathrm{r}$ the disk layers of low resistivity are given more weight (cf. \equ{eq:def_averaged_gas_velocity}).
This means that for a more massive disk, magnetic flux transport is faster in the layers above the dead zone, which results in a larger inclined large-scale field (cf. (d) of \fig{fig:cmp_Mdot}).

In the inner disk, $r_\mathrm{in}$ also shifts towards smaller radii for a larger $\dot M$ due to increased ram pressure \citep[for details cf.][cf. also (a), (c) and (e) of \fig{fig:cmp_Mdot}]{Steiner21}.
Our model with $\dot M = 10^{-9} \, \mathrm{M_\odot \, yr^{-1}}$ shows a large increase in $\bbar \eta$ (compared to the other models with higher $\dot M$) at the transition from an optically thick to an optically thin disk (cf. \sref{sec:reference_model} and panel (b) of \fig{fig:fid_temperature_profile}), because the disk is colder, both due to less viscous heating and its larger radial distance from the star. The strong diffusion of $\Brs$ in this region is superimposed with the radial contribution of $B_\mathrm{r,\star}^\mathrm{s}$ (cf. \equ{eq:stellar_Br} results in $\Brs$ changing sign in a very narrow region. The spike in $\Brs / \Bz$ is not visible for a disk with low $\dot M$ because the disk is truncated before $T_\mathrm{gas,0}$ rises high enough to decrease $\bbar \eta$ sufficiently (cf. (e) of \fig{fig:cmp_Mdot}.
In contrast, for larger disk masses the dip in $T_\mathrm{gas,0}$ is less distinct due to its proximity to the star, which also results in a less pronounced spike in inclination compared to our reference model (cf. (c) of \fig{fig:cmp_Mdot}).

\section{Discussion}
\label{sec:Discussion}

In this paper we have presented long-term, hydrodynamic simulations of a PPD with a large-scale magnetic field until the disk and large-scale magnetic field become stationary. This allows us to iterate further on the work done by other authors \citep[cf. e.g.][]{Lubow94, Guilet2014}, who have evolved a large-scale field threading a simplified, stationary disk, but whose models lack the inclusion of laminar, non-ideal effects, which in our case are limited to OD and AD. Additionally, due to the implicit time integration scheme of TAPIR, we are able to include the very inner disk region and therefore include the impact of a stellar dipole both on the inner disk and on the large-scale field by using the hydrodynamic disk evolution model of e.g. \cite{Steiner21, Gehrig2022}. 

Our results show that the large-scale field develops on the magnetic diffusion timescale, which differs between the DZ and the region further outwards, causing a distinction between the field inside and outside the DZ. Furthermore, the large-scale field is significantly bent in the very inner disk and in the outer disk (outside the DZ), but not in the dead zone close to the outer dead zone rim, which is in agreement with simulations carried out by other authors \citep[cf. e.g.][]{Bai2013, zhu2018}. Additionally we show that varying stellar parameters such as stellar magnetic field strength $|\vec B_\star|$, X-ray luminosity $\LX$, and stellar rotation period $P_\star$ can influence the large-scale magnetic field topology by altering the large-scale field strength and inclination. 
Our study also reveals which parameters influence the large-scale field in the very inner disk, the DZ, or the outer regions of the disk.

\subsection{Comparison with recent studies}\label{sec:comparison}

Our approach is novel in that it is able to evolve a PPD, including the very inner disk, threaded by a large-scale field magneto-hydrodynamically up to the point where the model becomes stationary without imposing any external stationarity constraint on it. Comparing our model to recent studies using other methods to determine the large-scale field structure in a PPD is therefore important to provide additional justification for our approach.

\cite{dudorov2014} use a standard, geometrically thin, $\alpha$-viscosity accretion disk of \cite{shakura73} to describe the disk and also include stellar irradiation as an additional heating source apart from viscous heating. A significant difference compared to our model is that they neglect radial magnetic flux transport, which makes the calculation of the external magnetic field solution at each time step unnecessary \citep[cf. e.g.][]{Ogilvie1997, Guilet2014} and a constant fossil field description suffices. By contrast, our model incorporates a changing external field due to radial dragging, which also influences the vertical profile of $\Br$ inside the disk \citep[cf.][]{Guilet2012, leung2019} and therefore leads to changes in the magnetic field profiles compared to \cite{dudorov2014}. The radial profiles of $\Bz$ and $\Brs$ compare qualitatively well with our results. Also the dependency of dust grain size and $\LX$ agrees well with our findings; however, in the very inner disk there are differences: (i) The disk of \cite{dudorov2014} is everywhere optically thick, which is not the case close to the inner boundary for a more realistic inner disk boundary, which leads to a different temperature profile close to the inner rim and hence influences $\xe$ as well as $\eta$. (ii) No stellar dipole has been modeled in their work, which would alter the disk structure around the corotation radius \citep[cf.][]{Steiner21} and also influences the large-scale disk field close to the inner rim. Notably, we obtain less inclined field lines than \cite{dudorov2014} in the inner disk, which is a combined effect of an optically thin inner disk and a more realistic inner boundary. Additionally, we don't include a toroidal field component $\Bphis$ in the disk exterior above and below the disk, as this would add the complexity of launching magnetically induced outflows.
\cite{dudorov2014} include $\Bphis$, but do not consider outflows in their model.

We also want to compare our results to both \cite{mohanty2018} and \cite{delage2022}, as the authors of both papers use an alternative approach to determine the large-scale magnetic field structure. The primary focus of those studies is to determine the effect of the magnetic field strength and non-ideal MHD effects on the $\alpha$-viscosity and therefore on the disk structure. Nevertheless, the determination of the ionization fraction $\xe$ is implemented analogously: \cite{mohanty2018} use thermal ionization in the inner disk, whereas non-thermal ionization in our models is analogous to \cite{delage2022}. In both papers, a standard diffusion disk model is used to describe a stationary disk, which especially in the inner disk leads to differences between the disk structure in \cite{mohanty2018} and our model. A diffusion approach cannot take into account any pressure gradients occurring in the inner disk and is not able to account for the effect of a stellar dipole \citep[cf.][]{Steiner21}, therefore the $\Sigma$-profile in our disk differs from the model in \cite{mohanty2018}. Moreover, they use a constant opacity $\kappa_\mathrm{R} = 10 \, \mathrm{cm^2 \, g^{-1}}$, which is different from $\kappa_\mathrm{R}$ in our models, as we obtain a small dusty ring just outside the optically thin region closest to the inner disk rim for most of our models. This opacity feature, together with the ionized, strongly heated innermost disk, leads in our simulations to a large inclination of the large-scale disk field very close to the star. However, \cite{mohanty2018} uses an $\alpha$-viscosity, which is dependent on the magnetic field strength. This results in  $\alpha \approx 0.1$ in the very inner disk and is a factor 10 higher than our adopted value of $\alpha_\mathrm{MRI} = 0.01$ (cf. \tab{tab:ref_model}). Therefore, the disk structure in the inner disk is accreting more effectively, which yields a less steep increase in $\Sigma$ for larger $r$ and a DZ inner boundary further outwards. The most notable difference in both \cite{mohanty2018} and \cite{delage2022} with respect to the determination of the large-scale field strength in the disk is their use of an an iterative approach: To obtain a stationary disk the condition $\dot M = \mathrm{const.}$ has to be fulfilled. The $\alpha$-viscosity at every radius determines $\dot M$ and is dependent on the large-scale field strength \citep[root-mean-square (r.m.s.) field, cf.][]{sano2004} under the assumption of a maximally efficient MRI. Then, iteratively, a disk structure can be obtained with $\dot M = \mathrm{const.}$ and a corresponding $\alpha$ and magnetic field. This approach cannot account for an external fossil field or a stellar dipole field threading the inner disk, since the field is solely determined by the condition of maximally efficient MRI. There is no evidence that MRI determines the large-scale field strength and not vice versa \citep[also mentioned and discussed by][]{delage2022}. The results for the radial profile of $|\vec B|$ in \cite{delage2022} differ from our results especially in the DZ, where their magnetic field strength follows a power-law over the whole disk, contrary to our results, where a DZ has a clear impact on the radial profile of $\Bz$ and $\Brs$.

\subsection{Implications of large-scale field inclination}
\label{sec:implications_inclination}

\begin{figure}[ht!]
    \centering
    \resizebox{\hsize}{!}{\includegraphics{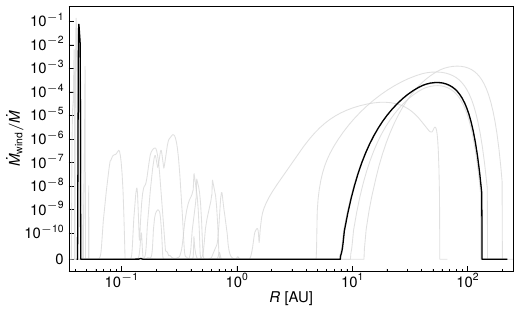}}
    \caption{We show the MCW mass loss $\dot M_\mathrm{wind}$ according to \equ{eq:wind_mass_loss} for our reference model (black). All other models discussed in this paper are shown as light-gray lines to further illustrate the two dominant wind regimes in the innermost disk and in the outer disk. }
    \label{fig:cmp_wind}
\end{figure}

The inclination angle of the large-scale field lines is a crucial parameter in determining the existence and strength of magneto-centrifugally accelerated outflows \citep[cf. e.g.][]{ogilvie1998, Campbell1999, ogilvie2001}).
We adopt the simple wind model of \cite{ogilvie1998} for a rough estimate of the wind mass loss $\dot M_\mathrm{wind}$, which we could expect for a MCW launching from the disk surface.
In accordance with our 2-zone model of \cite{Guilet2012}, the large-scale field is assumed to be force-free above the disk, which results in rigidly rotating field lines up to the Alfvénic surface (the surface above the disk where the wind velocity surpasses the Alfvén-velocity),
\begin{align}
    B_\mathrm{r}(z > \zB) &= \Bz \, \tan(i) \;,
\end{align}
where $i$ denotes the inclination angle at the disk surface with respect to the midplane normal. We note that we only assume the Alfvénic surface to be located somewhere above the sonic surface and make no effort to determine its actual position in this work. In theory, the Alfvén point along a field line determines how much angular momentum can be transported by the wind, whereas the sonic point determines the mass of the wind flow. Hence, for an estimate of the wind mass, we use that it has to pass through the sonic point at height $z_\mathrm{sonic}$ and assume that $\zB < z_\mathrm{sonic} \ll r$. The last assumption is needed for the approximation of a rigid field line. Then the MCW mass loss $\dot M_\mathrm{wind}$ is determined through \citep[for a much more comprehensive explanation, cf.][]{ogilvie1998,ogilvie2001},
\begin{align}
    \Omega_1^\mathrm{s} &\equiv \frac{u_\mathrm{\varphi}(z=\zB)}{r} - \Omega_\mathrm{K}  \,, \\
    \Delta \Phi &\equiv \frac{\Omega_\mathrm{K} \, \zB - 2 \, \Omega_1^\mathrm{s} \, r \, \tan(i)}{2( 3 \, \tan^2(i) - 1)} \;, \\
    \dot M_\mathrm{wind} &= \rho(z=\zB) \, \cs \, \cos(i) \exp\left( \frac{-\Delta \Phi}{\cs^2} - \frac{1}{2} \right) \;. \label{eq:wind_mass_loss}
\end{align}
Here, $\Delta \Phi$ denotes the effective centrifugal potential and $\Omega_1^\mathrm{s}$ the deviation from Keplerian rotation $\Omega_\mathrm{K}$ at the disk surface. We further assume that the MCW is launched at $z = z_\mathrm{B}$. The resulting $\dot M_\mathrm{wind}$ for our reference model is shown in \fig{fig:cmp_wind} and reveals that the most favorable regions for a MCW are very close to the inner disk boundary $r_\mathrm{in}$ and in the outer, AD-dominated disk for $r > r_\mathrm{DZ,out}$. 
These regions of increased $\dot M_\mathrm{wind}$ have also been obtained through simulations performed by other authors \citep[cf. e.g.][]{armitage11,suzuki2016,lesur2021} and observations \citep[cf.,][]{guedel2018}.
Nevertheless, we want to stress the point that this analysis of MCW mass loss is preliminary and serves the purpose of showing that purely magneto-centrifugally driven winds are possible in disks for realistic $\alpha$-values and fossil field strengths. 
We will interpret these results as a promising starting point for upcoming simulations, including the effects of MHD outflows.
A realistic disk model with MHD outflows has to include the effects of angular momentum loss from the wind, which would lead to increased mass transport and magnetic flux transport, and most likely would lead to the opening of a gap \citep[cf. ][]{armitage11}.
Additionally, some models suggest an inverted $\Sigma$-profile in the case of effective winds \citep[cf.][]{suzuki2016}, which further confirms the need that winds to be included in our MHD disk evolution model as a next step.

\subsection{Variation of the fossil field}
\label{sec:variation_field_properties}

The fossil field strength $B_\infty$ in PPDs is a topic of active studies \citep[cf. e.g.][]{crutcher2004, masson2016,tsukamoto2023} and lies in the range of [$10^{-5}$~G, $10^{-1}$~G]. Especially the upper end of this range is intriguing with respect to disk evolution, since such strong magnetic fields can potentially influence disk evolution. Also, with respect to magnetic field evolution in a dis,k stronger fields are expected to behave differently. A stronger field causes a larger effective resistivity, especially in the outer disks, since there AD is dominant (recall that $\etaA \propto |\vec B|^{2}$, cf.~\equ{eq:parametrized_ambipolar_resistivity}). 

\begin{figure}[ht!]
    \centering
    \resizebox{\hsize}{!}{\includegraphics{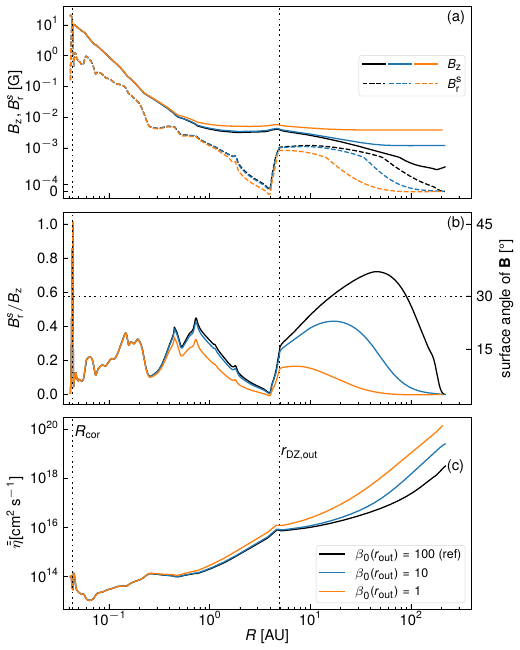}}
    \caption{Stationary magnetic field topology for different initial fossil field strengths. The reference model uses a fossil field with $\beta_0 = 10^2$ (black) and is compared to two models with a stronger initial field ($\beta_0 = 10$ (blue) and $\beta_0 = 1$ (orange)). Same panels as in \fig{fig:cmp_Bstar}. The vertical dashed lines correspond to $R_\mathrm{cor}$ and $r_\mathrm{DZ,out}$.}
    \label{fig:cmp_beta_out}
\end{figure}

An increased $\bbar \eta$ then causes less effective magnetic flux transport and a less inclined field geometry in the outer disk (cf. (b) of \fig{fig:cmp_beta_out}). Interestingly, the field strength for the two models with $\beta_0(r_\mathrm{out}) = 10$ and $\beta_0(r_\mathrm{out}) = 100$ is almost indistinguishable for all $r \lesssim r_\mathrm{DZ, out}$. The model with $\beta_0(r_\mathrm{out}) = 1$ yields a larger $\Bz$ until the stellar dipole dominates the large-scale field strength. Summarizing the results shows that a stronger fossil field $B_\infty$ leads to a less inclined outer disk field, but has no effect on the inner magnetic field topology due to the dominance of the stellar dipole field.

We want to emphasize that these results should be taken with a grain of salt, since a basic assumption of our large-scale field model is the two-zone model of \cite{Guilet2012,leung2019}, which assumes a passive magnetic field in the disk ($\beta_0 \gg 1$), which is clearly not the case in our model with $\beta_0(r_\mathrm{out}) = 1$ )(cf. \sref{sec:vertical_vel_profile}). A large-scale field with $\beta_0 \leq 1$ would influence the vertical disk structure \citep[cf. e.g.][]{Ogilvie1997} and therefore should be taken as a rough estimate of the outer magnetic field structure.

\subsection{Model limitations}
\label{sec:model_limitations}

We attempt in this work a long-term MHD simulation of the whole disk while including the very inner disk region $r~<~0.1~$~AU, a stellar dipole field, and a large-scale disk field. Covering sufficient time to allow the disk and large-scale field to become stationary while simultaneously including the inner disk makes it necessary to make various simplifications and assumptions:
\begin{itemize}
    \item We use a 1+1D model to conduct our simulations, hence the disk and the large-scale field are assumed to be axisymmetric. This constrains us to treat some intrinsically 3D-natured phenomena in a simplified way. One example is the inner boundary: The disk is magnetically disrupted by the stellar magnetic field at some radius close to the star and is accreted along magnetic funnels \citep[cf. e.g.][]{Romanova04}. However, such funnels are highly non-axisymmetric, and we therefore use the results of \cite{bessolaz08} to construct a physically motivated inner boundary \citep[cf. ][]{Steiner21}. The disk in the vertical direction is assumed to be hydrostatic, and the disk gas temperature is assumed constant for the hydrodynamic evolution of the disk; however, for the approximation of the ionization fraction, a temperature profile is used to obtain a better model of the vertical ionization profile. This is slightly inconsistent, but is not believed to change our results, as the temperature for the hydrodynamic evolution would be included as a density-weighted vertical average and hence would be very close to the mid-plane value of $\Tgas$ (cf. \sref{sec:non_ideal_resistivity}).
    % \newline
    \item In this work, we focus on studying the poloidal magnetic field structure and do not include toroidal fields. 
    Although we consider the effects of toroidal fields inside the disk, the toroidal field in the disk exterior is assumed to vanish. 
    However, the large-scale field very likely winds up corresponding to a disk field $\Bphi^\mathrm{s}$ component in the disk exterior above the disk surface. 
    This, in turn, would imply some outflow coupling to the large-scale field. 
    The launching mechanism could either be purely magneto-centrifugally, purely thermally (photo-evaporation), or a combination of both mechanisms \citep[cf. e.g.][]{lesur2021}. 
    The addition of magnetically and thermally induced outflows and their effects on the long-term disk evolution are the focus of subsequent studies using this model, as it is not constrained to steady-state solutions. 
    This work focuses on the effects of a stellar dipole and stellar rotation on the large-scale field topology, while the disk is self-consistently simulated.
    % \newline
    \item We do not include FUV radiation in our model, which is expected to strongly ionize the upper atmosphere \citep[cf. e.g.][]{rab2017} and can be important in the study of magnetic field evolution, as the large-scale field is radially transported in the disk surface layer in large parts of the disk. 
    Additionally, an ionized upper atmosphere is important for the study of magnetically induced outflows and should be included in upcoming papers aiming to investigate disk evolution, including MCWs.
    \item We do not include the effects of EUV heating of the thin, upper disk atmosphere due to secondary processes following the ionization of the gas. 
    As proposed by \cite{glassgold2004, glassgold2012, guedel2015}, up to 50\% of the energy deposited in the upper atmosphere is used to heat the gas. 
    For future work, including photoevaporative winds and/or magnetically driven outflows, these effects should be taken into consideration.
    \item The Hall effect is not included in our studies; however, it is the dominant non-ideal, laminar diffusion process in a large radial range of PPDs \citep[cf. e.g.][]{wardle2007, bai2017, leung2019}. The HE is linked to magnetically induced outflows \citep[cf.][]{leung2019}, which strongly indicates that this effect should be integrated in an upcoming model including magnetized winds. Nevertheless, the treatment of the HE keeps being challenging due to its anisotropic character, which only allows for an approximate treatment in a 1+1D disk model.
    \item Dust is treated in our model very simply. The dust-to-gas ratio is assumed to be constant radially and in time, which is very unlikely to ever occur in a realistic disk. Recent work shows that dust can accumulate in rings, which influences the vertical temperature equilibrium and can trigger FU Ori-like outbursts, if those rings occur in the inner disk \citep[cf. e.g.][]{testi2014, vorobyov2020, vorobyov2022}. Additionally, we do not include any dust evolution model, ignoring coagulation, fragmentation, or settling of dust grains, which can lead to different ionization fractions for different dust grain sizes. \citep[cf. e.g.][]{dudorov2014,delage2022}.
\end{itemize}

\section{Summary and Conclusion}
\label{sec:conclusion2}
In this study, we have extended the 1+1D disk evolution model of \cite{ragossnig20,Steiner21} with a large-scale magnetic field evolution model (cf. \sref{sec:large_scale_disk_field}). 
An implicit time integration scheme allows us to self-consistently include the very inner disk while performing long-term (of the order of the disk viscous timescale) simulations.
Including a large-scale field makes it necessary to carefully model the ionization in the disk both in radial and vertical direction for being able to determine the magnetic resistivity in the disk (cf. \sref{sec:non_ideal_resistivity}).
In our model, we include non-ideal MHD effects, for example, OD and AD.
The scientific objectives of this work are twofold.
First, we determine the large-scale field topology in a steady-state disk truncated by a stellar dipole.
The transition toward a stationary field yields the following conclusions:
\begin{itemize}
    \item Using the conductivity-weighted averaging for the radial magnetic transport velocity $\bbar u_\mathrm{\psi}$ and resistivity $\bbar \eta$ proposed by \cite{Guilet2012}, we can include the effects of the vertical disk profile and do not have to ignore magnetic advection at the disk surface (or layers at intermediate heights) in the framework of an 1+1D simulation. 
    \item Similar to the results of \cite{Guilet2014, dudorov2014}, the field is advected inwards effectively in the outer disk and in the inner, MRI-active disk. 
    The inclination of the large-scale field lines in the outer DZ is reduced, whereas in the inner DZ, magnetic flux transport in the layers above the DZ leads to a non-negligible inclination.
    \item The vertical large-scale field component $\Bz$ in the inner disk is dominated by the stellar dipole field, whereas the radial field $\Brs$ depends on the detailed ionization structure in the disk.
    \item The large-scale field inclination in the outer disk is dominated by AD. We can reproduce the same power-law dependencies for $\Bz$ and $\beta_0$ as e.g. \cite{dudorov2014}, which indicates the validity of our approach.
    \item The self-consistent hydrodynamical treatment of a magnetically truncated, irradiated disk shows that the radial profiles of $\Sigma$, $\vec u$ and $T_\mathrm{gas}$ also lead to a complex large-scale field structure in the inner disk. This is a unique property of our model as we are not limited by small time steps in the inner disk, hence long-term simulations of the large-scale field (even though conducted with the goal of constructing a stationary model) have not been covered yet \citep[cf.][]{dudorov2014,mohanty2018,delage2022}.
\end{itemize}
The second scientific objective is a parameter study to investigate how stellar and disk parameters influence the large-scale field topology, namely $P_\star$, $B_\star$, $L_\mathrm{X}$, $\Dot{M}$, and the strength of the fossil field:
\begin{itemize}
    \item The stellar dipole strength $|\vec B_\star|$ influences the large-scale field in the inner disk ($r < 1$~AU) in two ways: First, a weaker dipole field results in disc truncation closer to the star \citep[cf.][]{hartmann16} where the disk is hotter and therefore has a larger ionization level. Second, a stronger $|\vec B_\star|$ also increases AD in the region dominated by the dipole and therefore results in a large-scale field with less inclined field lines.
    \item A faster/slower rotating star with the rotation period $P_\star$ also affects the large-scale field by $R_\mathrm{cor}$ being closer to/further away from the star. It only affects the inner disk $r < 0.3$~AU such that a slower rotating star (larger $P_\star$) leads to less distinct increase of the field inclination very close to the star.
    \item The large-scale field in the DZ and the outer disk is strongly dependent on the X-ray luminosity. A larger $L_\mathrm{X}$ results in a larger ionization fraction at the disk surface above the DZ and in the outer disk surface layers. In those disk layers, the large-scale field is advected radially inwards most effectively in those disk regions.
    \item Changing $\dot M$ changes the disk structure and therefore also alters the large-scale field topology. In summary, a larger $\dot M$ results in a larger inclined field in the dead zone and $r_\mathrm{in}$ shifting towards smaller radii, which in turn leads to a pronounced spike in field inclination very close to the star.
\end{itemize}

This study shows that a large-scale field threading a disk is also dependent on stellar parameters and has non-negligible inclined field lines. 
Therefore, the next step in this series of papers will be to include a wind model dependent on the large-scale field topology and the disk structure.
Our simulation framework TAPIR \citep[cf.][]{ragossnig20} has an implicit time integration scheme, therefore, we can conduct long-term simulations covering the whole disk lifetime while including the inner disk, the interaction with a stellar dipole, a large-scale field, and a wind model. 
This will help to answer the question of how disks evolve under the influence of magnetically induced winds in combination with viscous accretion. 
It would further allow us to provide estimates on the mass loss and angular momentum loss rates that such winds would produce.
An exciting scientific question is whether MCWs lead to an unstable configuration in the disk: A strongly inclined field triggers a strong MCW, which then removes angular momentum and further increases the inclination. In such a scenario, multiple gaps could be opened in the disk. 
% alternative
Equally interesting is the interaction of photoevaporative flows with a large-scale field, and the reaction of the large-scale field to high accretion rates during an accretion outburst.

\begin{acknowledgements}
We thank the referee for the constructive feedback that helped to improve and clarify the manuscript.
\end{acknowledgements}

\bibliographystyle{aa_url}
\bibliography{literature/library}

\begin{thebibliography}{87}
\expandafter\ifx\csname natexlab\endcsname\relax\def\natexlab#1{#1}\fi

\bibitem[{Armitage(2011)}]{armitage11}
Armitage, P.~J. 2011, \href{http://dx.doi.org/10.1146/annurev-astro-081710-102521}{\color{magenta}Annual Review of Astronomy and Astrophysics}, 49, 195

\bibitem[{{Armitage} {et~al.}(2001){Armitage}, {Livio}, \& {Pringle}}]{Armitage01}
{Armitage}, P.~J., {Livio}, M., \& {Pringle}, J.~E. 2001, \href{http://dx.doi.org/10.1046/j.1365-8711.2001.04356.x}{\color{magenta}\mnras}, \href{https://ui.adsabs.harvard.edu/abs/2001MNRAS.324..705A}{324, 705}

\bibitem[{{Asplund} {et~al.}(2021){Asplund}, {Amarsi}, \& {Grevesse}}]{asplund2021}
{Asplund}, M., {Amarsi}, A.~M., \& {Grevesse}, N. 2021, \href{http://dx.doi.org/10.1051/0004-6361/202140445}{\color{magenta}\aap}, \href{https://ui.adsabs.harvard.edu/abs/2021A&A...653A.141A}{653, A141}

\bibitem[{{Bai} \& {Goodman}(2009)}]{bai2009}
{Bai}, X.-N. \& {Goodman}, J. 2009, \href{http://dx.doi.org/10.1088/0004-637X/701/1/737}{\color{magenta}\apj}, \href{https://ui.adsabs.harvard.edu/abs/2009ApJ...701..737B}{701, 737}

\bibitem[{{Bai} \& {Stone}(2011)}]{baistone2011}
{Bai}, X.-N. \& {Stone}, J.~M. 2011, in AAS/Division for Extreme Solar Systems Abstracts, Vol.~2, AAS/Division for Extreme Solar Systems Abstracts, \href{https://ui.adsabs.harvard.edu/abs/2011ESS.....2.3603B}{36.03}

\bibitem[{Bai \& Stone(2013)}]{Bai2013}
Bai, X.-N. \& Stone, J.~M. 2013, \href{http://dx.doi.org/10.1088/0004-637X/769/1/76}{\color{magenta}Astrophys. Journal, Vol. 769, Issue 1, Artic. id. 76, 21 pp. (2013).}, 769 [\eprint[arXiv]{1301.0318}]

\bibitem[{{Bai} \& {Stone}(2017)}]{bai2017}
{Bai}, X.-N. \& {Stone}, J.~M. 2017, \href{http://dx.doi.org/10.3847/1538-4357/836/1/46}{\color{magenta}\apj}, \href{https://ui.adsabs.harvard.edu/abs/2017ApJ...836...46B}{836, 46}

\bibitem[{{Balbus} \& {Hawley}(1991)}]{balbus91}
{Balbus}, S.~A. \& {Hawley}, J.~F. 1991, \href{http://dx.doi.org/10.1086/170270}{\color{magenta}ApJ}, \href{http://adsabs.harvard.edu/abs/1991ApJ...376..214B}{376, 214}

\bibitem[{{Baraffe} {et~al.}(2015){Baraffe}, {Homeier}, {Allard}, \& {Chabrier}}]{Baraffe15}
{Baraffe}, I., {Homeier}, D., {Allard}, F., \& {Chabrier}, G. 2015, \href{http://dx.doi.org/10.1051/0004-6361/201425481}{\color{magenta}\aap}, \href{http://adsabs.harvard.edu/abs/2015A\%26A...577A..42B}{577, A42}

\bibitem[{{Bell} \& {Lin}(1994)}]{bell94}
{Bell}, K.~R. \& {Lin}, D.~N.~C. 1994, \href{http://dx.doi.org/10.1086/174206}{\color{magenta}ApJ}, \href{http://adsabs.harvard.edu/abs/1994ApJ...427..987B}{427, 987}

\bibitem[{{Ben} {et~al.}(2025){Ben}, {Jose}, \& {Hern{\'a}ndez}}]{gregory2025}
{Ben}, G.~M., {Jose}, J., \& {Hern{\'a}ndez}, J. 2025, \href{http://dx.doi.org/10.1093/mnras/staf1089}{\color{magenta}\mnras}, \href{https://ui.adsabs.harvard.edu/abs/2025MNRAS.541.2246B}{541, 2246}

\bibitem[{{Bertrang} {et~al.}(2017){Bertrang}, {Flock}, \& {Wolf}}]{bertrang2017}
{Bertrang}, G.~H.~M., {Flock}, M., \& {Wolf}, S. 2017, \href{http://dx.doi.org/10.1093/mnrasl/slw181}{\color{magenta}\mnras}, \href{https://ui.adsabs.harvard.edu/abs/2017MNRAS.464L..61B}{464, L61}

\bibitem[{{Bessolaz} {et~al.}(2008){Bessolaz}, {Zanni}, {Ferreira}, {Keppens}, \& {Bouvier}}]{bessolaz08}
{Bessolaz}, N., {Zanni}, C., {Ferreira}, J., {Keppens}, R., \& {Bouvier}, J. 2008, \href{http://dx.doi.org/10.1051/0004-6361:20078328}{\color{magenta}\aap}, \href{https://ui.adsabs.harvard.edu/abs/2008A&A...478..155B}{478, 155}

\bibitem[{{Blandford} \& {Payne}(1982)}]{blandford1982}
{Blandford}, R.~D. \& {Payne}, D.~G. 1982, \href{http://dx.doi.org/10.1093/mnras/199.4.883}{\color{magenta}\mnras}, \href{https://ui.adsabs.harvard.edu/abs/1982MNRAS.199..883B}{199, 883}

\bibitem[{{Brandenburg} {et~al.}(1995){Brandenburg}, {Nordlund}, {Stein}, \& {Torkelsson}}]{brandenburg1995}
{Brandenburg}, A., {Nordlund}, A., {Stein}, R.~F., \& {Torkelsson}, U. 1995, \href{http://dx.doi.org/10.1086/175831}{\color{magenta}\apj}, \href{https://ui.adsabs.harvard.edu/abs/1995ApJ...446..741B}{446, 741}

\bibitem[{{Brauer} {et~al.}(2017){Brauer}, {Wolf}, \& {Flock}}]{brauer2017}
{Brauer}, R., {Wolf}, S., \& {Flock}, M. 2017, \href{http://dx.doi.org/10.1051/0004-6361/201731140}{\color{magenta}\aap}, \href{https://ui.adsabs.harvard.edu/abs/2017A&A...607A.104B}{607, A104}

\bibitem[{Campbell \& G.(1999)}]{Campbell1999}
Campbell, C.~G. \& G., C. 1999, \href{http://dx.doi.org/10.1046/j.1365-8711.1999.03073.x}{\color{magenta}Mon. Not. R. Astron. Soc.}, 310, 1175

\bibitem[{{Cecil} {et~al.}(2024){Cecil}, {Gehrig}, \& {Steiner}}]{cecil2024}
{Cecil}, M., {Gehrig}, L., \& {Steiner}, D. 2024, \href{http://dx.doi.org/10.1051/0004-6361/202348397}{\color{magenta}\aap}, \href{https://ui.adsabs.harvard.edu/abs/2024A&A...687A.136C}{687, A136}

\bibitem[{Courant {et~al.}(1928)Courant, Friedrichs, \& Lewy}]{Courant28}
Courant, R., Friedrichs, K., \& Lewy, H. 1928, \href{http://dx.doi.org/10.1007/BF01448839}{\color{magenta}Math. Ann.}, 100, 32

\bibitem[{{Crutcher} \& {Kemball}(2019)}]{crutcher2019}
{Crutcher}, R.~M. \& {Kemball}, A.~J. 2019, \href{http://dx.doi.org/10.3389/fspas.2019.00066}{\color{magenta}Frontiers in Astronomy and Space Sciences}, \href{https://ui.adsabs.harvard.edu/abs/2019FrASS...6...66C}{6, 66}

\bibitem[{{Crutcher} {et~al.}(2004){Crutcher}, {Nutter}, {Ward-Thompson}, \& {Kirk}}]{crutcher2004}
{Crutcher}, R.~M., {Nutter}, D.~J., {Ward-Thompson}, D., \& {Kirk}, J.~M. 2004, \href{http://dx.doi.org/10.1086/379705}{\color{magenta}\apj}, \href{https://ui.adsabs.harvard.edu/abs/2004ApJ...600..279C}{600, 279}

\bibitem[{{Delage} {et~al.}(2022){Delage}, {Okuzumi}, {Flock}, {Pinilla}, \& {Dzyurkevich}}]{delage2022}
{Delage}, T.~N., {Okuzumi}, S., {Flock}, M., {Pinilla}, P., \& {Dzyurkevich}, N. 2022, \href{http://dx.doi.org/10.1051/0004-6361/202141689}{\color{magenta}\aap}, \href{https://ui.adsabs.harvard.edu/abs/2022A&A...658A..97D}{658, A97}

\bibitem[{{Dong} {et~al.}(2016){Dong}, {Vorobyov}, {Pavlyuchenkov}, {Chiang}, \& {Liu}}]{dong2016}
{Dong}, R., {Vorobyov}, E., {Pavlyuchenkov}, Y., {Chiang}, E., \& {Liu}, H.~B. 2016, \href{http://dx.doi.org/10.3847/0004-637X/823/2/141}{\color{magenta}\apj}, \href{https://ui.adsabs.harvard.edu/abs/2016ApJ...823..141D}{823, 141}

\bibitem[{Dorfi \& Drury(1987)}]{Dorfi1987}
Dorfi, E. \& Drury, L. 1987, \href{http://dx.doi.org/10.1016/0021-9991(87)90161-6}{\color{magenta}J. Comput. Phys.}, 69, 175

\bibitem[{{Dudorov} \& {Khaibrakhmanov}(2014)}]{dudorov2014}
{Dudorov}, A.~E. \& {Khaibrakhmanov}, S.~A. 2014, \href{http://dx.doi.org/10.1007/s10509-014-1900-4}{\color{magenta}\apss}, \href{https://ui.adsabs.harvard.edu/abs/2014Ap&SS.352..103D}{352, 103}

\bibitem[{{Ferguson} {et~al.}(2005){Ferguson}, {Alexander}, {Allard}, {Barman}, {Bodnarik}, {Hauschildt}, {Heffner-Wong}, \& {Tamanai}}]{ferguson05}
{Ferguson}, J.~W., {Alexander}, D.~R., {Allard}, F., {et~al.} 2005, \href{http://dx.doi.org/10.1086/428642}{\color{magenta}\apj}, \href{https://ui.adsabs.harvard.edu/abs/2005ApJ...623..585F}{623, 585}

\bibitem[{{Ferreira} {et~al.}(2006){Ferreira}, {Dougados}, \& {Cabrit}}]{Ferreira06}
{Ferreira}, J., {Dougados}, C., \& {Cabrit}, S. 2006, \href{http://dx.doi.org/10.1051/0004-6361:20054231}{\color{magenta}\aap}, \href{https://ui.adsabs.harvard.edu/abs/2006A&A...453..785F}{453, 785}

\bibitem[{{Gammie}(1996)}]{gammie96}
{Gammie}, C.~F. 1996, \href{http://dx.doi.org/10.1086/176735}{\color{magenta}ApJ}, \href{http://adsabs.harvard.edu/abs/1996ApJ...457..355G}{457, 355}

\bibitem[{{Gehrig} {et~al.}(2022){Gehrig}, {Steiner}, {Vorobyov}, \& {G{\"u}del}}]{Gehrig2022}
{Gehrig}, L., {Steiner}, D., {Vorobyov}, E.~I., \& {G{\"u}del}, M. 2022, \href{http://dx.doi.org/10.1051/0004-6361/202243549}{\color{magenta}\aap}, \href{https://ui.adsabs.harvard.edu/abs/2022A&A...667A..46G}{667, A46}

\bibitem[{{Gehrig} \& {Vorobyov}(2023)}]{gehrig2023spin}
{Gehrig}, L. \& {Vorobyov}, E.~I. 2023, \href{http://dx.doi.org/10.1051/0004-6361/202345916}{\color{magenta}\aap}, \href{https://ui.adsabs.harvard.edu/abs/2023A&A...673A..54G}{673, A54}

\bibitem[{{Glassgold} {et~al.}(2012){Glassgold}, {Galli}, \& {Padovani}}]{glassgold2012}
{Glassgold}, A.~E., {Galli}, D., \& {Padovani}, M. 2012, \href{http://dx.doi.org/10.1088/0004-637X/756/2/157}{\color{magenta}\apj}, \href{https://ui.adsabs.harvard.edu/abs/2012ApJ...756..157G}{756, 157}

\bibitem[{{Glassgold} {et~al.}(2004){Glassgold}, {Najita}, \& {Igea}}]{glassgold2004}
{Glassgold}, A.~E., {Najita}, J., \& {Igea}, J. 2004, \href{http://dx.doi.org/10.1086/424509}{\color{magenta}\apj}, \href{https://ui.adsabs.harvard.edu/abs/2004ApJ...615..972G}{615, 972}

\bibitem[{{G{\"u}del}(2015)}]{guedel2015}
{G{\"u}del}, M. 2015, in European Physical Journal Web of Conferences, Vol. 102, European Physical Journal Web of Conferences, \href{https://ui.adsabs.harvard.edu/abs/2015EPJWC.10200015G}{00015}

\bibitem[{{G{\"u}del} {et~al.}(2018){G{\"u}del}, {Eibensteiner}, {Dionatos}, {Audard}, {Forbrich}, {Kraus}, {Rab}, {Schneider}, {Skinner}, \& {Vorobyov}}]{guedel2018}
{G{\"u}del}, M., {Eibensteiner}, C., {Dionatos}, O., {et~al.} 2018, \href{http://dx.doi.org/10.1051/0004-6361/201834271}{\color{magenta}\aap}, \href{https://ui.adsabs.harvard.edu/abs/2018A&A...620L...1G}{620, L1}

\bibitem[{Guilet \& Ogilvie(2012)}]{Guilet2012}
Guilet, J. \& Ogilvie, G.~I. 2012, \href{http://dx.doi.org/10.1111/j.1365-2966.2012.21361.x}{\color{magenta}Mon. Not. R. Astron. Soc. Vol. 424, Issue 3, pp. 2097-2117.}, 424, 2097

\bibitem[{Guilet \& Ogilvie(2014)}]{Guilet2014}
Guilet, J. \& Ogilvie, G.~I. 2014, \href{http://dx.doi.org/10.1093/mnras/stu532}{\color{magenta}Mon. Not. R. Astron. Soc. Vol. 441, Issue 1, p.852-868}, 441, 852

\bibitem[{Hartmann {et~al.}(2016)Hartmann, Herczeg, \& Calvet}]{hartmann16}
Hartmann, L., Herczeg, G., \& Calvet, N. 2016, \href{http://dx.doi.org/10.1146/annurev-astro-081915-023347}{\color{magenta}Annual Review of Astronomy and Astrophysics}, 54, 135

\bibitem[{{Herbst} {et~al.}(2002){Herbst}, {Bailer-Jones}, {Mundt}, {Meisenheimer}, \& {Wackermann}}]{herbst02}
{Herbst}, W., {Bailer-Jones}, C.~A.~L., {Mundt}, R., {Meisenheimer}, K., \& {Wackermann}, R. 2002, \href{http://dx.doi.org/10.1051/0004-6361:20021362}{\color{magenta}\aap}, \href{https://ui.adsabs.harvard.edu/abs/2002A&A...396..513H}{396, 513}

\bibitem[{{Hubeny}(1990)}]{hubeny1990}
{Hubeny}, I. 1990, \href{http://dx.doi.org/10.1086/168501}{\color{magenta}\apj}, \href{https://ui.adsabs.harvard.edu/abs/1990ApJ...351..632H}{351, 632}

\bibitem[{{Hueso} \& {Guillot}(2005)}]{hueso2005}
{Hueso}, R. \& {Guillot}, T. 2005, \href{http://dx.doi.org/10.1051/0004-6361:20041905}{\color{magenta}\aap}, \href{https://ui.adsabs.harvard.edu/abs/2005A&A...442..703H}{442, 703}

\bibitem[{{Irwin} \& {Bouvier}(2009)}]{irwin2009}
{Irwin}, J. \& {Bouvier}, J. 2009, in IAU Symposium, Vol. 258, The Ages of Stars, ed. {Mamajek}, E.~E., {Soderblom}, D.~R., \& {Wyse}, R. F.~G., \href{https://ui.adsabs.harvard.edu/abs/2009IAUS..258..363I}{363--374}

\bibitem[{Jackson(1999)}]{Jackson1999}
Jackson, J.~D. 1999, Classical electrodynamics, 3rd edn. (New York, {NY}: Wiley)

\bibitem[{{Jankovic} {et~al.}(2021){Jankovic}, {Owen}, {Mohanty}, \& {Tan}}]{jankovic2021}
{Jankovic}, M.~R., {Owen}, J.~E., {Mohanty}, S., \& {Tan}, J.~C. 2021, \href{http://dx.doi.org/10.1093/mnras/stab920}{\color{magenta}\mnras}, \href{https://ui.adsabs.harvard.edu/abs/2021MNRAS.504..280J}{504, 280}

\bibitem[{{Johns-Krull} {et~al.}(1999){Johns-Krull}, {Valenti}, \& {Koresko}}]{johnskrull1999}
{Johns-Krull}, C.~M., {Valenti}, J.~A., \& {Koresko}, C. 1999, \href{http://dx.doi.org/10.1086/307128}{\color{magenta}\apj}, \href{https://ui.adsabs.harvard.edu/abs/1999ApJ...516..900J}{516, 900}

\bibitem[{{Johnstone} {et~al.}(2014{\natexlab{a}}){Johnstone}, {Jardine}, {Gregory}, {Donati}, \& {Hussain}}]{Johnstone14}
{Johnstone}, C.~P., {Jardine}, M., {Gregory}, S.~G., {Donati}, J.~F., \& {Hussain}, G. 2014{\natexlab{a}}, \href{http://dx.doi.org/10.1093/mnras/stt2107}{\color{magenta}\mnras}, \href{https://ui.adsabs.harvard.edu/abs/2014MNRAS.437.3202J}{437, 3202}

\bibitem[{{Johnstone} {et~al.}(2014{\natexlab{b}}){Johnstone}, {Jardine}, {Gregory}, {Donati}, \& {Hussain}}]{johnstone2014}
{Johnstone}, C.~P., {Jardine}, M., {Gregory}, S.~G., {Donati}, J.~F., \& {Hussain}, G. 2014{\natexlab{b}}, \href{http://dx.doi.org/10.1093/mnras/stt2107}{\color{magenta}\mnras}, \href{https://ui.adsabs.harvard.edu/abs/2014MNRAS.437.3202J}{437, 3202}

\bibitem[{{Khaibrakhmanov} \& {Dudorov}(2022)}]{khaibrakhmanov22}
{Khaibrakhmanov}, S.~A. \& {Dudorov}, A.~E. 2022, \href{http://dx.doi.org/10.1134/S1063772922100079}{\color{magenta}Astronomy Reports}, \href{https://ui.adsabs.harvard.edu/abs/2022ARep...66..872K}{66, 872}

\bibitem[{{Lankhaar} \& {Teague}(2023)}]{lankhaar2023}
{Lankhaar}, B. \& {Teague}, R. 2023, \href{http://dx.doi.org/10.1051/0004-6361/202345840}{\color{magenta}\aap}, \href{https://ui.adsabs.harvard.edu/abs/2023A&A...678A..17L}{678, A17}

\bibitem[{{Lee}(2020)}]{lee2020}
{Lee}, C.-F. 2020, \href{http://dx.doi.org/10.1007/s00159-020-0123-7}{\color{magenta}\aapr}, \href{https://ui.adsabs.harvard.edu/abs/2020A&ARv..28....1L}{28, 1}

\bibitem[{{Lesur} {et~al.}(2023){Lesur}, {Flock}, {Ercolano}, {Lin}, {Yang}, {Barranco}, {Benitez-Llambay}, {Goodman}, {Johansen}, {Klahr}, {Laibe}, {Lyra}, {Marcus}, {Nelson}, {Squire}, {Simon}, {Turner}, {Umurhan}, \& {Youdin}}]{lesur2023}
{Lesur}, G., {Flock}, M., {Ercolano}, B., {et~al.} 2023, in Astronomical Society of the Pacific Conference Series, Vol. 534, Protostars and Planets VII, ed. {Inutsuka}, S., {Aikawa}, Y., {Muto}, T., {Tomida}, K., \& {Tamura}, M., \href{https://ui.adsabs.harvard.edu/abs/2023ASPC..534..465L}{465}

\bibitem[{{Lesur} {et~al.}(2014){Lesur}, {Kunz}, \& {Fromang}}]{Lesur2014}
{Lesur}, G., {Kunz}, M.~W., \& {Fromang}, S. 2014, \href{http://dx.doi.org/10.1051/0004-6361/201423660}{\color{magenta}AAP}, \href{https://ui.adsabs.harvard.edu/abs/2014A&A...566A..56L}{566, A56}

\bibitem[{{Lesur}(2021)}]{lesur2021}
{Lesur}, G. R.~J. 2021, \href{http://dx.doi.org/10.1051/0004-6361/202040109}{\color{magenta}\aap}, \href{https://ui.adsabs.harvard.edu/abs/2021A&A...650A..35L}{650, A35}

\bibitem[{{Leung} \& {Ogilvie}(2019)}]{leung2019}
{Leung}, P. K.~C. \& {Ogilvie}, G.~I. 2019, \href{http://dx.doi.org/10.1093/mnras/stz1620}{\color{magenta}\mnras}, \href{https://ui.adsabs.harvard.edu/abs/2019MNRAS.487.5155L}{487, 5155}

\bibitem[{{LeVeque} {et~al.}(1997){LeVeque}, {Mihalas}, {Dorfi}, \& {Müller}}]{SaasFee}
{LeVeque}, R.~J., {Mihalas}, D., {Dorfi}, E.~A., \& {Müller}, E. 1997, Astrophysics of Planet Formation (Saas-Fee Advanced Course 27. Lecture Notes 1997 Swiss Society for Astrophysics and Astronomy)

\bibitem[{{Lubow} {et~al.}(1994){Lubow}, {Papaloizou}, \& {Pringle}}]{Lubow94}
{Lubow}, S.~H., {Papaloizou}, J.~C.~B., \& {Pringle}, J.~E. 1994, \href{http://dx.doi.org/10.1093/mnras/268.4.1010}{\color{magenta}\mnras}, \href{https://ui.adsabs.harvard.edu/abs/1994MNRAS.268.1010L}{268, 1010}

\bibitem[{{Masson} {et~al.}(2016){Masson}, {Chabrier}, {Hennebelle}, {Vaytet}, \& {Commer{\c{c}}on}}]{masson2016}
{Masson}, J., {Chabrier}, G., {Hennebelle}, P., {Vaytet}, N., \& {Commer{\c{c}}on}, B. 2016, \href{http://dx.doi.org/10.1051/0004-6361/201526371}{\color{magenta}\aap}, \href{https://ui.adsabs.harvard.edu/abs/2016A&A...587A..32M}{587, A32}

\bibitem[{{Mohanty} {et~al.}(2018){Mohanty}, {Jankovic}, {Tan}, \& {Owen}}]{mohanty2018}
{Mohanty}, S., {Jankovic}, M.~R., {Tan}, J.~C., \& {Owen}, J.~E. 2018, \href{http://dx.doi.org/10.3847/1538-4357/aabcd0}{\color{magenta}\apj}, \href{https://ui.adsabs.harvard.edu/abs/2018ApJ...861..144M}{861, 144}

\bibitem[{{Nakamoto} \& {Nakagawa}(1994)}]{nakamoto1994}
{Nakamoto}, T. \& {Nakagawa}, Y. 1994, \href{http://dx.doi.org/10.1086/173678}{\color{magenta}\apj}, \href{https://ui.adsabs.harvard.edu/abs/1994ApJ...421..640N}{421, 640}

\bibitem[{Ogilvie(1997)}]{Ogilvie1997}
Ogilvie, G.~I. 1997, \href{http://dx.doi.org/10.1093/mnras/288.1.63}{\color{magenta}Mon. Not. R. Astron. Soc.}, 288, 63

\bibitem[{{Ogilvie} \& {Livio}(1998)}]{ogilvie1998}
{Ogilvie}, G.~I. \& {Livio}, M. 1998, \href{http://dx.doi.org/10.1086/305636}{\color{magenta}\apj}, \href{https://ui.adsabs.harvard.edu/abs/1998ApJ...499..329O}{499, 329}

\bibitem[{{Ogilvie} \& {Livio}(2001)}]{ogilvie2001}
{Ogilvie}, G.~I. \& {Livio}, M. 2001, \href{http://dx.doi.org/10.1086/320637}{\color{magenta}\apj}, \href{https://ui.adsabs.harvard.edu/abs/2001ApJ...553..158O}{553, 158}

\bibitem[{{Ohashi} {et~al.}(2018){Ohashi}, {Kataoka}, {Nagai}, {Momose}, {Muto}, {Hanawa}, {Fukagawa}, {Tsukagoshi}, {Murakawa}, \& {Shibai}}]{ohashi2018}
{Ohashi}, S., {Kataoka}, A., {Nagai}, H., {et~al.} 2018, \href{http://dx.doi.org/10.3847/1538-4357/aad632}{\color{magenta}\apj}, \href{https://ui.adsabs.harvard.edu/abs/2018ApJ...864...81O}{864, 81}

\bibitem[{{Pollack} {et~al.}(1985){Pollack}, {McKay}, \& {Christofferson}}]{pollack85}
{Pollack}, J.~B., {McKay}, C.~P., \& {Christofferson}, B.~M. 1985, \href{http://dx.doi.org/10.1016/0019-1035(85)90069-7}{\color{magenta}\icarus}, \href{https://ui.adsabs.harvard.edu/abs/1985Icar...64..471P}{64, 471}

\bibitem[{{Preibisch} {et~al.}(2005){Preibisch}, {Kim}, {Favata}, {Feigelson}, {Flaccomio}, {Getman}, {Micela}, {Sciortino}, {Stassun}, {Stelzer}, \& {Zinnecker}}]{preibisch2005}
{Preibisch}, T., {Kim}, Y.-C., {Favata}, F., {et~al.} 2005, \href{http://dx.doi.org/10.1086/432891}{\color{magenta}\apjs}, \href{https://ui.adsabs.harvard.edu/abs/2005ApJS..160..401P}{160, 401}

\bibitem[{{Rab} {et~al.}(2017){Rab}, {G{\"u}del}, {Padovani}, {Kamp}, {Thi}, {Woitke}, \& {Aresu}}]{rab2017}
{Rab}, C., {G{\"u}del}, M., {Padovani}, M., {et~al.} 2017, \href{http://dx.doi.org/10.1051/0004-6361/201630241}{\color{magenta}\aap}, \href{https://ui.adsabs.harvard.edu/abs/2017A&A...603A..96R}{603, A96}

\bibitem[{{Ragossnig} {et~al.}(2020){Ragossnig}, {Dorfi}, {Ratschiner}, {Gehrig}, {Steiner}, {St{\"o}kl}, \& {Johnstone}}]{ragossnig20}
{Ragossnig}, F., {Dorfi}, E.~A., {Ratschiner}, B., {et~al.} 2020, \href{http://dx.doi.org/10.1016/j.cpc.2020.107437}{\color{magenta}Computer Physics Communications}, \href{https://ui.adsabs.harvard.edu/abs/2020CoPhC.25607437R}{256, 107437}

\bibitem[{{Ragossnig} {et~al.}(2019){Ragossnig}, {Dorfi}, {Ratschiner}, {Gehrig}, {Steiner}, {Stökl}, \& {Johnstone}}]{Ragossnig2019}
{Ragossnig}, F., {Dorfi}, E.~A., {Ratschiner}, B., {et~al.} 2019, 1+1D implicit disk computations, 3rd edn. (Computer Physics and Communications)

\bibitem[{{Reyes-Ruiz} {et~al.}(2004){Reyes-Ruiz}, {P{\'e}rez-Tijerina}, \& {S{\'a}nchez-Salcedo}}]{reyes04}
{Reyes-Ruiz}, M., {P{\'e}rez-Tijerina}, E., \& {S{\'a}nchez-Salcedo}, F.~J. 2004, in Revista Mexicana de Astronomia y Astrofisica, vol.~27, Vol.~22, Revista Mexicana de Astronomia y Astrofisica Conference Series, ed. {Garcia-Segura}, G., {Tenorio-Tagle}, G., {Franco}, J., \& {Yorke}, H.~W., \href{http://adsabs.harvard.edu/abs/2004RMxAC..22..113R}{113--116}

\bibitem[{{Romanova} {et~al.}(2002){Romanova}, {Ustyugova}, {Koldoba}, \& {Lovelace}}]{Romanova02}
{Romanova}, M.~M., {Ustyugova}, G.~V., {Koldoba}, A.~V., \& {Lovelace}, R.~V.~E. 2002, \href{http://dx.doi.org/10.1086/342464}{\color{magenta}\apj}, \href{https://ui.adsabs.harvard.edu/abs/2002ApJ...578..420R}{578, 420}

\bibitem[{{Romanova} {et~al.}(2004){Romanova}, {Ustyugova}, {Koldoba}, \& {Lovelace}}]{Romanova04}
{Romanova}, M.~M., {Ustyugova}, G.~V., {Koldoba}, A.~V., \& {Lovelace}, R.~V.~E. 2004, \href{http://dx.doi.org/10.1086/421867}{\color{magenta}\apj}, \href{https://ui.adsabs.harvard.edu/abs/2004ApJ...610..920R}{610, 920}

\bibitem[{{Sano} {et~al.}(2004){Sano}, {Inutsuka}, {Turner}, \& {Stone}}]{sano2004}
{Sano}, T., {Inutsuka}, S.-i., {Turner}, N.~J., \& {Stone}, J.~M. 2004, \href{http://dx.doi.org/10.1086/382184}{\color{magenta}\apj}, \href{https://ui.adsabs.harvard.edu/abs/2004ApJ...605..321S}{605, 321}

\bibitem[{{Shakura} \& {Sunyaev}(1973)}]{shakura73}
{Shakura}, N.~I. \& {Sunyaev}, R.~A. 1973, A\&A, \href{http://cdsads.u-strasbg.fr/abs/1973A%26A....24..337S}{24, 337}

\bibitem[{{Simon} {et~al.}(2013){Simon}, {Bai}, {Armitage}, {Stone}, \& {Beckwith}}]{simon2013}
{Simon}, J.~B., {Bai}, X.-N., {Armitage}, P.~J., {Stone}, J.~M., \& {Beckwith}, K. 2013, \href{http://dx.doi.org/10.1088/0004-637X/775/1/73}{\color{magenta}\apj}, \href{https://ui.adsabs.harvard.edu/abs/2013ApJ...775...73S}{775, 73}

\bibitem[{{Spitzer}(1978)}]{spitzer1978}
{Spitzer}, L. 1978, {Physical processes in the interstellar medium} (Wiley-Interscience)

\bibitem[{{Steiner} {et~al.}(2021){Steiner}, {Gehrig, L.}, {Ratschiner, B.}, {Ragossnig, F.}, {Vorobyov, E. I.}, {G\"udel, M.}, \& {Dorfi, E. A.}}]{Steiner21}
{Steiner}, D., {Gehrig, L.}, {Ratschiner, B.}, {et~al.} 2021, \href{http://dx.doi.org/10.1051/0004-6361/202140447}{\color{magenta}A\&A}, 655, A110

\bibitem[{{Suzuki} {et~al.}(2016){Suzuki}, {Ogihara}, {Morbidelli}, {Crida}, \& {Guillot}}]{suzuki2016}
{Suzuki}, T.~K., {Ogihara}, M., {Morbidelli}, A., {Crida}, A., \& {Guillot}, T. 2016, \href{http://dx.doi.org/10.1051/0004-6361/201628955}{\color{magenta}\aap}, \href{https://ui.adsabs.harvard.edu/abs/2016A&A...596A..74S}{596, A74}

\bibitem[{{Tang} {et~al.}(2023){Tang}, {Dutrey}, {Koch}, {Guilloteau}, {Yen}, {di Folco}, {Pantin}, {Muto}, {Kataoka}, \& {Brauer}}]{tang2023}
{Tang}, Y.-W., {Dutrey}, A., {Koch}, P.~M., {et~al.} 2023, \href{http://dx.doi.org/10.3847/2041-8213/acc45b}{\color{magenta}\apjl}, \href{https://ui.adsabs.harvard.edu/abs/2023ApJ...947L...5T}{947, L5}

\bibitem[{{Testi} {et~al.}(2014){Testi}, {Birnstiel}, {Ricci}, {Andrews}, {Blum}, {Carpenter}, {Dominik}, {Isella}, {Natta}, {Williams}, \& {Wilner}}]{testi2014}
{Testi}, L., {Birnstiel}, T., {Ricci}, L., {et~al.} 2014, in Protostars and Planets VI, ed. {Beuther}, H., {Klessen}, R.~S., {Dullemond}, C.~P., \& {Henning}, T., \href{https://ui.adsabs.harvard.edu/abs/2014prpl.conf..339T}{339}

\bibitem[{{Tsukamoto} {et~al.}(2023){Tsukamoto}, {Maury}, {Commercon}, {Alves}, {Cox}, {Sakai}, {Ray}, {Zhao}, \& {Machida}}]{tsukamoto2023}
{Tsukamoto}, Y., {Maury}, A., {Commercon}, B., {et~al.} 2023, in Astronomical Society of the Pacific Conference Series, Vol. 534, Protostars and Planets VII, ed. {Inutsuka}, S., {Aikawa}, Y., {Muto}, T., {Tomida}, K., \& {Tamura}, M., \href{https://ui.adsabs.harvard.edu/abs/2023ASPC..534..317T}{317}

\bibitem[{{Vorobyov} \& {Basu}(2009)}]{Vorobyov09}
{Vorobyov}, E.~I. \& {Basu}, S. 2009, \href{http://dx.doi.org/10.1111/j.1365-2966.2008.14376.x}{\color{magenta}\mnras}, \href{https://ui.adsabs.harvard.edu/abs/2009MNRAS.393..822V}{393, 822}

\bibitem[{{Vorobyov} {et~al.}(2017){Vorobyov}, {Elbakyan}, {Hosokawa}, {Sakurai}, {Guedel}, \& {Yorke}}]{Vorobyov17c}
{Vorobyov}, E.~I., {Elbakyan}, V., {Hosokawa}, T., {et~al.} 2017, \href{http://dx.doi.org/10.1051/0004-6361/201630356}{\color{magenta}\aap}, \href{https://ui.adsabs.harvard.edu/abs/2017A&A...605A..77V}{605, A77}

\bibitem[{{Vorobyov} {et~al.}(2020){Vorobyov}, {Khaibrakhmanov}, {Basu}, \& {Audard}}]{vorobyov2020}
{Vorobyov}, E.~I., {Khaibrakhmanov}, S., {Basu}, S., \& {Audard}, M. 2020, \href{http://dx.doi.org/10.1051/0004-6361/202039081}{\color{magenta}\aap}, \href{https://ui.adsabs.harvard.edu/abs/2020A&A...644A..74V}{644, A74}

\bibitem[{{Vorobyov} {et~al.}(2019){Vorobyov}, {Skliarevskii}, {Elbakyan}, {Pavlyuchenkov}, {Akimkin}, \& {Guedel}}]{Vorobyov19}
{Vorobyov}, E.~I., {Skliarevskii}, A.~M., {Elbakyan}, V.~G., {et~al.} 2019, \href{http://dx.doi.org/10.1051/0004-6361/201935438}{\color{magenta}\aap}, \href{https://ui.adsabs.harvard.edu/abs/2019A&A...627A.154V}{627, A154}

\bibitem[{{Vorobyov} {et~al.}(2022){Vorobyov}, {Skliarevskii}, {Molyarova}, {Akimkin}, {Pavlyuchenkov}, {K{\'o}sp{\'a}l}, {Liu}, {Takami}, \& {Topchieva}}]{vorobyov2022}
{Vorobyov}, E.~I., {Skliarevskii}, A.~M., {Molyarova}, T., {et~al.} 2022, \href{http://dx.doi.org/10.1051/0004-6361/202141932}{\color{magenta}\aap}, \href{https://ui.adsabs.harvard.edu/abs/2022A&A...658A.191V}{658, A191}

\bibitem[{{Wardle}(2007)}]{wardle2007}
{Wardle}, M. 2007, \href{http://dx.doi.org/10.1007/s10509-007-9575-8}{\color{magenta}\apss}, \href{https://ui.adsabs.harvard.edu/abs/2007Ap&SS.311...35W}{311, 35}

\bibitem[{{Zhu} {et~al.}(2007){Zhu}, {Hartmann}, {Calvet}, {Hernandez}, {Muzerolle}, \& {Tannirkulam}}]{zhu07}
{Zhu}, Z., {Hartmann}, L., {Calvet}, N., {et~al.} 2007, \href{http://dx.doi.org/10.1086/521345}{\color{magenta}ApJ}, \href{http://adsabs.harvard.edu/abs/2007ApJ...669..483Z}{669, 483}

\bibitem[{{Zhu} \& {Stone}(2018)}]{zhu2018}
{Zhu}, Z. \& {Stone}, J.~M. 2018, \href{http://dx.doi.org/10.3847/1538-4357/aaafc9}{\color{magenta}\apj}, \href{https://ui.adsabs.harvard.edu/abs/2018ApJ...857...34Z}{857, 34}

\end{thebibliography}

\begin{appendix}

\section{Numerical details}
\label{sec:numerical_details}

A fully implicit time integration requires an initial model to start with, and its basics are explained in \cite{Steiner21}. As explained in detail in \cite{ragossnig20}, the hydrodynamical equations can be represented by local interactions and the resulting Jacobian can be efficiently inverted. The addition of a large-scale magnetic field adds a long-range acting component, which cannot be represented by a localized set of variables anymore and therefore is unsuitable for adding as a fully implicit component.

The timestep $\Delta t$ in our model is not restricted by the CFL-condition \citep[see also][]{Dorfi1987, Ragossnig2019, Steiner21}.
Thus, an evolution from an initial model toward a steady state solution can be achieved over $\sim t_\mathrm{visc}$ using the full set of hydrodynamic equations.
However, very large timesteps can result in an unphysical "damping" of processes occurring on smaller timescales.
To avoid this issue, we start with a small timestep (100 seconds) and perform a Newton-Raphson iteration for obtaining a new solution \citep{SaasFee,ragossnig20}. After a successful iteration, the timestep is adapted by applying the following criteria: 
\begin{itemize}
    \item The number of iterations, $n_\mathrm{it}$, needed to obtain a new solution determines if the new timestep is increased or decreased.
\end{itemize}
The timestep control is implemented in the following way for the determination of a new timestep $\Delta t^\mathrm{(new)}$:
\begin{align}
    \Delta t^\mathrm{(new)} &= 
        \begin{cases}
            1.5 \, \Delta t^\mathrm{(old)}  & \text{if } \qquad n_\mathrm{it} = 1 \\
            1.2 \, \Delta t^\mathrm{(old)}  & \text{if } \qquad n_\mathrm{it} = 2 \\
            1.0 \, \Delta t^\mathrm{(old)}  & \text{if } \qquad n_\mathrm{it} = 3 \\
            0.8 \, \Delta t^\mathrm{(old)}  & \text{if } \qquad n_\mathrm{it} = 4 \\
            0.5 \, \Delta t^\mathrm{(old)}  & \text{otherwise}
         \end{cases} \;, \label{eq:timestep_criteria_1}
\end{align}
This implementation of an adaptive timestep control allows TAPIR to react to perturbations occurring during disk evolution: 
If a perturbation causes the solution to differ much between two timesteps, more iterations are needed to find a new solution, and the timestep is reduced for the next iteration. 
It is also possible that the Newton-Raphson iteration does not converge anymore for a chosen timestep, which also leads to the maximum timestep reduction. 
As the solution approaches a stationary state, the variation between two iterations becomes small, and the timestep increases toward large values. In theory, the values of $\Delta t$ have no upper limit for a true stationary solution.
We assume our solution to be stationary when we have reached 10 times $t_\mathrm{visc}(r_\mathrm{out})$.

However, we double-checked every simulation run by taking the final stationary model and restarting the simulation with the initial small timestep of $\Delta t = 100$~seconds.
If there is no variation during the second simulation between the initial and final model, we can rule out that a very large $\Delta t$ toward the end of a simulation run would dampen any perturbations.
This is the case for all models presented in this work.

The external large-scale fossil field connects the field inside the disk to the force-free field in the vacuum above and below the disk, and therefore acts as non-local dependency all over the disk. The relation between the toroidal currents and the corresponding flux function of the disk field at the midplane $\psi_\mathrm{d}$ is described in \equ{eq:biot_savart} and is calculated at every grid point (index $i$) in the following way,
\begin{align}
    \psi_\mathrm{d}(r) &= \int_{r_\mathrm{in}}^{r_\mathrm{out}} \mathcal{L}(r,r') \, B_\mathrm{r}^\mathrm{s}(r') \,dr' \;,
\end{align}
which can be discretized in the following way \citep[cf.][]{Guilet2014},
\begin{align}
     \psi_\mathrm{d}(r_\mathrm{i}) \approx \sum_\mathrm{j=1}^\mathrm{n} L_\mathrm{ij} \, B_\mathrm{r}^\mathrm{s}(r_\mathrm{j}) \;,
\end{align}
where we have used \equ{eq:stokes} to connect $B_\mathrm{r}^\mathrm{s}$ to the currents in the disk $J_\mathrm{\varphi}^\mathrm{s}$. The summation maximum $n$ is the number of numerical cells (the radial grid size), whereas $L_\mathrm{ij}$ represents the linear operator $\mathcal{L}$ (cf. \sref{sec:large_scale_disk_field}) in cylindrical coordinates and is defined as,
\begin{align}
    L_\mathrm{ij} &\equiv \frac{2}{\pi} \, \int_{r_\mathrm{j-1/2}}^{r_\mathrm{j+1/2}} \frac{r_\mathrm{i} \, r'}{r_\mathrm{i} + r'} \left[ \frac{(2 - k^2) \, K(k) - 2\,E(k)}{k^2}  \right] \, dr' \;.
\end{align}
The radial grid points $r_\mathrm{j \pm 1/2}$ represent radii in between two neighboring grid points \citep[cf.][]{Guilet2014}. The integral is calculated for every grid cell using a Romberg integrator for sufficient accuracy. Then $L_\mathrm{ij}$ is inverted to yield $B_\mathrm{r}^\mathrm{s}$ at every grid point $r_\mathrm{i}$,
\begin{align}
    B_\mathrm{r}^\mathrm{s}(r_\mathrm{i}) &\approx \sum_\mathrm{j=1}^\mathrm{n} L_\mathrm{ij}^{-1} \, \psi_\mathrm{d}(r_\mathrm{j}) \;, \label{eq:inversion_L}
\end{align}
A narrow bandwidth is required for the Jacobian of our implicit integration scheme for efficient inversion, which requires a local formulation of the hydrodynamical equations  \citep[cf.][]{ragossnig20}. Since \equ{eq:inversion_L} yields a global dependency at every grid point, the following approximation can be made,
\begin{align}
        B_\mathrm{r}^\mathrm{s}(r_\mathrm{i}) &\approx \sum_\mathrm{j=i-l}^\mathrm{i+u} L_\mathrm{ij}^{-1} \, \psi_\mathrm{d}(r_\mathrm{j}) +  \sum_\mathrm{j=1}^\mathrm{i-l-1} L_\mathrm{ij}^{-1} \, \psi_\mathrm{d}(r_\mathrm{j}^\mathrm{A}) +  \sum_\mathrm{j=i+u+1}^\mathrm{n} L_\mathrm{ij}^{-1} \, \psi_\mathrm{d}(r_\mathrm{j}^\mathrm{A}) \;, \label{eq:inversion_L_split}
\end{align}
where '$\mathrm{l}$' and '$\mathrm{u}$' are indices below and above the grid index '$\mathrm{i}$', respectively, and correspond to $\mathrm{l = u = 2}$ in our work. Furthermore, $r_\mathrm{j}^\mathrm{A}$ is the grid point at position 'j' at the old time step, which means that $B_\mathrm{r}^\mathrm{s}$ is calculated  semi-implicitly, in contrast to the full-implicit treatment of all other hydrodynamical quantities. This way the localized narrow-banded Jacobian can be maintained at the cost of an explicit part. Additionally, since we use an adaptive grid in TAPIR to adapt the inner rim as well as to achieve better numerical accuracy in regions of steep gradients, $L_\mathrm{ij}$ would need to be calculated and inverted at each time step anew. This would result in greatly increased numerical cost in obtaining a solution of the system of equations at a new time step. Since $L_\mathrm{ij}$ is a property of the numerical grid only, and the grid is varied slowly by means of the grid equation \citep[][]{ragossnig20}, the deviation of the grid between two time steps remains small. Therefore, it is a viable approximation to invert $L_\mathrm{ij}$ not at every time step but only after a certain maximum threshold grid deviation is exceeded with respect to the last grid configuration used for inversion.
\begin{align}
    \gamma_\mathrm{inv} &= \max\left( \frac{|r_\mathrm{i}(t_\mathrm{inv,new}) - r_\mathrm{i}(t_\mathrm{inv,old})|}{t_\mathrm{inv,new} - t_\mathrm{inv,old}}  \right) \;, 
\end{align}
where $\gamma_\mathrm{inv}$, $r_\mathrm{i}(t_\mathrm{inv,new(old)})$ and $t_\mathrm{inv,new(old)}$ are the threshold grid deviation triggering the inversion of $L_\mathrm{ij}$, the numerical grid and the time at the new (old) inversion time, respectively. The determination of $\gamma_\mathrm{inv}$ is complex and dependent on the actual time step because of the explicit component in \equ{eq:inversion_L}, which introduces numerical instabilities above a certain time step. It is therefore necessary to calibrate $\gamma_\mathrm{inv}$ for every calculation and time step in order to obtain stable solutions.

\section{Validation of the model describing the vertical disk structure}
\label{sec:validation}

\begin{figure}[ht!]
    \centering
    \resizebox{\hsize}{!}{\includegraphics{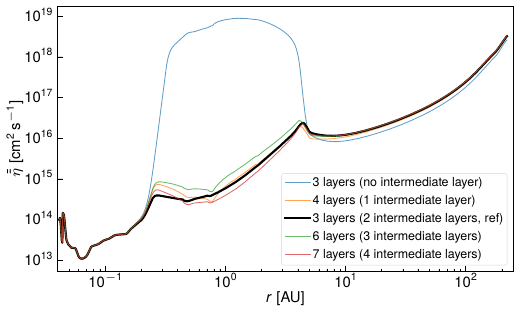}}
    \caption{Profiles of the vertically conductivity-averaged resistivity $\bbar \eta$ if calculated with a varying number of vertical layers. The three layers at $z = 0$, $z = \Hp$, and $z = \zB$ are always included, whereas the number of intermediate layers between $z = \Hp$ and $z = \zB$ is varied. The layers are distributed evenly in the vertical direction between $\Hp$ and $\zB$.}
    \label{fig:fid_layers}
\end{figure}

\paragraph{Thin disk approximation:}In this work we focus on Class II star-disk systems, i.e. we assume that a potential envelope has already accreted onto the disk and the central star. This means that the disk geometry of a disk without an envelope is much thinner in vertical direction. 
We therefore can assume that, even with an outer flared disk, a Class II disk maintains an approximate geometrically thin structure, which justifies our assumption and is furthermore also used by many other publications \citep[cf. e.g.][]{bell94,Lubow94,Armitage01,mohanty2018,delage2022,lesur2023}.
\paragraph{Assumption of a stationary vertical velocity- and magnetic field profile}: Perturbations of the vertical structure relax on the dynamical timescale $t_\mathrm{dyn}$. This is true for perturbations of both the large-scale magnetic field and the disk structure in vertical direction \citep[cf. e.g.][]{Guilet2012}. As $t_\mathrm{dyn}$ is much shorter than the viscous timescale $t_\mathrm{visc}$ and the magnetic diffusion timescale $t_\mathrm{diff}$, which are the respective timescales at which the disk and the large-scale magnetic field evolve radially, the assumption of stationarity for the vertical disk profile is justified (cf. timescales in \fig{fig:fid_timescales_stationarity} of \sref{sec:remarks_stationarity}).
\paragraph{Assumption of a dominant vertical magnetic field component:}A basic assumption for deriving the vertical velocity and large-scale magnetic field profile is a dominant vertical field component $B_\mathrm{z}$ \citep[in comparison to $\Brs$ and $\Bphis$, cf.][]{Guilet2012, leung2019}. 
This restriction is due to the fact that an analytic vertical profile can only be obtained if magnetic compression is negligible. 
However, \cite{Guilet2012} argue after comparison with numerical simulations of the vertical structure of a strongly magnetized disk, that in the absence of magnetically induced outflows the approximation remains well reasonable as long as: 
(i) the inclination angle does not exceed $i \approx 30^\circ$ and 
(ii) the magnetic field is sufficiently weak ($\beta_0 \gtrsim 10^2$). 
A stronger, large-scale magnetic field would lead to magnetic compression and therefore to a changed vertical disk structure, whereas a large inclination angle would very likely cause MCWs (cf. \sref{sec:implications_inclination} for an estimate and discussion of including MCWs).
Both conditions are almost everywhere fulfilled in our simulations, except a very narrow region in the innermost disk and for a region in the AD-dominated region outside of the DZ. However, the main topic of this work is to expand on former studies of the poloidal large-scale magnetic field structure \citep[cf.][]{dudorov2014,Guilet2012,mohanty2018,delage2022} by including a self-consistent MHD disk model and a stellar dipole. 
Magnetic compression would likely play a role in the innermost disk region. 
This effect will be included in future studies.
We therefore want to stress the point that the validity of the used vertical disk profile is limited in the innermost disk region (cf. \sref{sec:model_limitations} for a comprehensive list of our model limitations).
\paragraph{Choice of vertical layers for calculating the ionization degree and the magnetic resistivity:}Our 1+1D model self-consistently solves the MHD disk equations \equs{eq:cont}{eq:induction} only in the radial direction.
For being able to obtain an estimate of the vertical disk structure at every radius $r$, we calculate the ionization degree $\xe$ as well as OD and AD at the midplane and four different heights $z$ (cf. fifth step in \sref{sec:solution_procedure} and \sref{sec:reference_model}). 
The choices for the heights of the five layers are justified as follows:
\begin{itemize}
    \item $z = 0$: The disk midplane is important to capture the impact of a thermally ionized region in the inner disk on magnetic resistivity. 
    It further proves to be the dominant region of magnetic advection in the AD-dominated outer disk (cf. \fig{fig:fid_resistivity_profile}).
    \item $z = H_\mathrm{p}$: At this height, the disk is not likely to be able to maintain thermal ionization in the inner disk through viscous heating due to a low gas density at $z = \Hp$ \citep[cf.,][]{jankovic2021}. Non-thermal ionization becomes increasingly important in the layers above for $z > \Hp$.
    \item $z = \zB$: The magnetic surface serves as the boundary layer between a passive large-scale magnetic field inside the disk and a force-free magnetically dominated disk exterior \citep[cf.][]{Guilet2012,Guilet2014}. 
    The disk surface in our models is located at this height, which is subject to non-thermal ionization over most of the PPD's radial extent.
    \item $z = H_\mathrm{p} + \frac{1}{3}(\zB - H_\mathrm{p})$ and $z = H_\mathrm{p} + \frac{2}{3}(\zB - H_\mathrm{p})$: 
    Those two layers are evenly positioned in vertical direction between $z = H_\mathrm{p}$ and $z = \zB$ and are necessary to capture the effects of a MRI-active layer above the DZ. 
    This layer is located just above the DZ but below the disk surface, which is due to two competing physical processes: 
    (i) Non-thermal ionization decreases resistivity and allows MRI to operate, and 
    (ii) AD increases at larger $z$, because the gas density strongly decreases while the large-scale magnetic field maintains its strength \citep[cf. e.g.][]{delage2022}. 
    This leads to an intermediate region where magnetic advection is highly effective despite being in the radial region of the DZ. 
    Because our 1+1D model cannot calculate this MRI layer self-consistently and therefore its exact location (i.e. its height $z$) at some time $t$ is unknown, we chose to include two additional layers to capture the essentials of this MRI-active layer.
\end{itemize}
For validating if this choice of layers is suited reasonably well to capture the essential physics needed to model magnetic advection and diffusion of a large-scale magnetic field, we have varied the amount of intermediate layers used to calculate both $\xe$ and the resistivities $\etaO$ and $\etaA$ (cf. \fig{fig:fid_layers}). 
More layers lead to a more accurate description of the vertical disk structure. 
However, more layers lead to a vastly more complicated time integration scheme (an implicit time integration scheme requires calculating a Jacobian encapsulating all dependencies at every grid point) and to a potentially slower convergence of the Newton-Raphson iteration when finding a new solution \citep[cf. ][for an overview of the implicit time integration used in TAPIR]{SaasFee, ragossnig20}.
It can be seen in \fig{fig:fid_layers} that our approach with five layers (i.e., two intermediate layers between $\Hp$ and $\zB$) captures the MRI-active layer well while maintaining a manageable complexity in calculating the aforementioned Jacobian. 
We want to note that an approach without any intermediate layer is unsuited for describing magnetic advection in a MRI-active layer above the DZ: the corresponding profile for $\bbar \eta$ vastly overestimates the resulting effective resistivity in the DZ by only taking the values inside the DZ and at the AD-dominated disk surface into account (cf. blue curve in \fig{fig:fid_layers}). 
We further want to point out that intermediate layers are only altering the radial $\bbar \eta$-profile in the DZ, while in the inner disk and in the outer disk outside the DZ the number of intermediate layers does not change $\bbar \eta$ significantly. 
In the inner disk, this is due to the domination of thermal ionization close to the midplane, which then dominantly contributes to a conductivity-weighted $\bbar \eta$.
In the AD-dominated outer disk AD becomes increasingly important with increasing height $z$, which means that magnetic advection dominantly happens close to the disk midplane. 
\paragraph{Vertical temperature profile:}Our hydrodynamical equations \equs{eq:cont}{eq:induction} are formulated in using an isothermal vertical disk structure. 
Recent studies by e.g. \cite{jankovic2021} show that the vertical temperature structure in the inner disk is highly dependent on viscous heating and therefore is also dependent on the vertical density structure, but should also be dependent on heating via non-thermal ionization, making upper layers hotter than the midplane.
Hence, the calculation of the ionization degree by taking the midplane gas temperature $T_\mathrm{gas,0}$ and using it as the vertically isothermal gas temperature at every height $z$ would vastly overestimate the vertical extent of the thermally ionized disk region. 
We therefore use a physically motivated vertical gas temperature profile \citep[see \equ{eq:vertical_Tgas}, for the physical background cf. e.g.][]{hubeny1990}.
The thermally ionized region in our models agrees reasonably well in both the radial and vertical extent with the vertically self-consistent 2D models of \cite{jankovic2021} and are therefore valid to describe the thermally ionized region. 
With regard to the discrepancy between the use of an isothermal vertical disk for hydrodynamical simulations while applying a vertical temperature profile for calculating the thermal ionization degree, we argue as follows: 
The vertical disk structure (apart from the magnetic field) is described in a 1+1D simulation by density-weighted variables (e.g. $\Sigma$, $P_\mathrm{gas}$). 
Therefore, the disk region around the midplane dominantly contributes to describing the vertical disk structure, which justifies using $T_\mathrm{gas,0}$ as the temperature of an isothermal vertical disk when performing hydrodynamical simulations.

\end{appendix}

\end{document}